\documentclass{houches}
\usepackage{epsfig,pstricks}
\input{psfig.tex}

\newcommand{\gae}{$\stackrel{>}{\sim}$}
\newcommand{\beq}{\begin{equation}}
\newcommand{\eeq}{\end{equation}}
\newcommand{\beqa}{\begin{eqnarray}}
\newcommand{\eeqa}{\end{eqnarray}}

\newcommand{\laem}{\begin{array}{c} < \vspace{-.7em} \\ {\scriptstyle \sim}
\end{array}}
\newcommand{\gaem}{\begin{array}{c} > \vspace{-.7em} \\ {\scriptstyle \sim}
\end{array}}

\def\slashchar#1{\setbox0=\hbox{$#1$}           
   \dimen0=\wd0                                 
   \setbox1=\hbox{/} \dimen1=\wd1               
   \ifdim\dimen0>\dimen1                        
      \rlap{\hbox to \dimen0{\hfil/\hfil}}      
      #1                                        
   \else                                        
      \rlap{\hbox to \dimen1{\hfil$#1$\hfil}}   
      /                                         
   \fi}                                         %

\newcommand{\tr}{{\rm Tr}}
\newcommand{\im}{{\rm i}}

\newcommand{\mev}{{\rm \, MeV}}
\newcommand{\gev}{{\rm \, GeV}}
\newcommand{\tev}{{\rm \, TeV}}
\newcommand{\W}{{\rm\bf W}}
\newcommand{\B}{{\rm\bf B}}

\newcmykcolor{purp}{.2 1 0 .4}
\newcmykcolor{grnn}{1 .2 .5 .4}
\newcmykcolor{mag}{0 1 0 0}
\newcmykcolor{orngg}{.1 .1 1 .2}
\newrgbcolor{brwn}{.5 .35 .1}

\begin{document}
\setcounter{chapter}{0}
\setcounter{page}{1}
\author[R. S. Chivukula]{R. Sekhar Chivukula
\thanks{e-mail: {\tt sekhar@bu.edu}, www: {\tt http://physics.bu.edu/$\sim$sekhar}}}
\address{Department of Physics\\
Boston University\\
590 Commonwealth Ave.\\
Boston, MA 02215 USA}
\editor{}
\title{}
\chapter{Models of Electroweak Symmetry Breaking{\thanks{\tt BUHEP-98-5}}}

\section{Introduction}

Discovering the dynamics responsible for electroweak symmetry breaking
is the outstanding question facing particle physics today, and the
answer will be found in the next decade. In these lectures\footnote{The
  material presented here on the implications of triviality and on
  models of dynamical electroweak symmetry breaking draws heavily on
  previous reviews, in particular see
  \protect\cite{Chivukula:1996rz,Chivukula:1996uy}.} I discuss the range
of models which have been proposed to explain electroweak symmetry
breaking. I begin with an overview of Higgs models, with emphasis on the
naturalness/hierarchy and triviality problems, and then consider general
lessons which can be drawn about the symmetry breaking sector in
arbitrary scalar models.  Subsequently, I discuss the symmetry breaking
sector in supersymmetric models and then consider models of dynamical
electroweak symmetry breaking. I conclude with a brief review of the
open questions.

\section{The Standard Model Higgs Boson}

\subsection{The Standard Model Higgs Sector}

In the standard \cite{Weinberg:1967pk,Salam:1968rm} one-doublet Higgs
model, one introduces a fundamental scalar particle
\beq
{\phi=\left(\matrix{\phi^+ \cr \phi^0 \cr}\right)}
{}~~~,
\eeq
which transforms as a $2_{+{1\over 2}}$ under $SU(2)_W \times U(1)_Y$.
In order to break the electroweak interactions to electromagnetism,
one introduces the potential
\beq
V(\phi)=\lambda \left(\phi^{\dagger}\phi - {v^2\over 2}\right)^2
\label{eq:pot}
{}~~~,
\eeq
which is minimized for nonzero $\langle \phi \rangle$, breaking the
electroweak gauge symmetry appropriately.

The potential in eq. (\ref{eq:pot}) is necessarily $SU(2)_W \times U(1)_Y$
invariant. In fact, the potential has an additional symmetry
\cite{Weinstein:1973gj} as well. To see this, define
${\tilde{\phi}=i\sigma_2 \phi^* } $ and consider the $2\times 2$ matrix
\beq
{\Phi = \left( {\tilde{\phi}}\ {\phi} \right)}
\ \ \rightarrow \ \ 
\Phi^\dagger\Phi=\Phi\Phi^\dagger=
{({\phi^\dagger\phi})\, {\cal I}}~.
\label{eq:reality}
\eeq
Because of the pseudo-reality of the doublet representation of
$SU(2)$, under $SU(2)_L \times U(1)_Y$,  $\Phi \rightarrow {L} \Phi {R^\dagger}$,
where
\beq
{L} = \exp\left({i{w^a(x)} \sigma^a \over 2}\right)
\ \ \ \&  \ \ \ {R} = \exp\left({i{b(x)} \sigma^3 \over 2}\right)~,
\eeq
with $L \in SU(2)_W$, $R \in U(1)_Y$, and where the $\sigma^a$ are the
Pauli matrices. Note that $R$ acts as a $\sigma^3$ rotation in an
$SU(2)$ acting on $\Phi$ by multiplication on the right.  In terms of
this new field definition, the lagrangian for the Higgs sector can be
written
\beq
{1\over 2}{\rm Tr}\left( D^\mu \Phi D_\mu \Phi^\dagger \right)
+ {\lambda \over 4}\left( {\rm Tr}\left(\Phi\Phi^\dagger\right) 
- v^2\right)^2~,
\label{eq:phi}
\eeq
with
\beq
D_\mu \Phi=\partial_\mu \Phi +\im g\W_\mu 
\Phi  -\im \Phi g'\B_\mu~,
\label{eq:dmu}
\eeq
where $\W_\mu = W^a_\mu \sigma^a/2$ and $\B_\mu = B_\mu \sigma^3/2$.
Note that the {\it potential} in eq. (\ref{eq:phi}) has a manifest
$SU(2)_L \times SU(2)_R$ global symmetry, larger than required by the
$SU(2)_W \times U(1)_Y$ gauge symmetry.

The effect of symmetry breaking is easily seen using
the ``polar decomposition'' of $\Phi$. By eq.
(\ref{eq:reality}) we see that $\Phi$ may be written
as a real scalar field times a unitary matrix
\beq
\Phi(x) = {1\over\sqrt{2}}\,({H(x)} + v)\, {\Sigma(x)}~,
\eeq
with
\beq
{\Sigma(x)} = \exp(i{\pi^a(x)}\sigma^a/v)~.
\eeq
In unitary gauge we may set $\langle \Sigma \rangle = {\cal I}$.  This
symmetry breaking reduces the gauge symmetry to electromagnetism, and
breaks the global $SU(2)_L \times SU(2)_R$ symmetry discussed above to a
vectorial $SU(2)_V$ ``custodial'' \cite{Sikivie:1980hm} symmetry.  For
every linearly-independent spontaneously broken global symmetry, there
must be a Goldstone boson. From the three broken symmetries above we
obtain the $\pi^\pm$ and $\pi^0 $ which, by the Higgs mechanism, become
the longitudinal components, $W^\pm_L$ and $Z_L$, of the weak gauge
bosons. The remaining degree of freedom, $H(x)$, is the Higgs field.

The mass of the $W$ can be calculated directly from the lagrangian in
eq. (\ref{eq:phi}) to be
\beq
M_W = {g v \over 2} \rightarrow v \approx 246 {\rm GeV}~.
\eeq
The mass of the neutral gauge boson is somewhat more complicated
since there is mixing between the neutral $SU(2)_W$ gauge boson
and the $U(1)_Y$ gauge boson. A complete analysis requires
computing the full mass-squared matrix of all four gauge bosons
\beq
M^2 = {v^2\over 2} \left(\matrix {g^2 &&&\cr
                          & g^2 &&\cr
                          && g^2 & - g g^\prime \cr
                          && - g g^\prime & {g^\prime}^2 \cr} \right)~,
\eeq
which results in 
\beq
\rho \equiv {M^2_W \over M^2_Z \cos^2\theta_W} = 1~.
\label{eq:rho}
\eeq 
This prediction, one of the first in modern electroweak theory, insures
the equality of the strength of deep-inelastic charged- and
neutral-current interactions and, after properly accounting for small
corrections discussed below, its validity has been verified to a
fraction of a percent.

It is important to appreciate that the {\it form} of the electroweak
gauge boson mass matrix and hence the relation eq. (\ref{eq:rho}) is a
consequence of the residual {$SU(2)_V$ custodial symmetry}. This
symmetry requires that, in the limit $g^\prime \to 0$, the $SU(2)_L$
gauge bosons must be degenerate, since they form an irreducible
representation of the custodial group.  This, plus the constraint that
$m_\gamma = 0$, insures that the gauge boson mass matrix is proportional
to the matrix shown above.

\subsection{Violations of Custodial Symmetry}

Custodial $SU(2)_V$ is a symmetry of the Higgs potential, but {\it not} of Higgs
interactions. For example, the hypercharge interactions ({\it cf.} eq.
(\ref{eq:dmu})), and hence electromagnetism violate custodial symmetry.
Therefore there are contributions to $\Delta\rho \equiv \rho -1$
\beq
{\lower10pt\hbox{\epsfysize=1.25cm\epsfbox{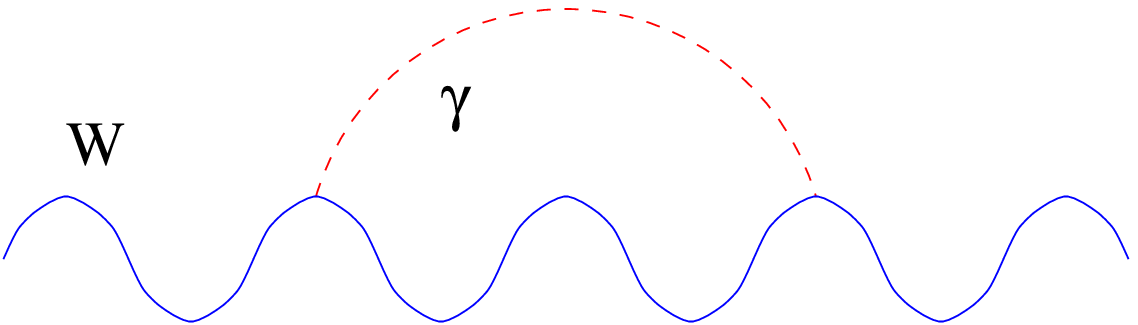}}}
\eeq
of ${\cal O}(\alpha)$. Furthermore, the Higgs boson must
couple to the ordinary fermions in order to give rise to their
observed masses. Written in terms of $\Phi$, the Yukawa couplings
of the scalar doublet to the third generation is
\beq
\left(\matrix{\bar{t}_L & \bar{b}_L \cr}\right) \Phi \left(\matrix{y_t&\cr &y_b\cr}\right) 
\left(\matrix{t_R\cr b_R}\right)~,
\eeq
and violates custodial $SU(2)_V$ since 
\beq
y_t \equiv {{\sqrt{2}m_t}\over v} \gg y_b \equiv {{\sqrt{2}m_b}\over v}
~.
\eeq
Contributions to the gauge-boson self-energies from
\beq
{\lower10pt\hbox{
\epsfysize=1.0cm\epsfbox{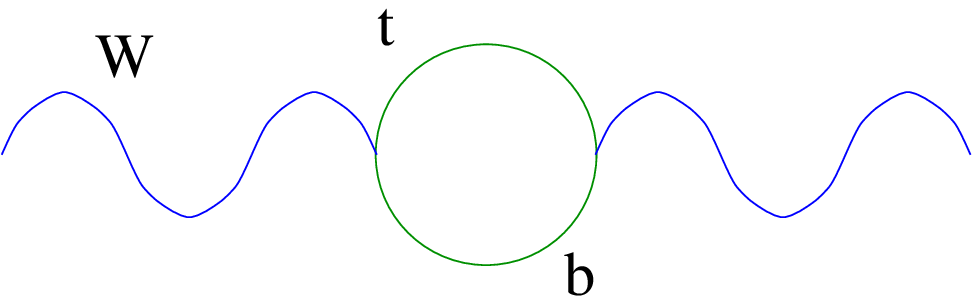}
\hskip0.5cm
\epsfysize=1.0cm\epsfbox{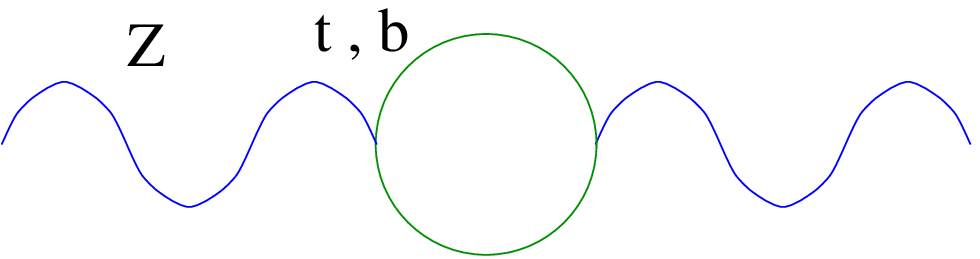}}}
\eeq
give rise to
\beq
\Delta \rho \approx {3 y^2_t \over 32\pi^2}\approx 1\%\, \left({m_t
  \over 175 {\rm GeV}}\right)^2\,.
\eeq

It is also important to note that custodial $SU(2)_V$ is an {\it accidental}
symmetry: it is a symmetry of all $SU(2)_L \times U(1)_Y$ invariant
terms of dimension 4 or less in the Higgs sector of the lagrangian in
the limit $g^\prime \to 0$.  It can be violated by terms of higher
dimension \cite{Buchmuller:1986jz,Grinstein:1991cd} arising from physics
at some higher scale, {\it e.g.}
\beq
(\phi^\dagger D^\mu \phi)(\phi^\dagger D_\mu \phi)
= {1\over 4} \left({\rm Tr}\,\sigma_3 \Phi^\dagger D^\mu \Phi \right)
\left({\rm Tr}\, \sigma_3 \Phi^\dagger D_\mu \Phi \right)~.
\eeq

\subsection{The Higgs Boson}

\begin{figure}
\epsfxsize 7cm \centerline
{\epsffile{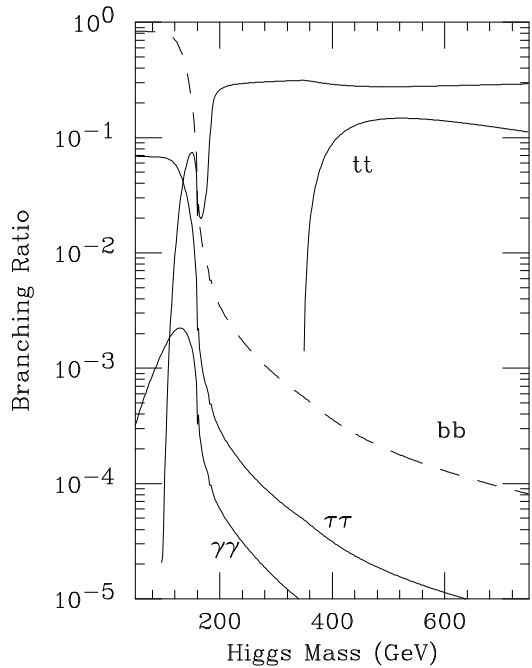}}
\caption{Higgs decay branching ratios as a function of Higgs
mass. All decays proceed through the couplings in eq. (\protect\ref{eq:higgscoup}),
with the exception of the $\gamma \gamma$ mode which occurs through
a fermion or gauge-boson loop. From ref. \protect\cite{Hinchliffe:1996we}.}
\label{fig:one}
\end{figure}

At tree-level, the Higgs Boson mass can be calculated directly
from the potential
\beq
m^2_H = 2 \lambda v^2~.
\eeq
As the masses of all of the standard model particles
arise from the vacuum expectation value of $\phi$,
the tree-level couplings of the Higgs are also immediately
determined to be
\beqa
{\cal L} \supset \left( 1 + {H\over v}\right)^2
& \left[M^2_W W^{\mu +}W_\mu^- +
  {1\over 2} M^2_Z Z^\mu Z_\mu   \right] \nonumber \\
& - \left( 1 + {H\over v}\right) 
\left[ \sum_i m_i \bar{\psi_i}\psi_i \right]~. 
\label{eq:higgscoup}
\eeqa
As a function of the Higgs boson mass, we then find the
branching ratios shown in fig. \ref{fig:one}.

\begin{figure}
\centering
\epsfig{file=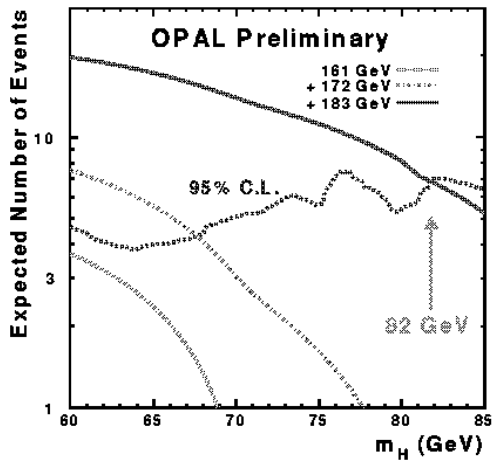,width=7cm}
\caption{OPAL limits on the standard model Higgs boson. As shown,
the preliminary bound from the run at 183 GeV imply that $m_H > 82$ GeV
at 95\% confidence level. From ref. \protect\cite{Honma:1997}.}
\label{fig:two}
\end{figure}

The best direct experimental limits on the Higgs boson come from the
non-observation of the process $e^+ e^-\to Z^* \to ZH$ at LEPII.  Recent
results, shown in fig. \ref{fig:two}, give a lower bound of 82 GeV on
the Higgs boson mass at 95\% confidence level. Future observations at
the LEPII and the Tevatron could discover a Higgs boson up to a mass of
order 120-130 GeV \cite{Bagley:1996ne,Haber:1996qb}.

\begin{figure}
\epsfxsize 7cm \centerline
{\epsffile{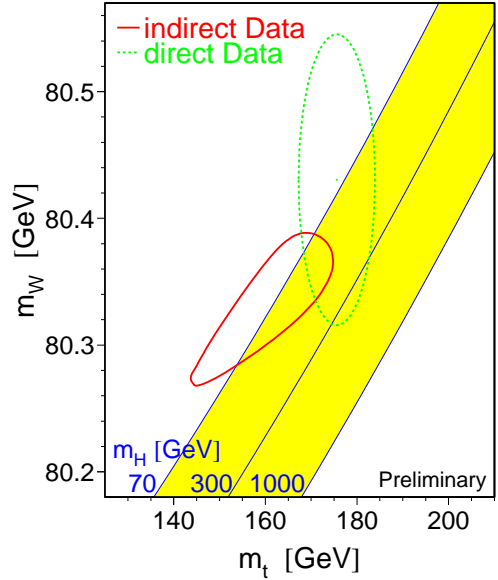}}
\caption{The allowed region in the  $(m_t,M_W)$ plane coming from precision electroweak
  tests at 68\% confidence level (solid curve) and from direct measurement
  (dashed curve).  Also shown is the standard model prediction of $M_W$
  as a function of $m_t$ for various Higgs boson masses from 60 GeV to
  1000 GeV. From ref. \protect\cite{LEPEWWG:1997}}
\label{fig:three}
\end{figure}

Precision measurements of electroweak quantities at LEP, SLC, and in
low-energy experiments can also, in principle, constrain the allowed
values of $m_H$. At one-loop, the relationship between $G_F$,
$\alpha_{em}$, $M_Z$ and any electroweak observable depends on the Higgs
mass and the top-quark mass. The collection of precision measurements
can then give an ``allowed'' region of top-quark and Higgs masses. This
allowed region can be illustrated in the $(m_t, M_W)$ plane, as shown in
fig. \ref{fig:three}. The solid curve gives the bounds at 68\%
confidence level coming from precision electroweak tests, the dashed
curve gives the 68\% bounds on $m_t$ coming from measurements at CDF and
D\O, and on $M_W$ coming from measurements at these experiments and
experiments at LEPII. The shaded band shows the predictions of the
standard model for Higgs masses between 60 and 1000 GeV.
The consistency of the standard model prediction and the indirect and
direct measurements of $m_t$ and $M_W$ is a remarkable triumph of the
standard model. Nonetheless, at 95\% or 99\% confidence level, these
measurements do not currently provide a significant constraint on the 
Higgs boson mass. 

\begin{figure}
\epsfxsize 7cm \centerline
{\epsffile{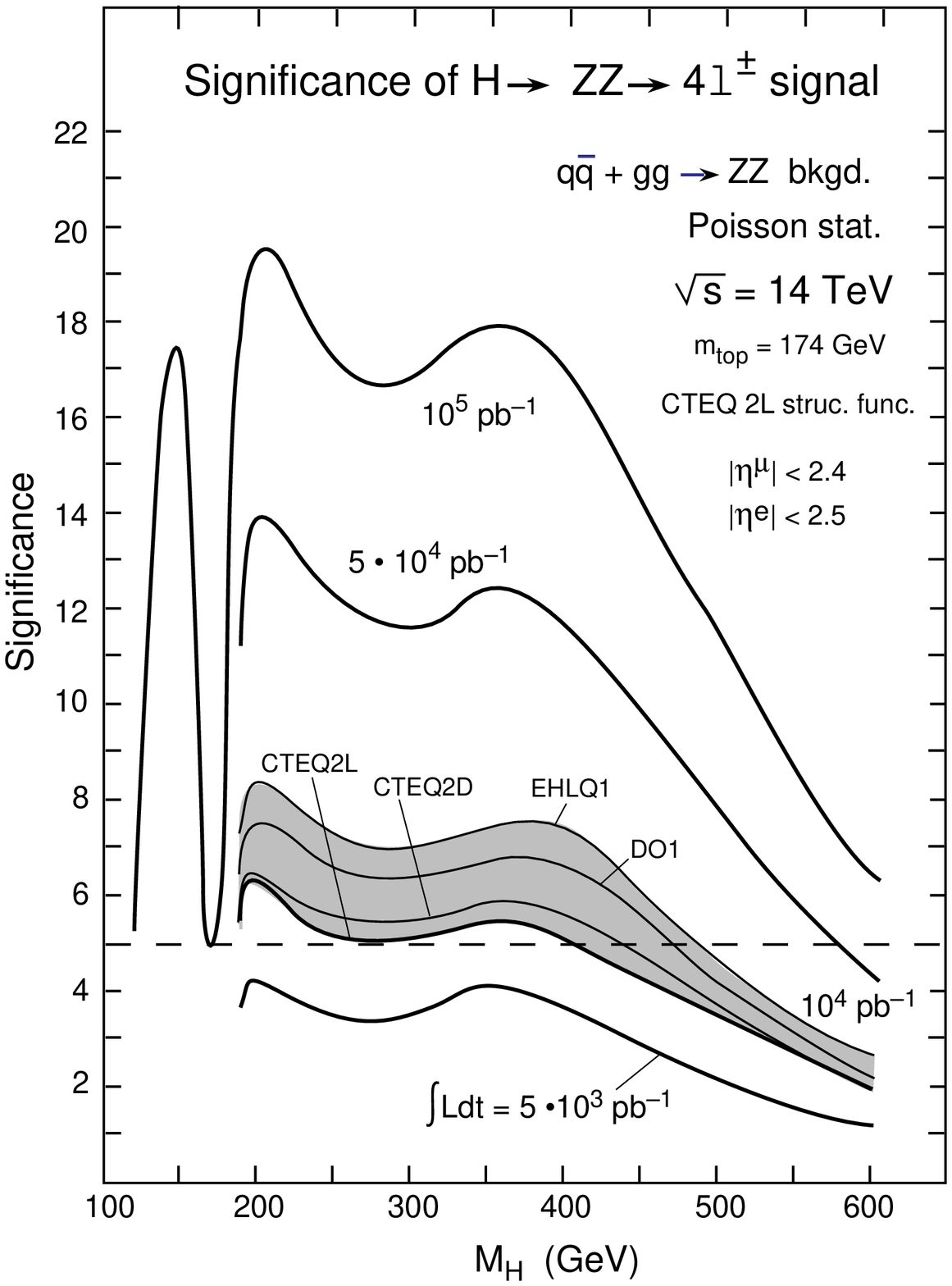}}
\caption{Significance of Higgs boson discovery signal in the channel
$H \to ZZ\to 4 \ell$ for various integrated luminosities ({\it n.b.}
$10^5$ pb$^{-1}$ equals one year at design luminosity).
From ref. \protect\cite{Bomestar:1995mu}.}
\label{fig:four}
\end{figure}

The {\it raison d'\^etre} of the LHC is to uncover the agent of
electroweak symmetry breaking. The experimental prospects for discovery
of a Higgs boson in the ``gold-plated'' $H \to ZZ\to 4\ell$ mode is
shown in fig. \ref{fig:four}.
When other modes, in particular $H \to ZZ \to 2\ell 2\nu$ and $H \to WW
\to \ell\nu jj$ for a heavy Higgs boson and $H \to \gamma \gamma$ for a
light Higgs boson, are considered, the LHC should be able to discover
a Higgs boson with a mass ranging from the ultimate LEP II/Tevatron
limit ($\sim$ 100 -- 130 GeV) to 800 GeV \cite{Haber:1996qb}.

We can obtain a theoretical {\it upper} bound on the Higgs
boson mass from unitarity \cite{Veltman:1977rt,Lee:1977yc}.
Consider (formally) the limit $\lambda \to \infty$. The Higgs
degree of freedom becomes heavy and may formally be ``integrated
out'' of the theory:
\beq
\Phi(x) = {1\over\sqrt{2}}\,({H(x)} + v)\, {\Sigma(x)}
\to {v\over\sqrt{2}}{\Sigma(x)}~.
\eeq
In this limit, the Higgs sector is equivalent to an
effective chiral theory for the symmetry breaking pattern $SU(2)_L
\times SU(2)_R$ $\to$ $SU(2)_V$. Allowing for custodial $SU(2)$
violation, the most general such effective lagrangian
\cite{Chanowitz:1987vj,Chanowitz:1986hu} at ${\cal O}(p^2)$ may be
written:
\beq
{  {v^2 \over4} \tr\left[D^\mu
   \Sigma^{\dagger}D_\mu \Sigma\right] +  {v^2 \over 4} ({1\over \rho} - 1) 
\left[ \tr \sigma_3 \Sigma^\dagger D^\mu \Sigma \right]^2}~.
\eeq
Setting $\Sigma=1$ in unitary gauge, we find
\beq
{g^2 v^2\over 4} W_-^\mu W_{\mu+} + {g^2 v^2\over{8 \rho \cos^2\theta}}
Z^\mu Z_\mu
~~~,
\eeq
as required.

Now consider what happens when these massive $W$ and $Z$ bosons scatter.
At {high-energies} we can use the {equivalence theorem}
\cite{Cornwall:1974km,Vayonakis:1976vz,Chanowitz:1985hj}
\beq
{\cal A}(W_L W_L) = {\cal A}(\pi \pi) + {\cal O}({M_W\over E})\, .
\eeq
to reduce the problem of longitudinal gauge boson ($W_L$) scattering to
the corresponding problem of the scattering of the Goldstone bosons
which would be present in the absence of the weak gauge interactions.
At ${\cal O}(p^2)$ results in \cite{Chanowitz:1987vj,Chanowitz:1986hu}
universal low-energy theorems:
\beqa 
{\cal M}[W^+_L W^-_L \to W^+_L W^-_L] =& {{\textstyle i u} 
\over {\textstyle {v^2 \rho}}}  \nonumber\\
{\cal M}[W^+_L W^-_L \to Z_L Z_L] =& 
{{\textstyle i s} \over {\textstyle v^2}} \left( 4 -
{{\textstyle 3}\over{\textstyle \rho}}\right) \nonumber \\
{\cal M}[Z_L Z_L \to Z_L Z_L] =& 0~~~.
\label{eq:universal}
\eeqa
Note that these amplitudes grow as the square of the center-of-mass
collision energy. Projecting onto the $I=J=0$ channel for $\rho=1$,
we find
\beq
{\cal A}_{00} = {s \over 16 \pi v^2} 
\approx \left({\sqrt{s}\over 1.8\, {\rm TeV}} \right)^2~.
\label{quigg}
\eeq
Thus, in the absence of additional contributions, the isosinglet
scalar scattering amplitude would violate unitarity at an energy scale of
approximately 1.8 TeV.  In the standard Higgs model, Higgs exchange
unitarizes the cross section. From this we conclude
\cite{Veltman:1977rt,Lee:1977yc} that {$m_H \laem 1.8$ TeV}.

We can also obtain a theoretical {\it lower bound} on the Higgs mass from
vacuum stability \cite{Weinberg:1976pe,Linde:1977mm,Witten:1981ez}.
To investigate this effect, we must compute the ``effective
potential'' \cite{Coleman:1973tz}, the sum of all one-particle
irreducible diagrams in the presence of a constant external field
background:
\beq
V(\phi) = V_0(\phi) + V_1(\phi) + \ldots
\eeq
with, for example,
\beq
V_1(\phi) = \lower4pt\hbox{\epsfxsize 1.0cm {\epsffile{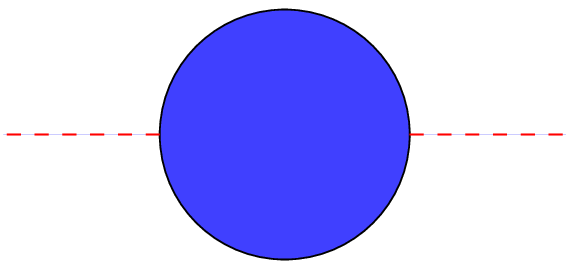}}}
+ \lower12pt\hbox{\epsfxsize 1.0cm {\epsffile{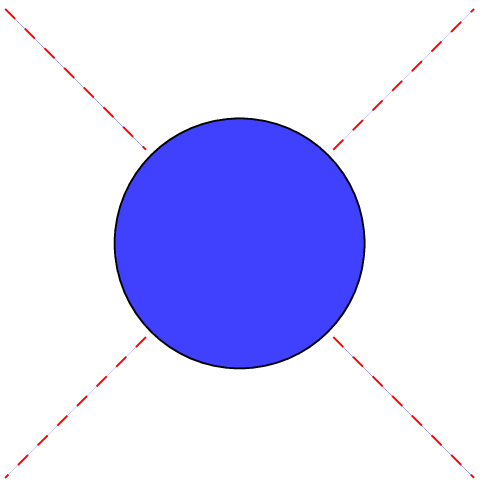}}}
+\ldots
\eeq
In the leading-log\footnote{Leading in $\log(\phi/M)$, where $M$ is an
  arbitrary renormalization point.} approximation the effective
potential at large field-values may be written
\cite{Coleman:1973tz,Yamagishi:1981qq}
\beq
V^{eff}(\phi,M) \approx \tilde{\lambda}(t)(\phi^\dagger \phi)^2
\exp{\left(4\int_0^t dw {\gamma(w)\over{1-\gamma(w)}}\right)}
\eeq
where $t=\log(\phi/M)$, $\gamma$ is the anomalous dimension of $\phi$, and
\beq
{d\tilde{\lambda}\over dt} = {\beta_\lambda \over {1-\gamma}}~.
\eeq
This equation allows for a nice geometrical description of the
Coleman-Weinberg mechanism \cite{Yamagishi:1981qq}.

\begin{figure}
\epsfxsize 7cm \centerline
{\epsffile{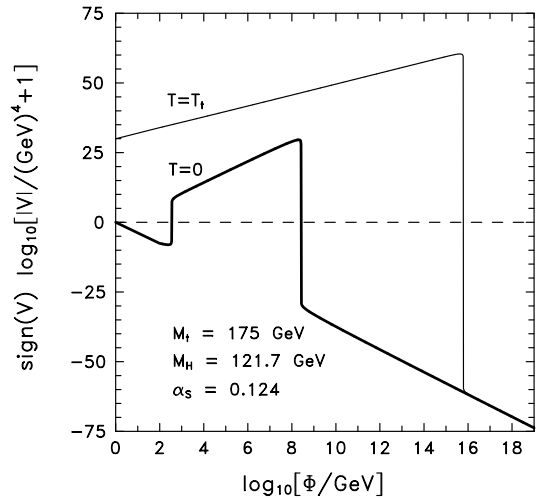}}
\caption{Effective potential in leading-log approximation
for $m_H= 121.7$ GeV.
From ref. \protect\cite{Quiros:1997vk}.}
\label{fig:five}
\end{figure}

\begin{figure}
\epsfxsize 7cm \centerline
{\epsffile{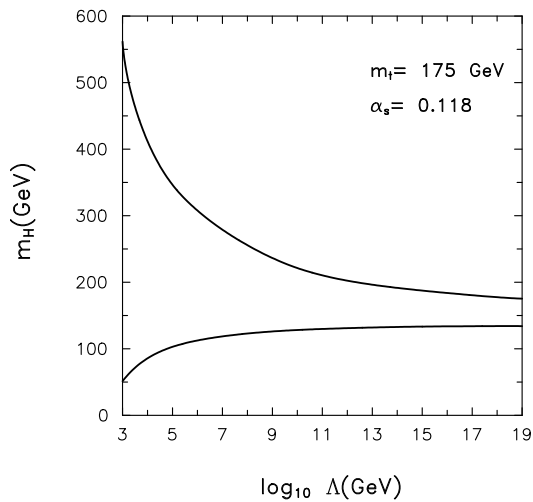}}
\caption{The lower curve graphs the lower bound on the
  Higgs mass as a function of energy scale $\Lambda$ obtained by
  requiring that the potential be stable out to field values of order
  $\Lambda$. The upper curve graphs the upper bound on the Higgs mass as
  a function of energy scale $\Lambda$ obtained by requiring that the
  Landau pole associated with the self-coupling $\lambda$ occur at a
  scale greater than $\Lambda$.  From ref.
  \protect\cite{Quiros:1997vk}.}
\label{fig:six}
\end{figure}

At one loop, neglecting terms involving $\lambda$ (since we are
investigating the possibility of a light Higgs boson) and light fermions
\beq
\beta_\lambda \approx {3 \over 128\pi^2}\left[3g^4 + 2 g^2 {g'}^2
+ {g'}^4  { - 16 y^4_t} \right]~.
\eeq
Because of the large top-quark mass (with Yukawa coupling $y_t$), for
small $m_H$ (and hence small $\lambda$) the perturbative vacuum is
unstable at large $\phi$.  This instability is illustrated in fig.
\ref{fig:five}.
If we require stability up to a scale $\Lambda$, where new-physics
enters and additional terms $\propto (\phi^\dagger
\phi)^n/\Lambda^{2n-4}$ can enter to stabilize the potential, we find
a {\it lower} bound on $m_H$ as a function of this new energy scale.
This bound is shown in in the lower curve in figure \ref{fig:six}.

\section{Triviality and its Implications}

While the standard model is simple and renormalizable, it has a number
of shortcomings\footnote{Much of this section appeared originally in
  \protect\cite{Chivukula:1996uy}.}. First, while the theory can be
constructed to accommodate the breaking of electroweak symmetry, it
provides no {\it explanation} for it. One simply assumes that the
potential is of the form in eq. (\ref{eq:pot}). In addition, in the
absence of supersymmetry, quantum corrections to the Higgs mass are
naturally of order the largest scale in the theory
\beq
{\lower5pt\hbox{\epsfysize=0.25 truein \epsfbox{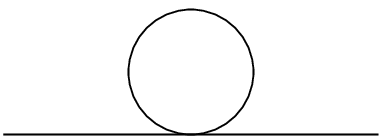}}}
\ \rightarrow  \ m_H^2 \propto \Lambda^2~,
\eeq
leading to the hierarchy and naturalness problems \cite{'tHooft:1980xb}.
Finally, the $\beta$ function for the self-coupling $\lambda$ is positive
\beq
{\lower7pt\hbox{\epsfysize=0.25 truein \epsfbox{figures/beta.eps}}}
\ \rightarrow \ \beta = {3\lambda^2 \over 2 \pi^2} \, > \, 0
{}~,
\eeq
leading to a ``Landau pole'' and triviality
\cite{Wilson:1971dh,Wilson:1974dg}.

\subsection{The Wilson Renormalization Group and Naturalness}

\begin{figure}[tbp]
\centering
\epsfysize=2in
\hspace*{0in}
\epsffile{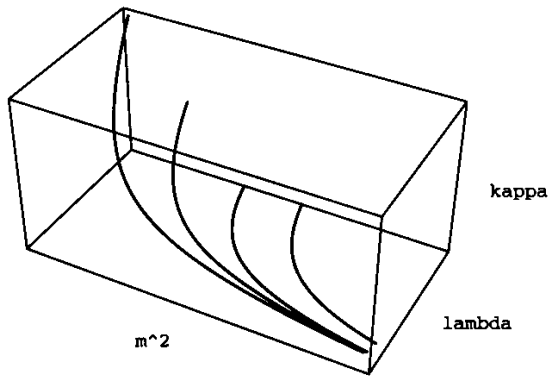}
\caption{Renormalization group flow of Higgs mass $m^2$, Higgs
self-coupling $\lambda$, and the coefficient of a representative
irrelevant operator $\hat{\kappa}$. The flows go from upper-left to
lower-right as one scales to the infrared.}
\label{fig:seven}
\end{figure}

The hierarchy,naturalness, and triviality problems can be nicely
summarized in terms of the Wilson renormalization group
\cite{Wilson:1971bg,Wilson:1971dh}.  Define the theory with a fixed
UV-cutoff:
\beqa
{\cal L}_\Lambda =  & D^\mu \phi^\dagger D_\mu \phi + 
m^2(\Lambda)\phi^\dagger \phi 
+ {\lambda(\Lambda)\over 4}(\phi^\dagger\phi)^2 \nonumber\\
& + {\hat{\kappa}(\Lambda)\over 36\Lambda^2}(\phi^\dagger\phi)^3+\ldots  
\label{eq:liz}
\eeqa
Here $\hat{\kappa}$ is the coefficient of a representative 
irrelevant operator, 
of dimension greater than four.
Next, integrate out states with $\Lambda^\prime < k < \Lambda$,
and construct a new lagrangian with the same {\it
low-energy} Green's functions:
\beqa
{\cal L}_\Lambda & \rightarrow & {\cal L}_{\Lambda^\prime} \nonumber\\
m^2(\Lambda)& \rightarrow & m^2(\Lambda^\prime) \nonumber \\
\lambda(\Lambda) & \rightarrow & \lambda(\Lambda^\prime) \nonumber \\
\hat{\kappa}(\Lambda) & \rightarrow & \hat{\kappa}(\Lambda^\prime)  
\eeqa
The low-energy behavior of the theory is then nicely summarized in terms
of the evolution of couplings in the infrared.\footnote{For convenience,
  we ignore the corrections due to the weak gauge interactions.  In
  perturbation theory, at least, the presence of these interactions does
  not qualitatively change the features of the Higgs sector.} A
three-dimensional representation of this flow in the
infinite-dimensional space of couplings is shown in Figure \ref{fig:seven}.

From Figure \ref{fig:seven}, we see that as we scale to the infrared the
coefficients of irrelevant operators, such as $\hat{\kappa}$, tend to
zero; {\it i.e.} the flows are attracted to the finite dimensional
subspace spanned (in perturbation theory) by operators of dimension four
or less; this is the modern understanding of {\it renormalizability}. 

On the other hand, the coefficient of the only {\it relevant} operator
(of dimension 2), $m^2$, tends to infinity. In the absence of a symmetry
that protects the scalar mass (such as supersymmetry, see section 6
below), it is natural for the mass to be proportional to the largest
scale present in the theory \cite{'tHooft:1980xb}. This is the
naturalness problem: since we want $m^2 \propto v^2$ at low energies we
must adjust the value of $m^2(\Lambda)$ to a precision of
\beq
{\Delta m^2(\Lambda) \over m^2(\Lambda)} \propto {v^2 \over \Lambda^2}~.
\eeq

We are sure that a large hierarchy of scales {\it does} exist between
the electroweak scale and the grand-unified or Planck scales. We expect
that, even in the presence of an extra symmetry which stabilizes the
Higgs mass, there should be some {\it dynamical} explanation for the
this large hierarchy. The construction of a model which does not suffer
from the naturalness and hierarchy problems will motivate the discussion
presented in sections 6 through 12.

\subsection{Implications of Triviality}

Central to our discussion here is the fact that the coefficient of the
only marginal operator, $\lambda$, tends to zero because of the positive
$\beta$ function.  If we try to take the continuum limit, $\Lambda \to
+\infty$, the theory becomes free (or trivial)
\cite{Wilson:1971dh,Wilson:1974dg}, and could not result in the observed
symmetry breaking.

The triviality of the scalar sector
of the standard one-doublet Higgs model implies that this theory is only
an effective low-energy theory valid below some cut-off scale $\Lambda$.
Physically this scale marks the appearance of new strongly-interacting
symmetry-breaking dynamics.  Examples of such high-energy theories
include ``top-mode'' standard models
\cite{Miranskii:1989ds,Miranskii:1989xi,Nambu:1989jt,Marciano:1989xd,Bardeen:1990ds,Hill:1991at,Cvetic:1997eb},
which we discuss in section 12,
and composite Higgs models
\cite{Kaplan:1984fs,Kaplan:1984sm,Dugan:1985hq}. As the Higgs mass
increases, the upper bound on the scale $\Lambda$ decreases.  An
estimate of this effect can be obtained by integrating the one-loop
$\beta$-function, which yields
\beq
\lambda(m_H) \stackrel{<}{\sim} {{2\pi^2}\over 3\log{\Lambda\over m_H}}\, .
\label{eq:est}
\eeq
Using the relation $m^2_H = 2\lambda(m_H) v^2$ we find 
\beq 
m^2_H \ln\left({\Lambda\over m_H}\right)\le {4\pi^2 v^2 \over 3}~.
\label{estimate}
\eeq 
Hence a lower bound\,\cite{Cabibbo:1979ay,Dashen:1983ts} on $\Lambda$
yields an upper bound on $m_H$. We must require that $m_H / \Lambda$ in
eq.~(\ref{estimate}) be small enough to afford the effective Higgs
theory some range of validity (or to minimize the effects of
regularization in the context of a calculation in the scalar theory).

Non-perturbative
\cite{Kuti:1988nr,Luscher:1989uq,Hasenfratz:1987eh,Hasenfratz:1989kr,Bhanot:1990zd,Bhanot:1991ai}
studies on the lattice using analytic and Monte Carlo techniques result
in an upper bound on the Higgs mass of approximately 700 GeV.  The
lattice Higgs mass bound is potentially ambiguous because the precise
value of the bound on the Higgs boson's mass depends on the (arbitrary)
restriction placed on $M_H / \Lambda$.  The ``cut-off'' effects arising
from the regulator are not universal: different schemes can give rise to
different effects of varying sizes and can change the resulting Higgs
mass bound.

On the other hand, we show below that, for models that reproduce the
standard one-doublet Higgs model at low energies, electroweak and flavor
phenomenology provide a lower bound on the scale $\Lambda$ of order 10
-- 20 TeV. This limit is regularization-independent (i.e.  independent
of the details of the underlying physics). Using eq.~(\ref{estimate}) we
estimate that this gives an {\it upper} bound of 450 -- 500 GeV on the
Higgs boson mass.  The discussion we will present is based on
perturbation theory and is valid in the domain of attraction of the
``Gaussian fixed point'' ($\lambda=0$).  In principle, however, the
Wilson approach can be used {\it non-perturbatively}, even in the
presence of nontrivial fixed points or large anomalous dimensions.  In a
conventional Higgs theory, neither of these effects is thought to occur
\cite{Kuti:1988nr,Luscher:1989uq,Hasenfratz:1987eh,Hasenfratz:1989kr,Bhanot:1990zd,Bhanot:1991ai}.

\subsection{Dimensional Analysis}

We will analyze the effects of the underlying physics by estimating the
sizes of various operators in a low-energy effective lagrangian
containing the (presumably composite) Higgs boson and the ordinary gauge
bosons and fermions. Since we are considering theories with a heavy
Higgs field, we expect that the underlying high-energy theory will be
strongly interacting. Borrowing a technique from QCD we will rely on
dimensional analysis\,\cite{Weinberg:1979kz,Manohar:1984md} to estimate
the sizes of various effects of the underlying physics.

A strongly interacting theory has no small parameters.  As noted by
Georgi \cite{Georgi:1993dw}, a theory\footnote{These dimensional
  estimates only apply if the low-energy theory, when viewed as a scalar
  field theory, is defined about the infrared-stable Gaussian
  fixed-point.  For a discussion of possible ``non-trivial'' theories,
  see \protect\cite{Chivukula:1996rz}.} with light scalar particles
belonging to a single symmetry-group representation depends on two
parameters: $\Lambda$, the scale of the underlying physics, and $f$ (the
analog of $f_\pi$ in QCD), which measures the amplitude for producing
the scalar particles from the vacuum. Our estimates will depend on the
ratio $\kappa = \Lambda / f$, which is expected to fall between 1 and
$4\pi$.

Consider the kinetic energy of a scalar bound-state in the appropriate
low-energy effective lagrangian. The properly normalized kinetic energy
is
\beq
\partial^\mu \phi^\dagger \partial_\mu \phi
= {\Lambda^2 f^2} \left({\partial^\mu \over { \Lambda}}\right)
\left({\phi^\dagger \over { f}}\right)
\left({\partial_\mu \over { \Lambda}}\right) 
\left({\phi \over { f}}\right)~.
\eeq
Here, because the fundamental scale of the interactions is $\Lambda$, we
ascribe a $\Lambda$ to each derivative, and we associate an $f$ with each
$\phi$ since $f$ measures the amplitude to produce the bound state. This
tells us that the overall magnitude of each term in the effective
lagrangian is ${\cal{O}}(f^2\Lambda^2)$.  We can next estimate the
``generic'' size of a mass term in the effective theory:
\beq
m^2 \phi^\dagger \phi = {\Lambda^2 f^2} \left({\phi^\dagger 
\over {f}}\right)
\left({\phi \over {f}}\right) \ \rightarrow\  
{m^2 \propto \Lambda^2}~.
\eeq
This is the hierarchy problem in a nutshell.  In the absence of some
other symmetry not accounted for in these rules, fine-tuning\footnote{We
  will not be addressing the solution of the hierarchy problem here; we
  will simply assume that some other symmetry or dynamics has produced
  the scalar state with a mass of order the weak scale.} is required to
obtain $m^2 \ll \Lambda^2$. Next, consider the size of scalar
interactions.  From the simplest interaction
\beq
\lambda (\phi^\dagger \phi)^2 \ \rightarrow \ { \lambda 
\propto \left({\Lambda \over f}\right)^2 = \kappa^2}~,
\eeq
we see that $\kappa$ will determine the size of coupling
constants. Similarly, for a higher-dimension interaction such as
the one in eq.~(\ref{eq:liz}) we find
\beq
{\hat{\kappa} \over { \Lambda^2}}(\phi^\dagger \phi)^3 \ \rightarrow\ 
{ \hat{\kappa} \propto \kappa^4}~.
\eeq

These rules are easily extended to include strongly-interacting fermions
self-consistently.  Again, we start with the properly normalized
kinetic-energy
\beq
\bar{\psi}\slashchar{\partial}\psi = 
{\Lambda^2 f^2} \left({\bar{\psi} \over {f\sqrt{\Lambda}}}\right)
\left({\slashchar{\partial} \over {\Lambda}}\right)
\left({\psi \over {f\sqrt{\Lambda}}}\right)~,
\eeq
and learn that $f\sqrt{\Lambda}$ is a measure of the amplitude
for producing a fermion from the vacuum. Next, consider
a Yukawa coupling of a strongly-interacting fermion to
our composite Higgs,
\beq
y (\bar{\psi}\phi \psi) \ \rightarrow\  { y \propto \kappa}~.
\label{eq:natyukawa}
\eeq
And finally, the natural size of a four-fermion operator is
\beq
{\nu \over { \Lambda^2}} (\bar{\psi}\psi)^2 \ \rightarrow\  
{ \nu \propto \kappa^2}~.
\label{eq:grhoex}
\eeq

We will rely on these estimates to derive bounds on the scale $\Lambda$.
By way of justification, we note that these estimates work in QCD for
the chiral lagrangian \cite{Weinberg:1979kz,Manohar:1984md}, with $f \to
f_\pi$, $\Lambda \to 1$ GeV, and $\kappa \approx {\cal{O}}(4 \pi)$.  For
example, four-nucleon operators of the form shown in
eq.~(\ref{eq:grhoex}) arise in the vector channel from $\rho$-exchange
and we obtain $\Lambda = m_\rho$ and $\kappa = g_\rho \approx 6$.  In a
QCD-like theory with $N_c$ colors and $N_f$ flavors one
expects\,\cite{Chivukula:1992nw} that
\beq
\kappa \approx \min \left({4\pi a\over N_c^{1/2}},
{4\pi b\over N_f^{1/2}}\right)~,
\eeq
where $a$ and $b$ are constants of order 1. In the results that
follow, we will display the dependence on $\kappa$ explicitly; when
giving numerical examples, we set $\kappa$ equal to the geometric mean
of 1 and $4\pi$, {\it i.e.} $\kappa \approx 3.5$.

\subsection{Isospin Violation and Bounds on $m_H$}

Because of the $SU(2)_W \times U(1)_Y$ symmetry of the low-energy
theory, all terms of dimension less than or equal to four respect
custodial symmetry\,\cite{Weinstein:1973gj,Sikivie:1980hm}.  The leading
custodial-symmetry violating operator is of dimension
six\,\cite{Buchmuller:1986jz,Grinstein:1991cd} and involves four Higgs
doublet fields $\phi$. According to the rules of dimensional analysis,
the operator
\beq
{\lower35pt\hbox{\epsfysize=1.0 truein \epsfbox{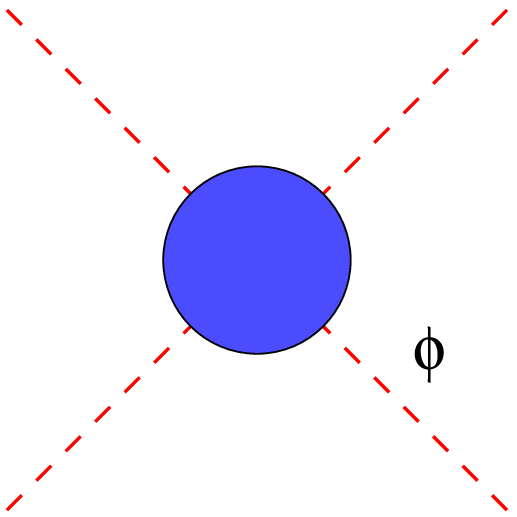}}}
\rightarrow
{\kappa^2 \over \Lambda^2} 
(\phi^\dagger D^\mu \phi)
(\phi^\dagger D_\mu \phi)~,
\label{eq:isoviol}
\eeq
should appear in the low-energy effective theory with a coefficient of
order one\,\cite{Grinstein:1991cd}. Such an operator will give rise to a
deviation
\beq
\Delta \rho_* = - {\cal O}\left(\kappa^2 {v^2 \over \Lambda^2}\right) ~,
\eeq
where $v \approx 246$ GeV is the expectation value of the Higgs
field. Imposing the constraint\,\cite{Barnett:1996hr,Chivukula:1995dc} 
that $|\Delta \rho_*| \le 0.4\%$, we find the lower bound
\beq
\Lambda \stackrel{>}{\sim} 4\, {\rm TeV} \cdot \kappa ~.
\eeq
For $\kappa \approx 3.5$, we find $\Lambda \stackrel{>}{\sim} 14$ TeV. 

Alternatively, it is possible that the underlying strongly-interacting
dynamics respects custodial symmetry. Even in this case, however, there
must be custodial-isospin-violating physics (analogous to
extended-technicolor\,\cite{Eichten:1979ah,Dimopoulos:1979es} interactions) which couples the
$\psi_L=(t,\ b)_L$ doublet and $t_R$ to the strongly-interacting
``preon'' constituents of the Higgs doublet in order to produce a top
quark Yukawa coupling at low energies and generate the top quark mass.
If, for simplicity, we assume that these new weakly-coupled
custodial-isospin-violating interactions are gauge interactions with
coupling $g$ and mass $M$, dimensional analysis allows us to estimate
the size of the resulting top quark Yukawa coupling.  The ``natural
size'' of a Yukawa coupling (eq.~(\ref{eq:natyukawa})) is $\kappa$ and
that of a four-fermion operator (eq.~(\ref{eq:grhoex})) is
$\kappa^2/\Lambda^2$; the ratio $(g^2/M^2)/(\kappa^2/\Lambda^2)$ is the
``small parameter'' associated with the extra flavor interactions and we
find
\beq {\lower35pt\hbox{\epsfysize=1.0 truein
    \epsfbox{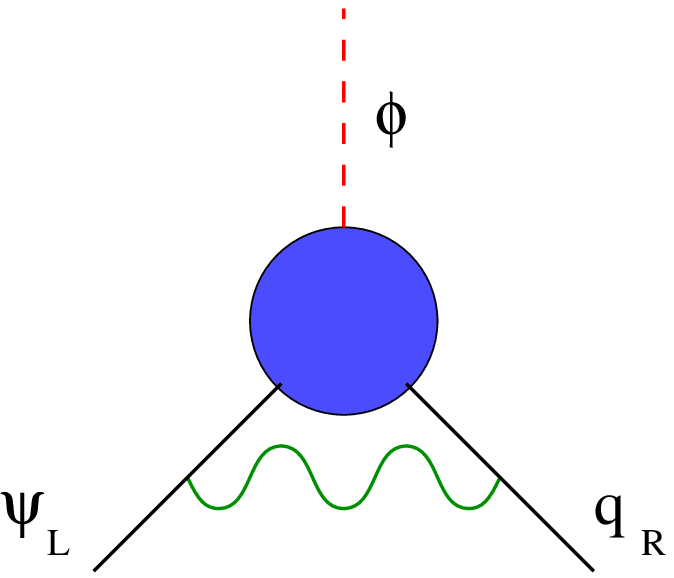}}} \rightarrow {g^2 \over M^2} {\Lambda^2 \over
  \kappa}\bar{q}_R \phi \psi_L ~.
\label{eq:quarkpreon}
\eeq
In order to give rise to a quark mass $m_q$, the 
Yukawa coupling must be equal to
\beq
{\sqrt{2} m_q \over v}
\eeq
where $v\approx 246$ GeV. This implies
\beq
\Lambda \stackrel{>}{\sim}{M \over g} \sqrt{\sqrt{2} \kappa {m_q \over v}}~.
\label{eq:yukawa}
\eeq 

These new gauge interactions will typically also give rise to
custodial-isospin-violating 4-preon interactions\footnote{These
  interactions have previously been considered in the context of
  technicolor theories.\cite{Appelquist:1984nc,Appelquist:1985rr}}
which, at low energies, will give rise to an operator of the same form
as the one in eq.~(\ref{eq:isoviol}). Using dimensional analysis, we
find
\beq
{\lower35pt\hbox{\epsfysize=1.0 truein \epsfbox{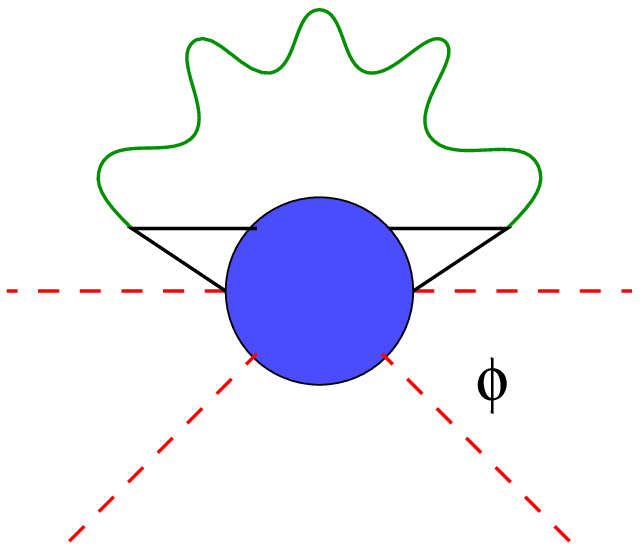}}}
\rightarrow 
\left[{ {g^2 \over M^2}}
\left({ {\kappa^2 \over \Lambda^2}}\right)^{-1}\right] 
{ \kappa^2 \over \Lambda^2} (\phi^\dagger D^\mu \phi) (\phi^\dagger
D_\mu \phi)~,
\label{eq:isoviola}
\eeq
which results in the bound  $M/g \stackrel{>}{\sim} 4$ TeV.
From eq.~(\ref{eq:yukawa}) with $m_t \approx 175$ GeV we then derive the limit
\beq
\Lambda \stackrel{>}{\sim} 4\, {\rm TeV} \cdot \sqrt{\kappa}~.
\eeq
For $\kappa \approx 3.5$, we find $\Lambda \stackrel{>}{\sim} 7.5$ TeV.

As previously discussed, a lower bound on the scale $\Lambda$ yields an
upper bound on the Higgs boson mass.  Here we provide an estimate of
this upper bound by naive extrapolation of the lowest-order perturbative
result\footnote{Though not justified, the naive perturbative bound has
  been remarkably close to the non-perturbative estimates derived from
  lattice Monte Carlo calculations
  \protect\cite{Kuti:1988nr,Luscher:1989uq,Hasenfratz:1987eh,Hasenfratz:1989kr,Bhanot:1990zd,Bhanot:1991ai}
  .} shown in eq. (\ref{estimate}). For $\Lambda \stackrel{>}{\sim}
7.5$ TeV, this results in the bound\footnote{If $\kappa \approx 4\pi$,
  $\Lambda$ would have to be greater than 14 TeV, yielding an upper
  bound on the Higgs boson's mass of 490 GeV. If $\kappa \approx 1$,
  $\Lambda$ would be greater than 4 TeV, yielding the upper bound $m_H
  \stackrel{<}{\sim} 670$ GeV.} $m_H \stackrel{<}{\sim} 550$ GeV.

\section{Two-Higgs Doublet Model}

\subsection{The Higgs Potential and Boson Masses}

Up to this point, we have discussed the {\it simplest} model
which can account for electroweak symmetry breaking, the one-doublet
Higgs model. In this case, the electroweak breaking sector consists
of only one field. In general, the symmetry breaking sector
can be more complicated. As a case study of an extended symmetry
breaking sector, we next consider a model with {\it two} Higgs
fields\footnote{For a review, see \protect\cite{Gunion:1989we}.}
\beq
\phi_{1}=\left(\matrix{\phi_{1}^+ \cr \phi_{1}^0 \cr}\right)
\ \ \ \& \ \ \ 
\phi_{2}=\left(\matrix{\phi_{2}^+ \cr \phi_{2}^0 \cr}\right)
{}~~~.
\eeq
The most general potential for such a model, with a softly broken $\phi_1
\to -\phi_1$ symmetry (the necessity of which will be 
discussed in the following), is
\beqa
\hskip-5pt V(\phi_1,\phi_2) &=&  \lambda_1(\phi^\dagger_1\phi_1-{ v_1^2}/2)^2
 + \lambda_2(\phi^\dagger_2\phi_2-{ v_2^2}/2)^2 \nonumber \\
&+&\lambda_3\left[(\phi^\dagger_1\phi_1-{ v_1^2}/2) 
 + (\phi^\dagger_2\phi_2-{ v_2^2}/2)\right]^2 \nonumber
 \\
&+& { \lambda_4}\left[(\phi^\dagger_1\phi_1)(\phi^\dagger_2\phi_2)
- (\phi^\dagger_1\phi_2)(\phi^\dagger_2\phi_1)\right] \nonumber \\
&+& { \lambda_5}\left[{\rm Re}(\phi^\dagger_1\phi_2) 
       - { v_1}{ v_2}\cos\xi/2\right]^2 \nonumber \\
&+& { \lambda_6}\left[{\rm Im}(\phi^\dagger_1\phi_2) 
       - { v_1}{ v_2}\sin\xi/2\right]^2  ~. 
\label{eq:twohiggpot}
\eeqa
For a range of $\lambda_i$ and for $v^2_1,\, v^2_2 >0$,
the potential is minimized when
\beq
\langle\phi_1\rangle=\left(\begin{array}{c} 0 \\ 
{{ v_1}/\sqrt{2}} \end{array}\right)
\ \ \ \&\ \ \ 
\langle\phi_2\rangle=\left(\begin{array}{c} 0 \\ 
{{ v_2} e^{i\xi}/ \sqrt{2}} \end{array}\right)~.
\eeq
In this vacuum, $SU(2)_W \times U(1)_Y \to U(1)_{em}$ with
\beq
M^2_W = {g^2\over 2} ({ v^2_1} + {  v^2_2})\ \ \ \& \ \ \ 
\rho = {M_W \over M_Z \cos\theta_W} \equiv 1 ~.
\eeq
From this we conclude that
\beq
v^2_1 + v^2_2 = v^2\approx (246\, {\rm GeV})^2.
\eeq

\begin{figure}
\centerline{
\hbox{\epsfxsize 3cm {\epsffile{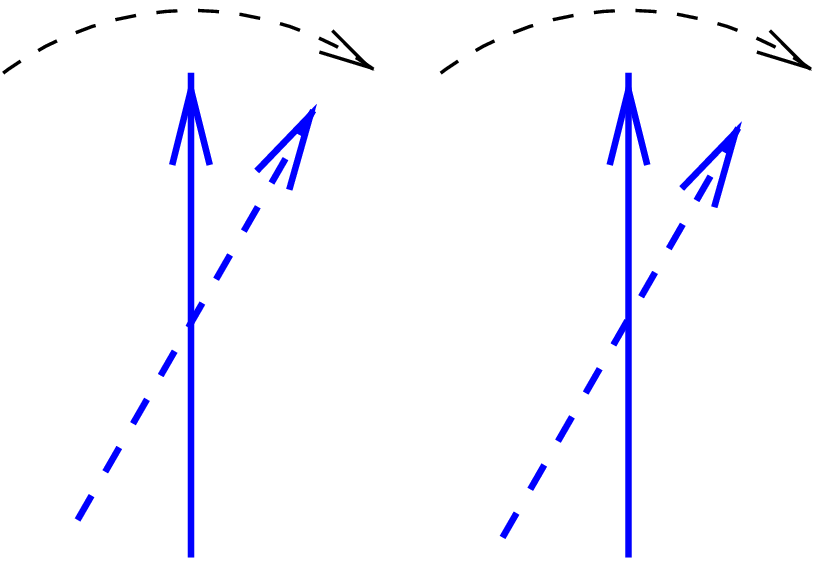}}
\hskip1cm
{\epsfxsize 3cm {\epsffile{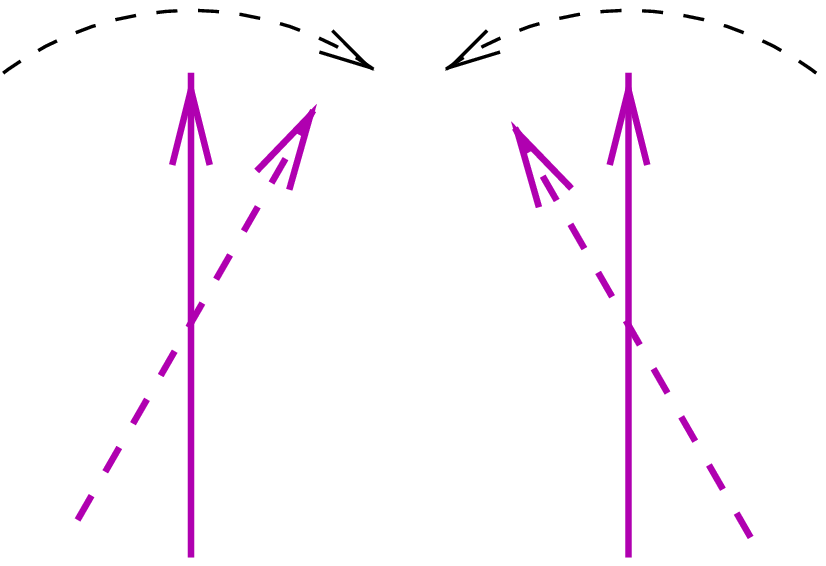}}}}}
\caption{Viewed as a system of coupled pendula, the Goldstone
bosons correspond to the symmetric mode at left, and the pseudo-Goldstone
bosons to the asymmetric mode at right.}
\label{fig:eight}
\end{figure}

Two complex Higgs doublets correspond to eight real degrees of freedom. The
eight mass eigenstates can readily be determined from the potential.
In what follows, we will make the simplifying assumption that $\sin\xi
= 0$, which avoids  CP-violation in the symmetry breaking sector.
One may view the mass eigenstates as normal modes of a system
of coupled pendula (see Fig. \ref{fig:eight}).
Defining ${\cos\beta = v_1/v}$ and ${\sin\beta = v_2/v}$, we find
that the three ``eaten'' Goldstone bosons (which become the longitudinal
components of the $W$ and $Z$) may be viewed as the ``symmetric oscillation
mode'':
\beq
G^{{\pm},0} = (\phi^\pm_1, \sqrt{2}\,{\rm Im}\,\phi^0_1)
{ \cos\beta} + (\phi^\pm_2, \sqrt{2}\,{\rm Im}\,\phi^0_2) { \sin\beta}~.
\eeq
The three fields orthogonal to the Goldstone modes are physical
pseudo-scalars
\beq
(H^{\pm},A^0) = - (\phi^\pm_1, \sqrt{2}\,{\rm Im}\phi^0_1)
{ \sin\beta} + (\phi^\pm_2, \sqrt{2}\,{\rm Im}\phi^0_2) { \cos\beta}~,
\eeq
and have masses
\beq
m^2_{H^\pm}={ \lambda_4}(v_1^2 + v_2^2)/2~~~{ \neq}~~~
m^2_{A^0}={ \lambda_6}(v_1^2 + v_2^2)/2~.
\eeq

In addition to the oscillation modes, there are two ``breathing'' modes
in which the lengths of the pendula (Higgs doublet fields) change.
These modes correspond to two neutral scalar fields. Defining
\beq
(H_1,H_2) = (\sqrt{2}\,{\rm Re}\, \phi^0_{1,2} - v_{1,2})~,
\eeq
we find the $2 \times 2$ mass matrix
\beq
\left(\begin{array}{cc}
2v_1^2(\lambda_1 + \lambda_3) + v_2^2\lambda_5/2 &
(4\lambda_3 + \lambda_5)v_1v_2/2 \\
(4\lambda_3 + \lambda_5)v_1v_2/2 &
2v_1^2(\lambda_2 + \lambda_3) + v_2^2\lambda_5/2 \end{array}
\right)
\eeq
The mass eigenstates define a mixing-angle $\alpha$
\beqa
H^0 &=& H_1\cos\alpha + H_2\sin\alpha \nonumber \\
h^0 &=& - H_1\sin\alpha + H_2\cos\alpha ~.
\eeqa

\subsection{Neutral Scalars}

As in the case of the one-doublet model, the high-energy scattering of
longitudinally-polarized electroweak gauge bosons is unitarized
by the exchange of neutral scalars. The coupling of the neutral
scalars to the $W$'s can be written
\beqa
{\cal L} & \supset & \left( 1 + 2 {{ H_1}\over v}{ \cos\beta} 
+ 2 {{ H_2}\over v}
{ \sin\beta}\right) \nonumber \\
& & \ \times \  \left[M^2_W W^{\mu +}W_\mu^- +
  {1\over 2} M^2_Z Z^\mu Z_\mu  \right]~.
\label{eq:nhcoup}
\eeqa
changing basis from $(H_1,H_2)$ to $(h^0, H^0)$ amounts to replacing
$\beta$ in eq. (\ref{eq:nhcoup}) with $(\beta-\alpha)$. Note that the
exchange of {\it both} $h^0$ and $H^0$ is required to maintain
unitarity, and from eq. (\ref{quigg}) we conclude that $m_{h^0,H^0}
\laem 1.8$ TeV.

In the one-doublet model, the couplings of the single Higgs boson to the
fermions were proportional to the fermion masses, eq. (\ref{eq:higgscoup}).
For this reason, the couplings were manifestly flavor-diagonal.  In the
most general two-Higgs model, it is possible for {\it each} fermion
species to acquire mass from the vacuum expectation value of both Higgs
fields. In this case, it is not possible to ensure that the couplings of
the $h^0$ and $H^0$ are flavor-diagonal, {\it i.e.} Higgs exchange could
give rise to {\bf flavor-changing neutral-currents}.

In order to avoid this possibility, it is necessary to ensure that each
species of fermion couples to one and only one Higgs-doublet field.
This can be done naturally \cite{Glashow:1977nt}. One conventional
choice (often referred to as ``model II'' in the literature) is to
impose the symmetry
\beq
\phi_1 \to -\phi_1\ \ \  \&\ \ \  (d_R,\, l)^i \to
  -(d_R,\, l)^i~,
\eeq
which implies that the Higgs doublet $\phi_1$ couples
only to down-quarks and leptons, and the doublet
$\phi_2$ couples to up-quarks. These constraints result
in the couplings
\beqa
-{\cal L} &\supset&  \left( 1 + 
{H_2\over v{ \sin\beta}}\right)\!
\sum_i   m^u_i \overline{u_i}u_i \, + \nonumber \\
&&\left( 1 + 
{H_1\over v{ \cos\beta}}\right)\!
 \sum_i  \left(m^d_i \overline{d_i}d_i+m^l_i \overline{l_i}l_i \right)~,
\eeqa
and {\it no} tree-level flavor-changing neutral-currents.

\subsection{Charged-Scalars and Pseudo-Scalar}

At tree-level, there are couplings between the weak gauge bosons
and pairs of scalars proportional to gauge-couplings times
sines or cosines of mixing angles.
The model II quark couplings are
\beq 
{iA^0 \over v} \sum_i \left( m^u_i {
  \cot\beta}\,\overline{u_i}\gamma_5 u_i + m^d_i {\tan\beta}\,
\overline{d_i}\gamma_5 d_i\right) 
\eeq
for the pseudo-scalar and 
\beq
{H^+ \over \sqrt{2}v}\sum_{ij}\overline{u_{Ri}}
\left[{ \cot\beta}\, m^u_i V_{ij} + 
{ \tan\beta}\, V_{ij}m^d_j\right]d_{Lj} + h.c. 
\eeq
for the charged scalars. We see that the couplings are proportional to
$m/\langle \phi_i\rangle$, generally {\it larger} than Higgs-couplings
in the one-doublet model. Furthermore, the discrete symmetry has ensured
that the tree-level couplings of the $A^0$ are flavor-diagonal and the
couplings of the $H^\pm$ have the usual KM ($V_{ij}$) suppression.

\subsection{Comments}

Although $\rho=1$ at tree-level (a general result, when only isospin-doublets
are used in the symmetry breaking sector), the two-Higgs
model has a custodial Symmetry only if ${ \lambda_4} 
= { \lambda_6} = { -\lambda_5}$. In general, $m^2_{H^\pm} \neq
m^2_{A^0}$ and there are one-loop corrections to $\Delta \rho$
\beq
{\lower 7pt \hbox{\epsfysize=0.4truein\epsfbox{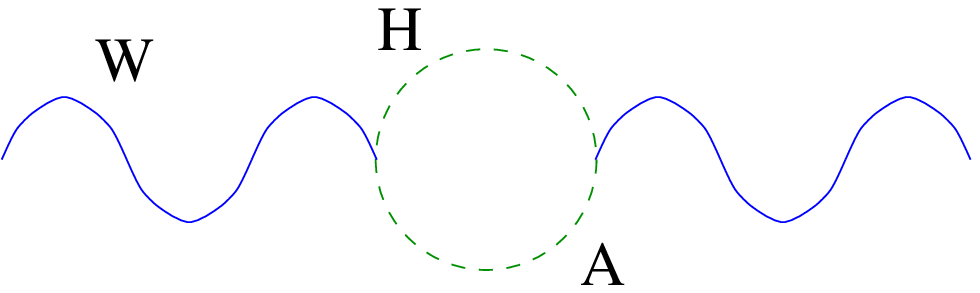}}}
\hskip0.5in
{\lower 7pt \hbox{\epsfysize=0.4truein\epsfbox{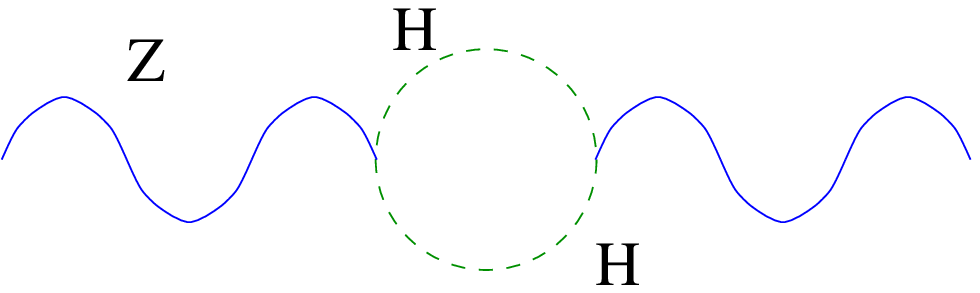}}}
\eeq
The hierarchy and naturalness problems of the one-doublet model remain
\beq
{\lower10pt\hbox{\epsfysize=0.4 truein \epsfbox{figures/msq.eps}}}
\ \rightarrow\  \ m_{\phi_i}^2 \propto \Lambda^2~.
\eeq
Finally, the various self-couplings in the two-Higgs model
have positive $\beta$-functions\footnote{The {\it only} asymptotically
free theories are non-abelian gauge theories \protect\cite{Coleman:1973sx}.}
\beq
{\lower15pt\hbox{\epsfysize=0.5 truein \epsfbox{figures/beta.eps}}}
\ \rightarrow \ \beta_i(\lambda_j) > \, 0
{}~~~.
\eeq
These theories are therefore also trivial and must be interpreted
as effective theories valid below some energy scale $\Lambda$. Requiring
that $\Lambda \gaem 2$ TeV, one finds \cite{Kominis:1993zc} the
bound $m_h \laem 470$ GeV.

\section{General Scalar Models}

\subsection{Lessons in Symmetry Breaking}

There are a number of lessons from our study of the
simplest Higgs models that apply to a model with an arbitrary number of
scalars.  While these issues are discussed in terms of fundamental
scalar models they are also relevant, as we shall see, to models of
dynamical electroweak symmetry breaking.

\medskip

\noindent\underline{Custodial Symmetry}

\medskip

In a general scalar model with arbitrary Higgs representations, the
tree-level {$\rho$} parameter is {\it not} equal to one. If there are
several scalars, the gauge-boson mass matrix may be written as
\beq
M^2_{ab} = \sum_i \langle \phi^\dagger_i \rangle 
T^a T^b \langle \phi_i \rangle~,
\eeq
in terms of the vacuum-expectation-values $\langle \phi_i \rangle$
of the various scalar fields. It can then be shown
that at tree-level
\beq
\rho = {{\sum_i \langle \phi^\dagger_i \rangle 
T^1 T^1 \langle \phi_i \rangle} \over
{\sum_i \langle \phi^\dagger_i \rangle 
T^3 T^3 \langle \phi_i \rangle}}
= {{\sum_i (I_i(I_i + 1) - I^2_{3i}) v^2_i}\over
{\sum_i 2I^2_{3i} v^2_i}}~,
\eeq
where $I_i$ and $I_{3i}$ denote the weak isospin of each scalar scalar
field and the $I_3$ of the (neutral) component which receives a vev. In
particular, this shows that $\rho \equiv 1$ at tree-level for a model
with {\it any number} of Higgs doublets and is not
automatically\footnote{For a model with triplet fields and $\rho=1$, see
  \protect\cite{Georgi:1985nv}.} 1 if other representations are
included.

\medskip

\noindent\underline{{Flavor-Changing} Higgs Couplings (FCHC)}

\medskip

If there are multiple weak-doublet scalars $H^\alpha$,
then in general each fermion
species can couple to {\it every} scalar doublet.
The most general Yukawa structure can be written
\beq
-{\cal L} \supset \sum_{\alpha i j} H^\alpha
\left( y^U_{\alpha i j} \bar{u}_i u_j
+ y^D_{\alpha i j} \bar{d}_i d_j +
y^L_{\alpha i j} \bar{l}_i l_j \right)~.
\eeq
As discussed in the case of the two-Higgs model, this will generically
give rise to {tree-level flavor-changing Higgs couplings}.  The only
natural \cite{Glashow:1977nt} solution is to ensure that only one scalar
contributes to the mass of each fermion species.

\medskip

\noindent\underline{{Pseudo-Goldstone Bosons}}

\medskip

Consider a two-Higgs model where the scalar self-couplings of eq.
(\ref{eq:twohiggpot}) satisfy $\lambda_{3,4,5,6} \ll \lambda_{1,2}$.  In
this case,
\beqa
\hskip-5pt V(\phi_1,\phi_2) &\approx&  \lambda_1(\phi^\dagger_1\phi_1-v_1^2/2)^2
 + \lambda_2(\phi^\dagger_2\phi_2-v_2^2/2)^2 ~.
\eeqa
so that the EWSB sector is approximately {\it two separate sectors}.
The mass eigenstates $(H,h)$ can approximately be
identified with the gauge eigenstates $(H_1,H_2)$ and satisfy
\beqa
m^2_{ h_1} &\approx& 2\lambda_1 { v^2_1}~, \nonumber \\
m^2_{ h_2} &\approx& 2\lambda_2 { v^2_2}~.
\eeqa
These two sectors have approximate {\it independent} symmetries for
$\phi_{1,2}$
\beq
(SU(2)_L \times U(1))^2 \to (U(1))^2~.
\label{eq:pgbsym}
\eeq
This symmetry breaking pattern results in {\it six} broken generators:
three corresponding to the exact gauge symmetries and three
corresponding to the extra approximate global symmetries.  The pseudo- scalars and
charged scalars have masses
\beq
m^2_{H^\pm} \propto \lambda_4 v^2\, \, , \, \, m^2_A \propto \lambda_6 v^2
\, \ll \, m^2_{h_{1,2}}~,
\eeq
and are {{\it pseudo}-Goldstone} bosons corresponding to the 
three (approximate) extra spontaneously broken symmetries in 
eq. (\ref{eq:pgbsym}).

\begin{figure}
\centerline{
\hbox{\epsfxsize 5cm \centerline{\epsffile{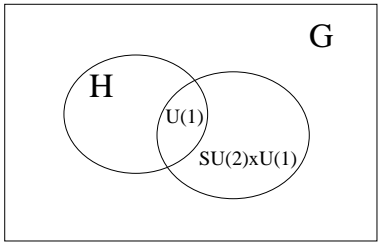}}}
}
\caption{Venn diagram representation of the symmetry structure
of an extended symmetry breaking sector. The symmetry of the electroweak
breaking sector is $G$ and breaks spontaneously to $H$. In order
to break the weak interactions to electromagnetism, $SU(2) \times U(1)$
must be embedded in $G$ in such a way that the $U(1)$ of electromagnetism
(and possibly an entire $SU(2)_C$ custodial symmetry) is in $H$. 
After \protect\cite{Weinberg:1976gm}.}
\label{fig:nine}
\end{figure}

These considerations can be generalized to extended or multiple
symmetry-breaking sectors.  The situation is nicely illustrated by the
Venn diagram \cite{Weinberg:1976gm} shown in fig. \ref{fig:nine}.  The
$SU(2) \times U(1)$ electroweak symmetry must be a subgroup of the full
symmetry group $G$ of the electroweak breaking sector.  In order to
break the weak interactions to electromagnetism, $SU(2)\times U(1)$ must
be embedded in $G$ in such a way that the $U(1)$ of electromagnetism
(and possibly an entire $SU(2)_C$ custodial symmetry) is in the {\it
  unbroken} subgroup $H$.

The diagram also allows one to visualize the remaining global symmetries
and the Goldstone bosons. The generators of $H$ orthogonal to
$U(1)_{em}$ correspond to {\it unbroken} global symmetries. Every
generator in $SU(2) \times U(1)$ orthogonal to $U(1)_{em}$ is
spontaneously broken and the three corresponding exact Goldstone bosons
are ``eaten'' by the $W^\pm$ and $Z$. 

To every generator in $G$ orthogonal to both $H$ and $SU(2)
\times U(1)$ there is a potentially massless Goldstone boson. There are
stringent limits on the existence of light or massless particles
\cite{Barnett:1996hr}. Therefore one must arrange that these be only
{\it approximate} symmetries leading to {\it pseudo-Goldstone bosons}.  The general
properties of the pseudo-Goldstone bosons are expected to be similar to
those of the extra states in the two-Higgs model.  Namely, we expect that:

\begin{narrower}
\begin{itemize}
\item the fermion couplings of the pseudo-Goldstone
bosons are  $\propto\, m_f/{ v_i}\, \ge\, m_f/v$(!),
\item they should have masses ${ m^2}\, \propto\, { v^2_i}$,
where $v_i$ represents the ``vev'' of the corresponding sector,
\item and  their couplings to $(W,Z)_L$ $\propto \, { v_i/v}$.
\end{itemize}
\end{narrower}

\noindent
Finally, as always one must be careful to avoid flavor-changing neutral
currents.

\subsection{The Axion}

There is a particularly dangerous Goldstone boson that appears in many
different models, the axion\footnote{The discussion presented in this
  section follows closely the exposition in
  \protect\cite{Georgi:1986df}.}
\cite{Peccei:1977hh,Peccei:1977ur,Weinberg:1978ma,Wilczek:1978pj}.
Consider a toy model, which results in the ``KSVZ''
\cite{Kim:1979if,Shifman:1980if} axion:
\beq
{\cal L} = \bar{Q}i\slashchar{D}Q + |\partial \phi|^2
+ {\tilde{\lambda}\over 4} (|\phi|^2-f^2)^2 
+y(\bar{Q}_L \phi Q_R + h.c.)\, ,
\eeq
where the $Q$ is a new color-triplet fermion and $\phi$ a complex
scalar.

The symmetries of this model are  $U(1)_Q \times U(1)_A$
where $U(1)_Q$ is ``Q-number'' and $U(1)_A$ acts as follows:
\beq
Q \to e^{{i \gamma_5 \alpha}\over 2} Q \ \ \ \& \ \ \ 
 \phi \to e^{i\alpha} \phi\, .
\eeq
The potential causes the spontaneous breaking of this symmetry at
a scale $f$, leaving
only $U(1)_Q$.  The model is more conveniently analyzed in terms of the
fields
\beq
\phi(x)=e^{i{a(x)\over f}}{(h(x) + f)\over \sqrt{2}} \ \ \  \& \ \ \ 
Q'(x) = e^{i\gamma_5{a(x)\over f}} Q(x)\, ,
\label{eq:change}
\eeq
yielding the spectrum: $m_{Q}=yf$, $m^2_h \propto \tilde{\lambda}f^2$,
and a Goldstone boson $m^2_a=0$. The field $a(x)$ is the axion.

However, $U(1)_A$ is anomalous \cite{Adler:1969av,Bell:1969re,Bardeen:1969md}. 
Therefore the low-energy effective theory for the Goldstone
boson $a$ is
\beq
{\cal L} = {1\over 2}(\partial a)^2 + { \left({\theta +{a(x)\over f}}\right)
{g^2\over 32\pi^2} G^a_{\mu\nu} \tilde{G^a}^{\mu\nu}} + \ldots~,
\eeq
where the second term arises from the Ward identity for an anomalous
transformation of the sort required in eq. (\ref{eq:change}).

Consider the effect of the axion in low energy QCD.
Above the chiral-symmetry breaking (and confinement) scale,
the lagrangian for QCD is
\beqa
{\cal L} & = & \bar{\psi}(i\slashchar{D}-M)\psi + 
{1\over 2}(\partial a)^2 \nonumber\\ 
&& + { \left({{\bar{\theta}}+{a(x)\over f}}\right) 
{g^2\over 32\pi^2} G^a_{\mu\nu} \tilde{G^a}^{\mu\nu}}
-{1\over 4} G^a_{\mu\nu} {G^a}^{\mu\nu}\, . 
\eeqa
Here $\psi$ is the $(u\, d)$ isodoublet and we have chosen a basis in
which ${\rm arg}\, \left[{\rm det}\right]\, M = 0$. We may redefine $a(x) +
\bar{\theta} \to a(x)$.  To eliminate the troublesome $a G \tilde G$
coupling, we may rotate $a(x)$ into the mass matrix
\beq M \to {\tilde{M}= e^{-ia(x)Q_A/f}\, M\, e^{-ia(x)Q_A/f}}\, , 
\eeq
where $Q_A$ is an arbitrary matrix with $\tr Q_A = {1\over 2}$.  
Below the chiral symmetry breaking scale, the effective chiral
theory\footnote{For a review, see the lectures by A. Pich in these
  proceedings.} reads:
\beq
{f^2_\pi \over 4}\tr \left(\partial^\mu \Sigma^\dagger
\partial_\mu \Sigma\right) + {f^2_\pi \over 2}
\mu \tr\left({ \tilde{M}}\Sigma^\dagger
+ h.c.\right) + \ldots~.
\label{eq:chirallag}
\eeq
A clever choice of $Q_A$
\beq
Q_A = {1\over 2}\, {M^{-1}\over \tr\, M^{-1}}
\eeq
eliminates $\pi$--$a$ mixing.

\begin{figure}
\centerline{
\hbox{\epsfxsize 7.0cm \centerline
{\epsffile{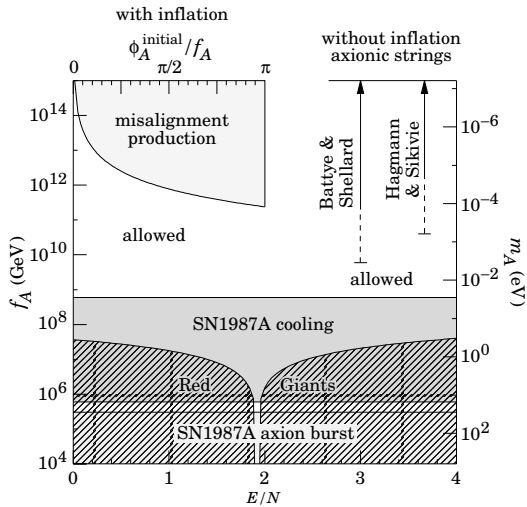}}}
}
\caption{Limits on the KSVZ axion coming from
  astrophysical arguments. $E/N$ measures extra, model-dependent,
  couplings to quarks and leptons. From \protect\cite{Barnett:1996hr}.}
\label{fig:ten}
\end{figure}

The potential for the axion can be read off from the second
term in eq. (\ref{eq:chirallag})
\beq
V(a)=-f^2_\pi \mu\, \tr\, M\, \cos\left({2a(x)Q_A \over f}\right)\, .
\eeq
The potential is minimized at $\langle a \rangle \equiv 0$. This implies
that $\bar{\theta}_{eff} \equiv 0$, solving the strong-CP problem. 
We can also compute the mass of the axion to be
\beqa
m^2_a & = & m^2_\pi {m_u\, m_d\over (m_u + m_d)^2} {f^2_\pi \over f^2}
\nonumber \\
&= &  {\cal O}\left(10^{-3}\, {\rm eV}\right)\left({10^{10}\, {\rm GeV}
\over f}\right)\, . 
\eeqa
Note that the couplings of the axion are suppressed by $1/f$.  Limits on
the cooling of neutron stars, shown in fig. \ref{fig:ten}, imply that $f
\gaem 10^9$ GeV.

While we have illustrated the axion in terms of the KSVZ model, an axion
arises whenever there is a classically exact but {\it anomalous} $U(1)$
symmetry that is spontaneously broken. It can occur in the two-Higgs
model, for example, in the limit that $\lambda_{5,6} \to 0$ and with
appropriate fermion couplings. This would result in a ``weak-scale''
axion with $f \approx v$, which is strongly forbidden.

\section{Solving the Naturalness/Hierarchy Problems}

As detailed in the last two sections, fundamental Higgs theories suffer
from the naturalness/hierarchy and triviality problems. These problems
follow from the seeming inability of fundamental Higgs theories to
naturally (in the technical and colloquial senses) maintain a hierarchy
between the weak scale and any fundamental higher-energy scale $\Lambda$
({\it e.g}, the grand-unified or Planck scales).  While scalar masses are
{\it susceptible} to ${\cal O}(\Lambda^2)$ mass corrections, their
masses can be protected by a symmetry.  There are essentially two
approaches to do this: supersymmetry and dynamical electroweak symmetry
breaking.

In supersymmetric\footnote{For a more complete review, see
  \protect\cite{Bagger:1996ka} and references therein.} models, one
introduces fermionic (super-)partners for the Higgs boson. The mass of
the scalar Higgs particles are then related by supersymmetry to the
masses of their fermionic partners.  These fermion masses, in turn, can
be protected by a {\it chiral symmetry}.  In a fully
supersymmetric theory all particles (including the ordinary fermions and
gauge bosons) must come in supermultiplets that include scalar
(sfermions) partners of the ordinary fermions and fermionic (gaugino)
partners of the gauge-bosons.  Supersymmetry also requires the existence
of (at least) {\it two} Higgs doublets: this is necessary both to cancel
a potential $SU(2)_W$ anomaly \cite{Witten:1982fp}, and to provide the
necessary $Y=\pm{1\over 2}$ multiplets to give mass to all of the
ordinary fermions.

Supersymmetry cannot be exact, as superpartners have not been observed.
Instead, supersymmetry is assumed to be {\it softly} broken\footnote{For
  a discussion of this point, see the lectures by G.~Ross in this
  volume.}. In practice the potentially problematic contributions to the
Higgs bosons masses largely cancel between ordinary and supersymmetric
contributions and are proportional to soft supersymmetry-breaking
masses, for example
\beq
{\lower20pt\hbox{
\epsfysize=0.5 truein \epsfbox{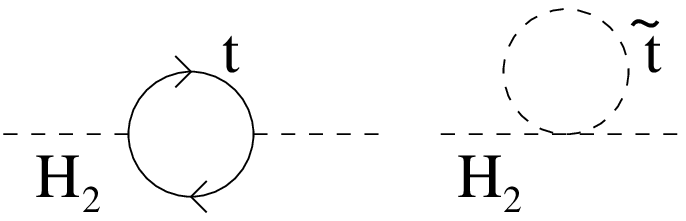}}
\ \ \rightarrow \ \ \delta m_H^2 \propto m^2_{\tilde{t}}\log \Lambda^2}~.
\eeq
For this reason, if supersymmetry is to be relevant to the hierarchy
problem, the masses of at least some \cite{Cohen:1996vb} of the
superpartners must be of order a TeV.  Note that this divergence is
proportional only to $\log\Lambda$, and the theory is no longer {\it
  technically} unnatural \cite{'tHooft:1980xb}:  if the supersymmetry
breaking scale is of order a TeV, one stabilizes the hierarchy.
An {\it explanation} of the hierarchy between the weak scale and the
grand-unified or Planck scale(s) requires a dynamical theory of
supersymmetry breaking.  

Supersymmetry also severely constrains the form of the electroweak
symmetry breaking sector. The theoretical and phenomenological
consequences of this are reviewed in the following section.

In models of dynamical electroweak symmetry breaking, chiral symmetry
breaking in an asymptotically-free gauge theory is assumed to be responsible for
breaking the electroweak symmetry.  The simplest models of this sort
rely on QCD-like ``technicolor'' interactions
\cite{Weinberg:1979bn,Susskind:1978ms}. The weak scale is then the
result of ``dimensional-transmutation,'' and because of asymptotic
freedom (as in the case of $\Lambda_{QCD}$) an exponentially large
hierarchy becomes natural. New dynamics arises at a scale of order a TeV
and, in this sense, the {\it hierarchy is eliminated}. Unlike models
with fundamental Higgs fields, where fermion masses can be provided by
Yukawa interactions, these models typically require additional
flavor-dependent dynamics \cite{Eichten:1979ah,Dimopoulos:1979es} to
give rise to the various masses of the ordinary quarks and leptons. The
theoretical and phenomenological consequences of these models are
reviewed in the last half of these lectures.

There is also an approach to the hierarchy/naturalness problem which
allows one to ``interpolate'' between a technicolor-like model, with new
dynamics at a scale of order a TeV, and a fundamental Higgs model.  In
composite Higgs models \cite{Kaplan:1984fs,Kaplan:1984sm,Dugan:1985hq},
one constructs a theory in which the {\it Higgs boson is a Goldstone
  boson} of a spontaneously-broken chiral symmetry. In this case, the
important dimensional parameter is the $f$-constant, the analog of $f_\pi$
in chiral-symmetry breaking in QCD.  If $f \approx v$, these theories
are technicolor-like with additional resonances at energies
of order a TeV, while if $f \gg v$ (if such can
be arranged naturally) the low-energy theory is essentially a
fundamental Higgs model\footnote{In this case the triviality bounds
  discussed previously apply.}. This approach, which may be quite
interesting in the regime $f \gaem v$, has not been fully explored.
However, even in this case, flavor-dependent dynamics will be required
to provide masses to the ordinary fermions. Many of the constraints and
lessons learned from technicolor-like models will apply.

\section{Electroweak Symmetry Breaking in Supersymmetric Theories}

\subsection{The Electroweak Potential and Higgs Boson Masses}

For the reasons discussed in the previous section, in the ``minimal
supersymmetric standard model'' (MSSM) one introduces superpartners for
all standard model particles, (sfermions and gauginos), and two Higgs
fields $H_1\vert_{+{1\over 2}}$ and $H_2\vert_{-{1\over 2}}$, and their
superpartners.  One further assumes that supersymmetry is broken softly.
In the minimal model, including only terms of dimension four or less
consistent with the softly broken symmetry, the form of the electroweak
potential is \cite{Gunion:1989we}
\beqa
V & = &({m^2_1} + { |\mu|^2}) H^\dagger_1 H_1
+ ({m^2_2} + { |\mu|^2}) H^\dagger_2 H_2 \nonumber \\
& &\mbox{} -{m^2_{12}}(H^T_1i\sigma_2 H_2 + h.c.) \nonumber \\
& & \hskip-20pt \mbox{} { +{1\over 8}(g^2 + {g^\prime}^2)\left[ H^\dagger_1 H_1
- H^\dagger_2 H_2\right]^2 + {g^2\over 2}|H^\dagger_1 H_2|^2}~. 
\label{eq:susypot}
\eeqa
The different terms in this potential arise from different sources:

\begin{narrower}
\begin{itemize}

\item ${m^2_1,\, m^2_2,\, m^2_{12}}$ are {soft supersymmetry breaking}
  terms. For SUSY to be relevant to the hierarchy problem we expect
  these mass terms to be less than ${\cal O}(1\, {\rm TeV})$.

\item ${\mu}$ comes from {superpotential} ``{$F$-terms}'' and {\it
    respects supersymmetry}. For electroweak symmetry to occur as
  required, we need $\mu$ to be ${\cal O}(1\, {\rm TeV})$. Additional
  dynamics is generally required to make this occur naturally.

\item The quartic terms arise from electroweak gauge symmetry ``{
    $D$-terms}.'' Their size is given in terms of the weak gauge
  couplings.

\end{itemize}
\end{narrower}

\noindent Note that this potential is a special case of the more general
two-Higgs potential in eq. (\ref{eq:twohiggpot}).

Only {three unknown parameters} (linear combinations of masses) appear
in the potential in eq. (\ref{eq:susypot}). Fixing $v \approx\, 246$ GeV, leaves
two free parameters, which may be taken as {$\tan\beta$} and {
  $m_{A^0}$}. At tree-level, the other masses are then determined:
\beqa
m^2_{H^\pm} & = & m^2_{A^0} + m^2_W \nonumber \\
m^2_{H^0,h^0} & = & {1\over 2}
(m^2_{A^0} + m^2_Z)\pm \nonumber \\
& &  \hskip-20pt {1\over 2} \sqrt{(m^2_{A^0} + m^2_Z)^2
-4m^2_Z m^2_{A^0}\cos^2 2 \beta}~.
\eeqa
This implies that ${m_{h^0} \le m_Z |\cos 2\beta| \le m_Z}$, {\it i.e.}
at tree level it is necessary that the lightest neutral scalar have a
mass less than $m_Z$. This conclusion will be modified in light of the
discussion in the next section.

The MSSM has a ``decoupling limit'', in which the
theory reduces to the standard model. For the Higgs sector as
the supersymmetry breaking scale becomes large, $m_{1,2,12}\to \infty$. 
In this case $m_{A^0,H^\pm,H^0} \to \infty$ and all of the extra
particles {decouple}. As required, $\cos(\beta-\alpha)$, which sets
the $HWW$ and $HZZ$ couplings, equals
\beq
{m^2_{h^0}(m^2_Z-m^2_{h^0})\over
(m^2_{H^0}-m^2_{h^0})(m^2_{H^0}+m^2_{h^0}-m^2_Z)}\, ,
\eeq
and goes to zero when $m_{H^0} \gg m_{h^0},\, m_Z$.  While the
decoupling limit is {\it not} the interesting one from the point of view
of solving the hierarchy problem, it does bear on the question of limits
on supersymmetric models arising from precision electroweak tests.  In
particular, to the extent that the {\it standard model} cannot be
excluded, neither can the minimal supersymmetric model -- all that can
be obtained are {\it lower} bounds on the superparticle masses.

\subsection{Radiative Corrections and $m_t$}

\begin{figure}
\centerline{
\hbox{\epsfxsize 7cm \centerline
{\epsffile{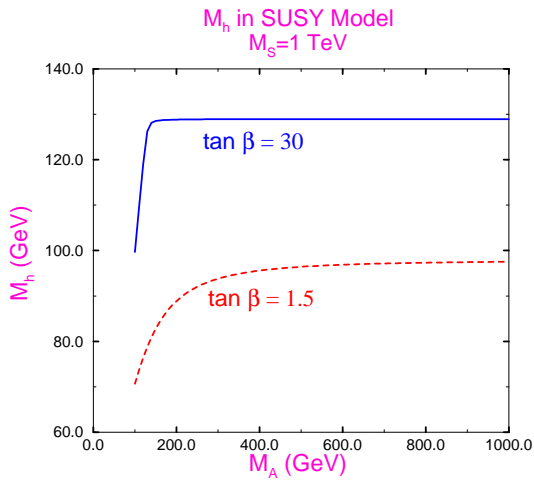}}}
}
\caption{$m_{h^0}$, including one-loop corrections, as a function of $m_A$
for two values of $\tan\beta$. From \protect\cite{Dawson:1996cq}.}
\label{fig:eleven}
\end{figure}

The analysis of the Higgs sector of the MSSM given above is true at {\it
  tree-level}. However, because of the heavy top-quark and the
correspondingly large Yukawa coupling, there are important corrections
at one-loop\footnote{For a complete review, see
  \protect\cite{Dawson:1996cq} and references therein.}. For example, at
one-loop the bound on the $h^0$ mass is modified
\beq
m^2_{h^0} \laem m^2_Z \cos^2\beta + { {3 G_F \over \sqrt{2}\pi^2}
m^4_t \log\left({\tilde{m}^2_t \over m^2_t}\right)}~.
\eeq
For $m_t \approx 175$ GeV and $\tilde m_t \approx 1$ TeV the bound on the
$h^0$ mass becomes $m_{h^0} \laem 130$ GeV, as shown in Fig.
\ref{fig:eleven}.  Similar bounds can also be derived in non-minimal
supersymmetric models, so long as one requires that all couplings in the
Higgs sector remain small up to the presumed grand-unified scale of
$10^{16}$ GeV.  In this case, we find the somewhat looser bound
{$m_{h^0} \laem 150$ GeV}.

\begin{figure}
\centerline{
\hbox{\epsfxsize 7cm \centerline
{\epsffile{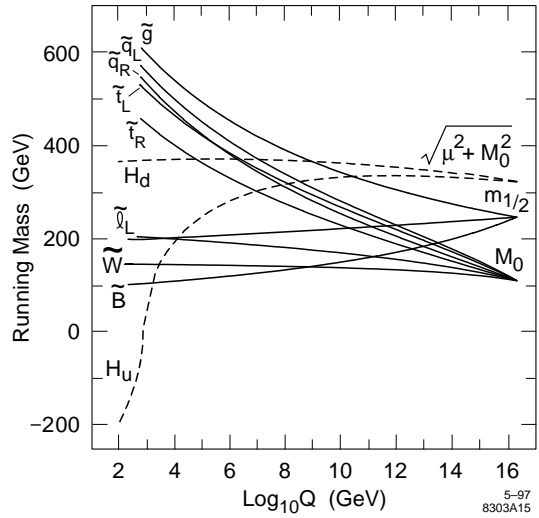}}}
}
\caption{Running of the scalar masses in constrained minimal
supersymmetry. Here $M_0$ is the assumed common soft-SUSY breaking scalar
mass and $m_{1/2}$ the corresponding gaugino masses. 
From \protect\cite{Kane:1994td}.}
\label{fig:twelve}
\end{figure}

Ultimately, supersymmetric models should {\it explain} the negative
mass-squared for Higgs and the absence of vevs for the sfermions. One
common approach is ``constrained'' SUSY breaking, which assumes {\it
  common} scalar masses ($m_0$) at SUSY breaking scale. Since we do not
want the QCD interactions to break, we must assume that $m^2_0 > 0$. If
this is the case, how does the electroweak symmetry break?  As shown in
Fig. \ref{fig:twelve}, the large top-quark Yukawa coupling drives
$M^2_{H_2}$ negative first!  This radiatively-induced origin for
electroweak symmetry breaking is elegant and successful, so long as
there is an explanation for why $\mu$ is of order 1 TeV.

\subsection{SUSY Higgs Phenomenology}

\begin{figure}
\centerline{
\epsfig{file=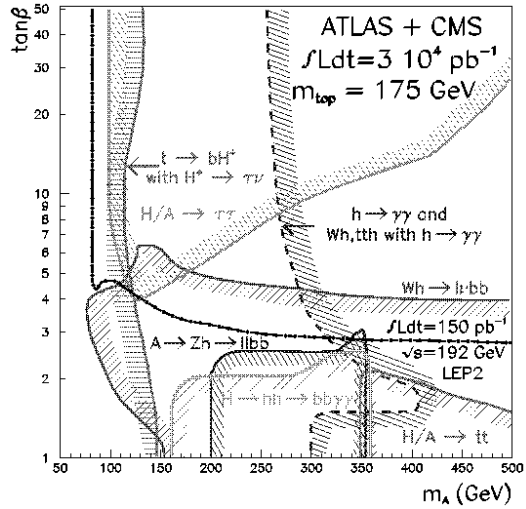,width=7cm}
}
\caption{Reach of LHC to discover one or more of the particles in the
minimal SUSY Higgs sector. Assumes an integrated luminosity of
$3 \times 10^4$ pb$^{-1}$ and 2-detectors. Note that there is a slight
``hole'' for $m_A \sim 250$ GeV and $\tan\beta \sim 6$. 
From \protect\cite{Froidevaux}.}
\label{fig:thirteen}
\end{figure}

\begin{figure}
\centerline{
\epsfig{file=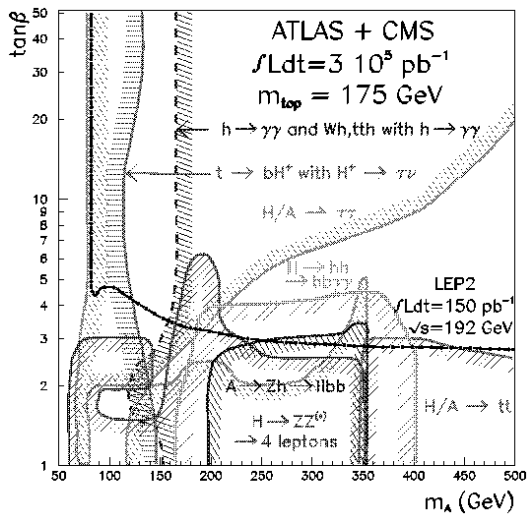,width=7cm}
}
\caption{Same as Fig. \protect\ref{fig:thirteen}, but assuming 
  an integrated luminosity of $3 \times 10^5$ pb$^{-1}$. Note that the
  ``hole'' for $m_A \sim 250$ GeV and $\tan\beta \sim 6$ in the previous
  figure is now covered.  From \protect\cite{Froidevaux}.}
\label{fig:fourteen}
\end{figure}

The reach of the LHC to discover one or more of the particles in the
minimal SUSY Higgs sector is shown in Fig. \ref{fig:thirteen}, for an
integrated luminosity of $3\times 10^4$ pb$^{-1}$, and in Fig.
\ref{fig:fourteen}, for an for an integrated luminosity of $3\times 10^5$
pb$^{-1}$. Note that one ``year'' ($10^7$ s) at design luminosity
corresponds to $10^5$ pb$^{-1}$.


While the results in Fig. \ref{fig:thirteen} and (especially) Fig.
\ref{fig:fourteen} are reassuring, the difficult issues will be to
distinguish this signal from that of the standard Higgs. This is
particularly a problem in the large-$m_A$ region where the
decoupling-limit insures that the couplings of the $h^0$ are identical
to those of the standard Higgs. Establishing the minimal SUSY model
will certainly require the discovery of superpartners, in addition
to an exploration of the symmetry breaking sector.

\section{Dynamical Electroweak Symmetry Breaking}

\subsection{Technicolor}

The simplest theory of dynamical electroweak symmetry
breaking is technicolor \cite{Weinberg:1979bn,Susskind:1978ms}. Consider
an $SU(N_{TC})$ gauge theory with fermions in the fundamental
representation of the gauge group
\beq
\Psi_L=\left(
\begin{array}{c}
U\\D
\end{array}
\right) _L\,\,\,\,\,\,\,\,
U_R,D_R
\eeq
The fermion kinetic energy terms
for this theory are
\beqa
{\cal L} &=& \bar{U}_L i\slashchar{D} U_L+
\bar{U}_R i\slashchar{D} U_R+\\
 & &\bar{D}_L i\slashchar{D} D_L+
\bar{D}_R i\slashchar{D} D_R~,
\nonumber
\eeqa
and, like QCD in the $m_u$, $m_d \to 0$ limit, they have
a chiral $SU(2)_L \times SU(2)_R$ symmetry.

As in QCD, exchange of technigluons in the spin zero, isospin zero
channel is attractive causing the formation of a condensate
\beq
{\lower15pt\hbox{
\epsfysize=0.5 truein \epsfbox{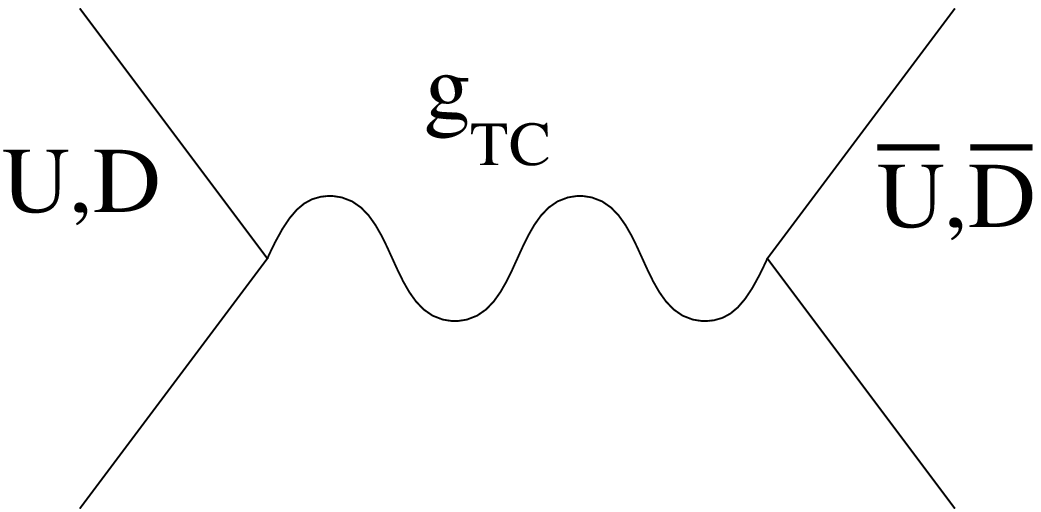}}
\ \ \rightarrow  \ \ \langle \bar U_LU_R\rangle
=\langle \bar D_LD_R\rangle \neq 0\, ,}
\eeq
which dynamically breaks $SU(2)_L \times SU(2)_R \to SU(2)_V$.  These
broken chiral symmetries imply the existence of three massless Goldstone
bosons, the analogs of the pions in QCD.

Now consider gauging $SU(2)_W \times U(1)_Y$ with the left-handed
fermions transforming as weak doublets and the right-handed ones as weak
singlets. To avoid gauge anomalies, in this one-doublet technicolor
model we will take the left-handed technifermions to have hypercharge
zero and the right-handed up- and down-technifermions to have
hypercharge $\pm 1/2$.  The spontaneous breaking of the chiral symmetry
breaks the weak-interactions down to electromagnetism. The would-be
Goldstone bosons become the longitudinal components of the $W$ and $Z$
\beq
\pi^\pm,\, \pi^0 \, \rightarrow\, W^\pm_L,\, Z_L~,
\eeq
which
acquire a mass
\beq
M_W = {g F_{TC} \over 2}~.
\eeq
Here $F_{TC}$ is the analog of $f_\pi$ in QCD. In order
to obtain the experimentally observed masses, we must have
that $F_{TC} \approx 246 {\rm GeV}$ and hence this model
is essentially QCD scaled up by a factor of
\beq
{F_{TC}\over f_\pi} \approx 2500\, .
\eeq

While I have described only the simplest model above, from the
discussion of section 5.1 (see fig. \ref{fig:nine}) it is
straightforward to generalize to other cases.  {\it Any} strongly
interacting gauge theory with a chiral symmetry breaking pattern $G \to
H$, in which $G$ contains $SU(2)_W \times U(1)_Y $ and breaks to a
subgroup $H \supset U(1)_{em}$ (with $SU(2)_W \times U(1)_Y \not\subset
H$) will break the weak interactions down to electromagnetism.  In order
to be consistent with experimental results, however, we must also
require that $H$ contain ``custodial'' $SU(2)_C$. This custodial
symmetry insures that the $F$-constant associated with the $W^\pm$ and
$Z$ are equal and therefore that the relation
\beq
\rho = {M_W \over M_Z \sin\theta_W} =1
\eeq
is satisfied at tree-level.  If the chiral symmetry is larger than
$SU(2)_L\times SU(2)_R$, theories of this sort will contain additional
(pseudo-)Goldstone bosons which are not ``eaten'' by the $W$ and $Z$.
For simplicity, in this lecture we will discuss the phenomenology of the
one-doublet model\footnote{For a review of the phenomenology of
  non-minimal models, see \protect\cite{Chivukula:1995dt}.}.

\subsection{The Phenomenology of Dynamical Electroweak Symmetry Breaking}

Of the particles that we have observed to date, the only ones directly
related to the electroweak symmetry breaking sector are the longitudinal
gauge-bosons\footnote{Except, possibly, for the third generation of
  fermions. See the discussion of topcolor in section 11.}. Therefore,
we expect the most direct signatures for electroweak symmetry breaking
to come from the scattering of longitudinally gauge bosons. As discussed
in section 2, at {\it high energies}, we may use the equivalence theorem
\cite{Cornwall:1974km,Vayonakis:1976vz,Chanowitz:1985hj}
\beq
{\cal A}(W_L W_L) = {\cal A}(\pi \pi) + {\cal O}({M_W\over E})\, .
\eeq
to reduce the problem of longitudinal gauge boson ($W_L$) scattering to
the corresponding (and generally simpler) problem of the scattering of
the Goldstone bosons ($\pi$) that would be present in the absence of the
weak gauge interactions.

In order to correctly describe the weak interactions, the symmetry
breaking sector must have an (at least approximate) custodial
symmetry \cite{Weinberg:1979bn,Susskind:1978ms,Sikivie:1980hm}, and the
most general effective theory describing the behavior of the Goldstone
bosons is an effective chiral lagrangian\footnote{For a review, see the
  lectures by A. Pich in this volume.} with an $SU(2)_L \times SU(2)_R
\to SU(2)_V$ symmetry breaking pattern. This effective lagrangian is
most easily written in terms of a field
\beq
\Sigma = \exp(i\pi^a\sigma^a/F_{TC})~,
\eeq
where the $\pi^a$ are the Goldstone boson
fields, the $\sigma^a$ are the Pauli matrices, and
where the field $\Sigma$ which transforms as 
\beq
\Sigma  \to L \Sigma R^\dagger
\eeq
under $SU(2)_L \times SU(2)_R$.

The interactions can then be ordered in a power-series in
momenta.  Allowing for custodial $SU(2)$ violation, the
lowest-order terms in the effective theory are
the gauge-boson kinetic terms
\beq
 -{1\over2}\tr\left[\W^{\mu\nu}\W_{\mu\nu}\right]- 
 {1\over2}\tr\left[\B^{\mu\nu} \B_{\mu\nu}\right]\, ,
\eeq
and the ${\cal O}(p^2)$ terms
\beq
{  {F_{TC}^2 \over4} \tr\left[D^\mu
   \Sigma^{\dagger}D_\mu \Sigma\right] +  {F_{TC}^2 \over 2} ({1\over\rho} - 1) 
\left[ \tr T_3 \Sigma^\dagger D^\mu \Sigma \right]^2}\, ,
\label{psquare}
\eeq
where
\beq
D_\mu \Sigma=\partial_\mu \Sigma+\im g\W_\mu 
\Sigma-\im \Sigma g'\B_\mu~.
\eeq
In unitary gauge, $\Sigma=1$ and the lowest-order
terms in eq. (\ref{psquare}) give rise to the
$W$ and $Z$ masses
\beq
{g^2 F_{TC}^2\over 4} W^{-\mu} W^+_{\mu} + {g^2 F_{TC}^2\over{8 \rho \cos^2\theta}}
Z^\mu Z_\mu
~.
\eeq

So far, the description we have constructed is valid in {\it any} theory
of electroweak symmetry breaking.  The interactions in eq.
(\ref{psquare}) result in {\it universal} low-energy
\cite{Chanowitz:1987vj,Chanowitz:1986hu} theorems shown in eq.
(\ref{eq:universal}).  These amplitudes increase with energy and, at
some point, this growth must stop \cite{Veltman:1977rt,Lee:1977yc}.
What dynamics cuts off growth in these amplitudes?  In general, there
are three possibilities:

\begin{narrower}
\begin{itemize}
\item new particles
\item the born approximation fails $\rightarrow$ strong interactions
\item both.
\end{itemize}
\end{narrower}

\begin{figure}[tbp]
\centering
\epsfxsize=7cm
\hspace*{0in}
\epsffile{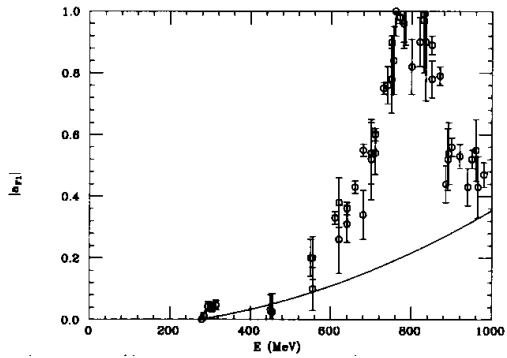}
\caption{QCD data \protect\cite{Donoghue:1988xa}and low-energy theorem
(solid line) prediction for the magnitude of the spin-1/isospin-1 
pion scattering amplitude $|a_{F1}|$.}
\label{Fig2}
\end{figure}

In the case of QCD-like technicolor, we take our inspiration
from the familiar strong interactions. The
data for $\pi\pi$ scattering in QCD in the
$I=J=1$ channel is shown in Figure \ref{Fig2}.
After correcting for the finite pion mass, we see that
the scattering amplitude follows the low-energy prediction
near threshold, but at higher energies the amplitude is
dominated by the $\rho$-meson whose appearance both
enhances the scattering cross-section and cuts-off
the growth of the scattering amplitude at higher energies.
In a QCD-like technicolor theory, then, we expect
the appearance of a vector meson whose mass we
estimate by scaling by $F_{TC}/f_\pi \approx 2500$. That is,
\beq
M_{\rho_{TC}} \approx 2 \tev \, \sqrt{3\over N_{TC}}~,
\eeq
where we have included large-$N_{TC}$ scaling to estimate the
effect of $N_{TC} \neq 3$ \cite{'tHooft:1974jz}.

\begin{figure}[tbp]
\centering
\hskip-5pt\hbox{
\epsfig{file=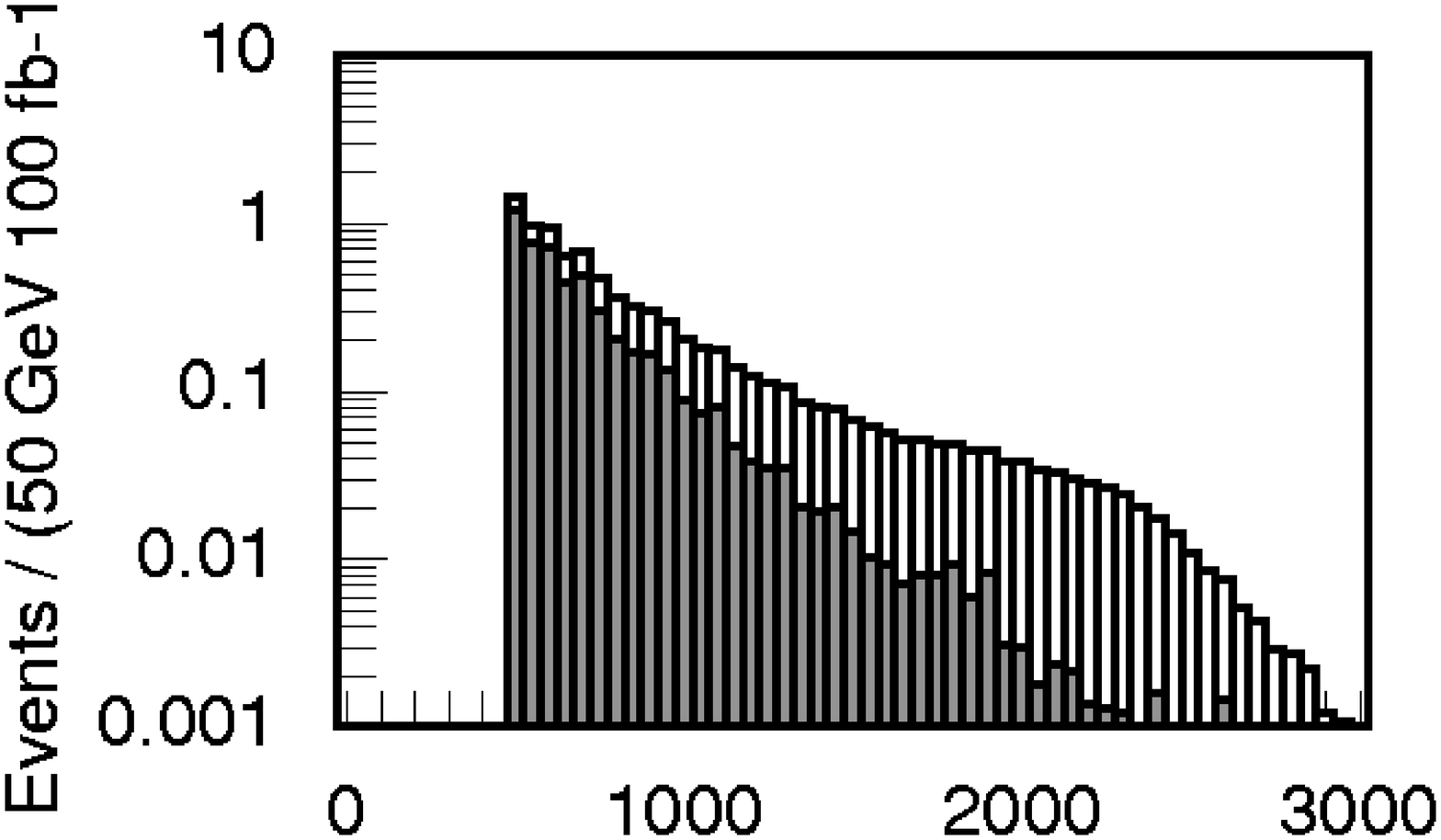,width=5cm}
\hskip12pt
\epsfig{file=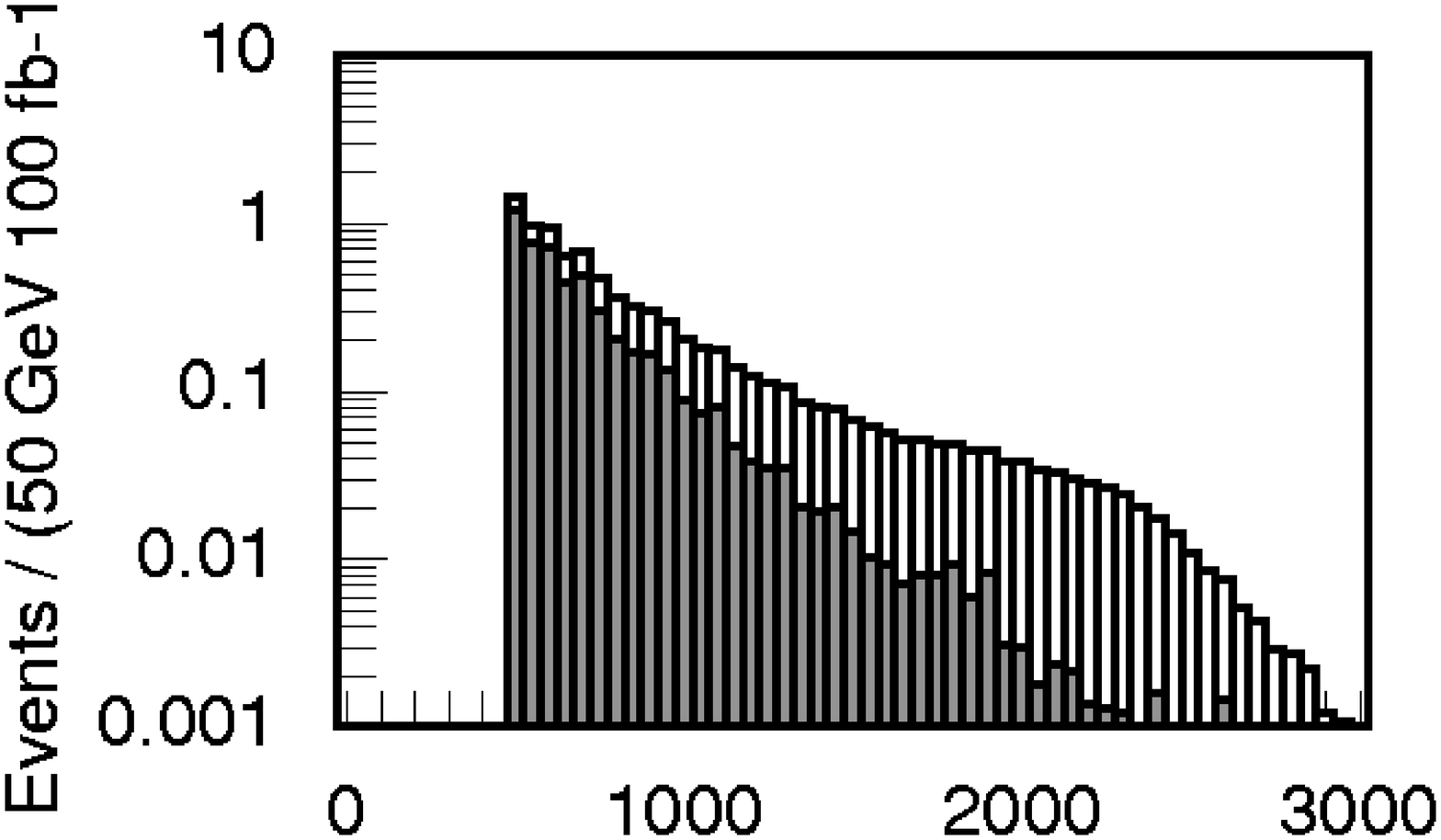,width=5cm}
}
\epsfig{file=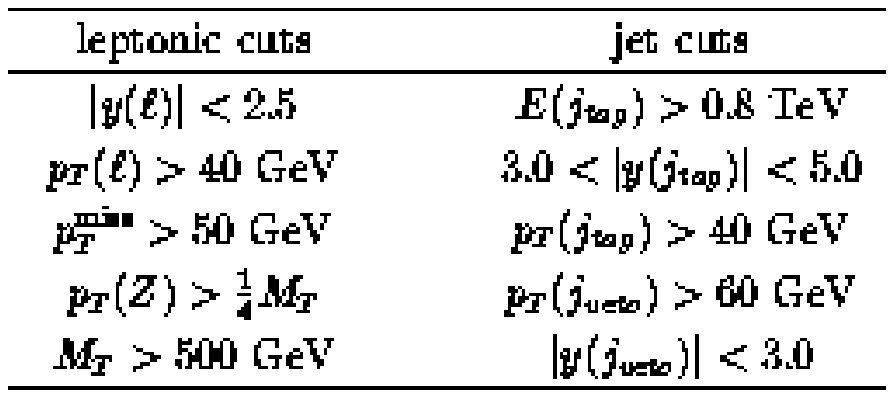, width=6cm}
\caption{Gauge boson scattering signal plus background (grey) and background (black) for
$W^\pm Z$ production \protect\cite{Bagger:1994zf,Bagger:1995mk} at LHC for technirho masses of 1.0 TeV and
2.5 TeV. Signal selection requirements shown in table above.}
\label{Fig3}
\end{figure}

The appearance of these technivector mesons would provide
the most direct experimental signature of dynamical
electroweak symmetry breaking. At the LHC, gauge boson
scattering occurs through the following process,
\beq
{\lower20pt\hbox{{\epsfxsize=2cm \epsfbox{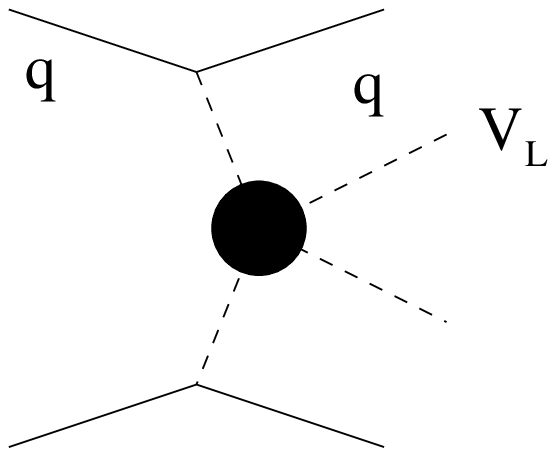}}}}~,
\eeq
and would receive contributions from technivector meson exchange.
Note that, in addition to high-$p_T$ gauge bosons, one
expects forward ``tag'' jets (with a typical transverse
momentum of order $M_W$) from the quarks which radiate
the initial gauge bosons.
The signal expected is shown \cite{Bagger:1994zf,Bagger:1995mk} in Figure \ref{Fig3}
for $M_{\rho_{TC}}= 1.0\,\tev, 2.5\,\tev$. Note the
scale: events per 50 GeV bin of transverse mass ($M_T$)
per 100 fb$^{-1}$!

\begin{figure}[tbp]
\centering
\epsfig{file=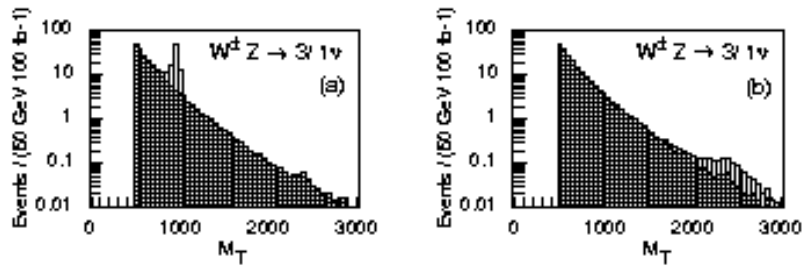,width=11cm}
\caption{Vector meson mixing signal plus background (grey) and background (black) for
$W^\pm Z$ production \protect\cite{Golden:1995xv} at LHC for technirho masses of (a) 1.0 TeV and
(b) 2.5 TeV.}
\label{Fig4}
\end{figure}

An additional signal is provided through the technicolor
analog of ``vector-meson dominance.'' In particular, the
$W$ and $Z$ can mix with the technirho mesons in a manner
exactly analogous to $\gamma$-$\rho$ mixing in QCD:
\beq
{\lower5pt\hbox{{\epsfxsize=2cm \epsfbox{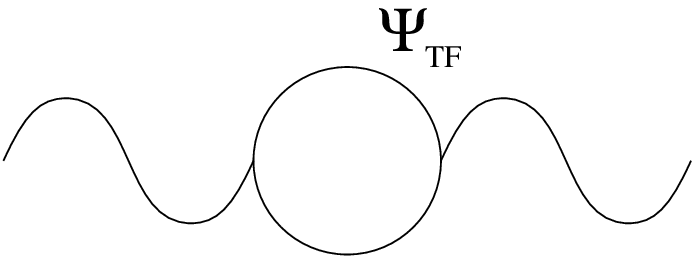}}}}
\hskip5pt 
\rightarrow
\hskip5pt
{\lower15pt\hbox{{\epsfxsize=2cm \epsfbox{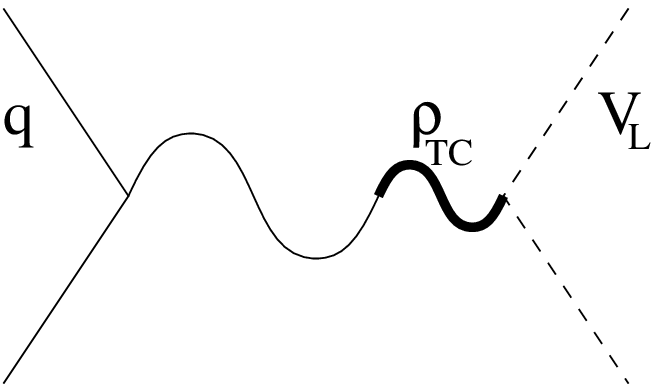}}}}~.
\eeq
Note that this process does {\it not} have a very
forward jet and is distinguishable from the gauge boson
scattering signal discussed above.
The vector-meson mixing signal \cite{Golden:1995xv} at the LHC is shown in
Figure \ref{Fig4} for $M_{\rho_{TC}}= 1.0\,\tev, 2.5\,\tev$.

\begin{figure}[tbp]
\centering
\hskip0.3cm\hbox{\epsfxsize=5cm\epsfbox{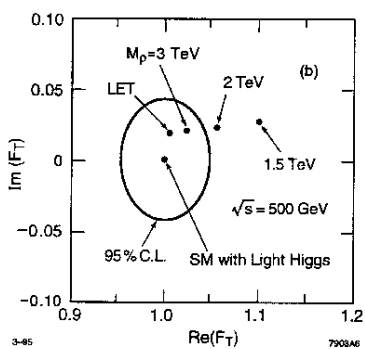}}
\caption{$ZWW$ form-factor measurement \protect\cite{Barklow1996} 
at a 500 GeV NLC with 80fb$^{-1}$. Predictions are shown for the standard model,
and for technicolor for various technirho masses.}
\label{Fig5}
\end{figure}

A dynamical electroweak symmetry breaking sector will
also have affect two gauge-boson production at a
high-energy $e^+ e^-$ collider such as the NLC.
For example, if gauge-boson re-scattering~\footnote{If
the technicolor theory satisfies a ``KSRF'' relation \cite{Kawarabayashi:1966kd,Riazuddin1966},
this ``re-scattering'' effect is exactly equivalent to the
vector-meson mixing effect discussed above \cite{Kroll:1967it}.}
\beq
{\hbox{{\epsfxsize=3cm \epsfbox{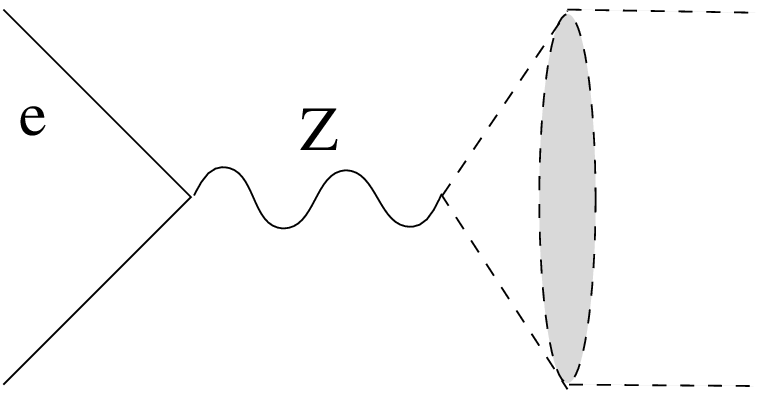}}}}
\eeq
is dominated by a technirho meson, it 
can be parameterized in terms of a $ZWW$ form-factor
\beq
 F_T =
         \exp\bigl[{1\over \pi} \int_0^\infty
          ds'\delta(s',M_\rho,\Gamma_\rho)
          \{{1\over s'-s-i\epsilon}-{1\over s'}\}
         \bigr]~,
\eeq
where
\beq
\delta(s) = {1\over 96\pi} {s\over v^2}
+ {3\pi\over 8} \left[ \tanh (
{
s-M_\rho^2
\over
M_\rho\Gamma_\rho
}
)+1\right]~.
\eeq
This two gauge-boson production mechanism interferes with
continuum production, and by an accurate measurement of 
the decay products it is possible \cite{Barklow1996} to reconstruct
the real and imaginary parts of the form-factor $F_T$.
The expected accuracy of a 500 GeV NLC with 80 fb$^{-1}$ is
shown in Figure \ref{Fig5}.

\subsection{Low-Energy Phenomenology}

Even though the most direct signals of a dynamical electroweak symmetry
breaking sector require (partonic) energies of order 1 TeV, there are
also effects which may show up at lower energies as well. While the
${\cal O}(p^2)$ terms in the effective lagrangian are universal, terms of
higher order are model-dependent.  At energies below $M_{\rho_{TC}}$,
there are corrections to 3-pt functions:
\beq
{\lower15pt\hbox{\epsfysize=0.75 truein \epsfbox{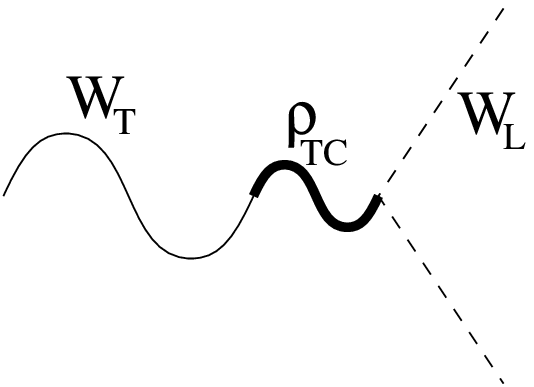}}}~,
\eeq
which, following Gasser and Leutwyler
\cite{Weinberg:1979kz,Manohar:1984md,Georgi1984,Gasser:1984yg,Gasser:1985gg},
give rise to the {\cal O}($p^4$) terms
\beq
 -\ i g {{\it l}_{9L} \over 16 \pi^2}\, \tr {\W^{\mu \nu} D_\mu
\Sigma D_\nu \Sigma^\dagger}~,
\eeq
and
\beq
-\ i g' {{\it l}_{9R} \over 16 \pi^2}\, \tr {\B^{\mu \nu}
D_\mu \Sigma^\dagger D_\nu\Sigma}~.
\eeq
There are also corrections to the 2-pt functions:
\beq
{\lower5pt\hbox{\epsfysize=0.3 truein \epsfbox{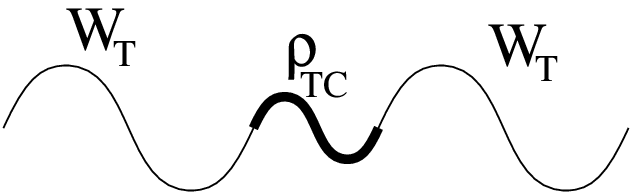}}}
\eeq
which give rise to
\beq
\  +\ g g' {{\it l}_{10}\over 16 \pi^2}\, \tr {\Sigma \B^{\mu \nu}
\Sigma^\dagger \W_{\mu \nu}}
~.
\eeq
In these expressions, the {\it l}'s are normalized to be {\cal O}(1).

The corrections to the 3-point functions are typically parameterized,
following Hagiwara, {\it et. al.} \cite{Hagiwara:1987vm},
parameterized as:
\beqa
{i\over e \cot\theta} & {\cal L}_{WWZ}  =  g_1
(W^\dagger_{\mu\nu} W^\mu Z^\nu - W^\dagger_\mu Z_\nu W^{\mu\nu}) \nonumber \\
& + \kappa_Z W^\dagger_\mu W_\nu Z^\mu\nu + {\lambda_Z\over
M_W^2}W^\dagger_{\lambda\mu}W^\mu_\nu Z^{\nu\lambda}~,
\eeqa
and
\beqa
{i\over e}  & {\cal L_{WW\gamma}} =  (W^\dagger_{\mu\nu} W^\mu A^\nu -
W_\mu^\dagger A_\nu W^{\mu\nu})\nonumber \\
&  + \kappa_\gamma W^\dagger_\mu W_\nu
F^\mu\nu + {\lambda_\gamma\over M_W^2} W^\dagger_{\lambda\mu}W^\mu_\nu
F^{\nu\lambda}~.
\eeqa
Re-expressing these coefficients in terms of 
the parameters in ${\cal L}_{p^4}$ given above, we find
\beq
\left.
\begin{array}{c}
g_1 - 1\\
\kappa_Z - 1\\
\kappa_\gamma - 1
\end{array}
\right\}
\ \approx\ {\alpha_* l_i \over 4\pi \sin^2\theta} = {\cal O}(10^{-2}-10^{-3})~,
\eeq
and $\lambda_{Z,\gamma}$ from ${\cal L}_{p^6}$ implying that
\beq
\lambda_{Z,\gamma} = {\cal O}(10^{-4}-10^{-5})\, .
\eeq

\begin{figure}[tbp]
\centering
\hskip-5pt\hbox{
{\epsfxsize=5cm \epsfbox{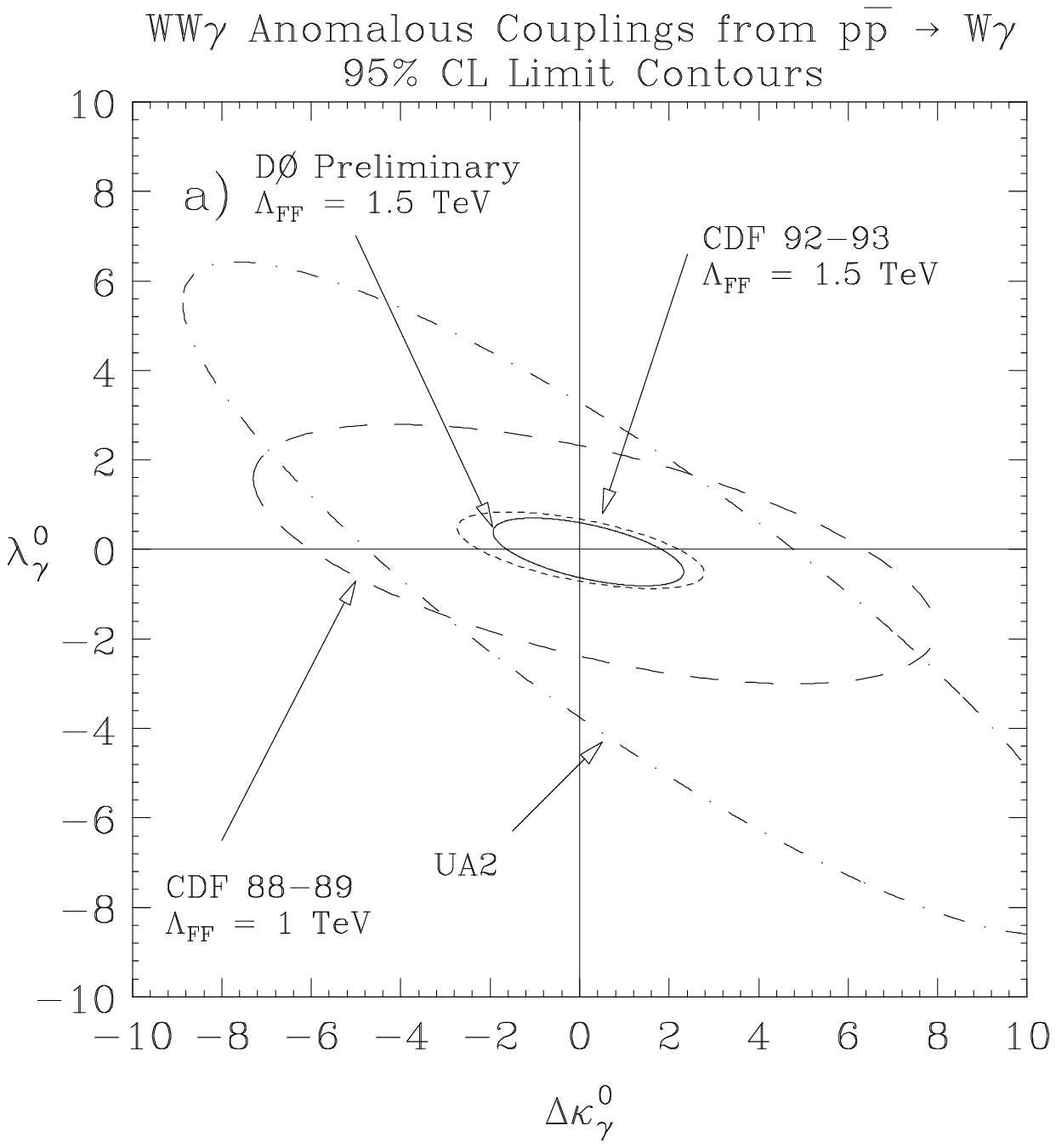}}
\hskip12pt
{\epsfxsize=5cm \epsfbox{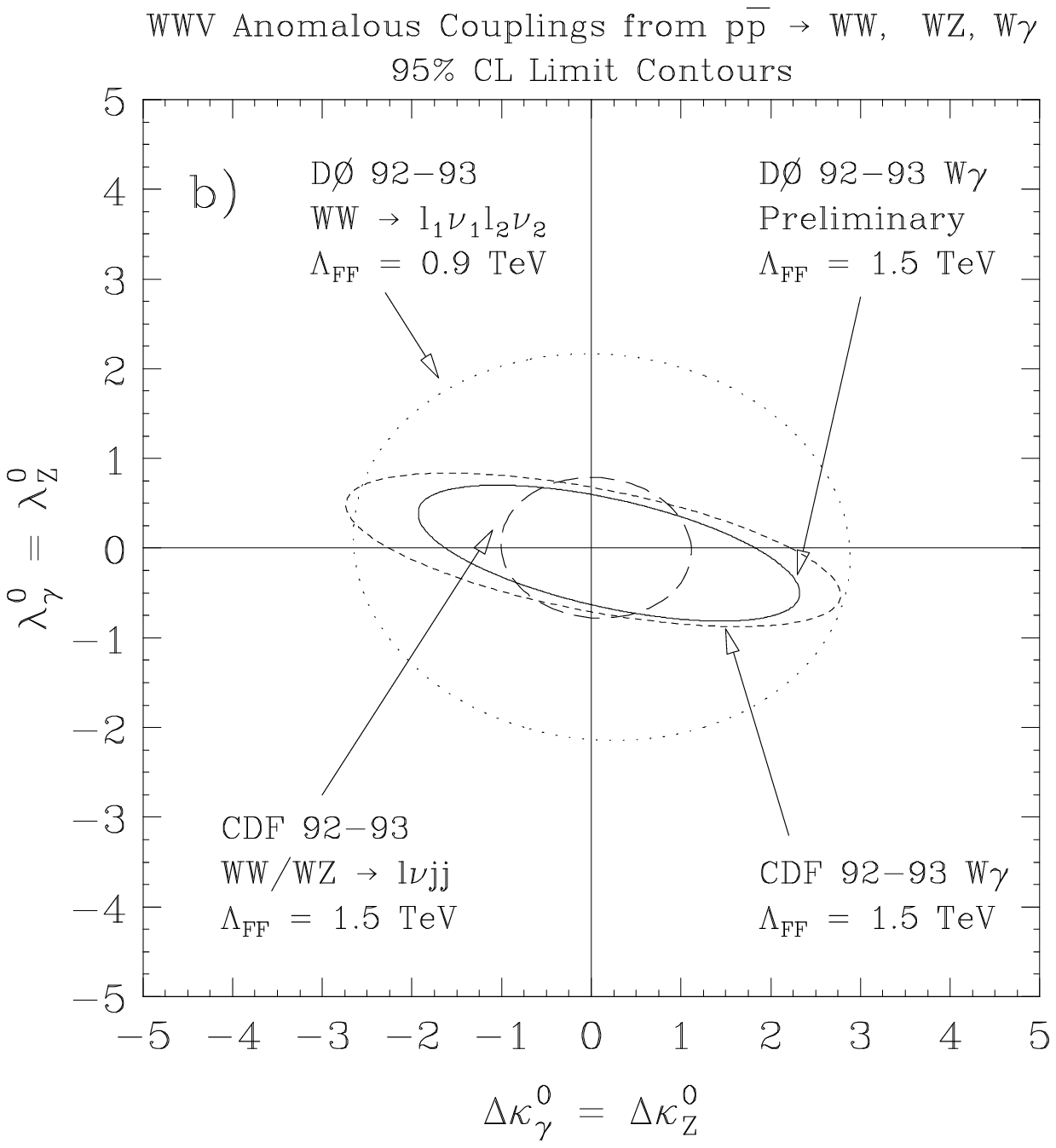}}
}
\caption{Current limits \protect\cite{Aihara:1995iq} on anomalous 
gauge-boson vertices from Tevatron data.}
\label{Fig6}
\end{figure}

\begin{figure}[tbp]
\centering
\epsfig{file=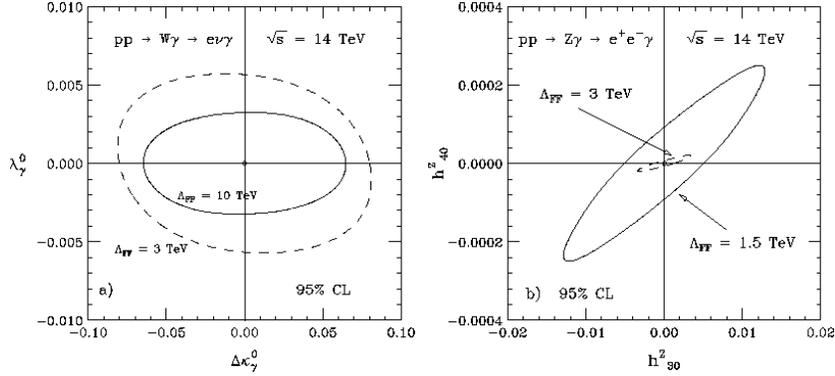, width=11cm}
\caption{Experimental \protect\cite{Aihara:1995iq} reach of LHC to probe
anomalous gauge-boson vertices given an integrated luminosity
of 100 fb$^{-1}$.}
\label{Fig7}
\end{figure}

\begin{figure}[tbp]
\centering
\epsfig{file=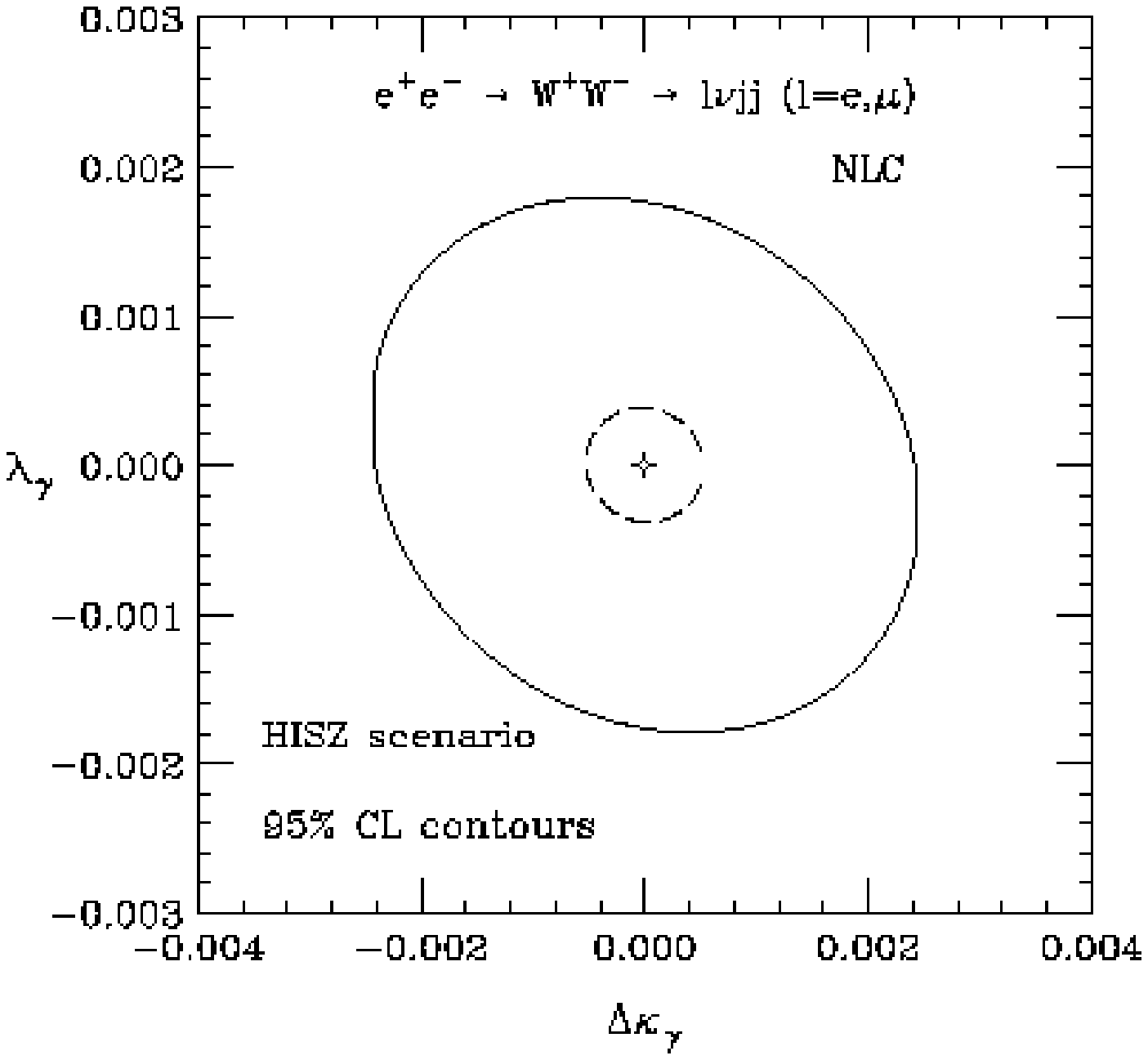,width=5cm}
\caption{Experimental \protect\cite{Aihara:1995iq} reach of a
500 GeV (solid) or 1500 GeV (dashed) NLC to probe
anomalous gauge-boson vertices, assuming 80 fb$^{-1}$ or
190 fb$^{-1}$ respectively.}
\label{Fig8}
\end{figure}

The best current limits \cite{Aihara:1995iq}, coming from Tevatron experiments,
are shown in Figure \ref{Fig6}. Unfortunately, they do not reach the
level of sensitivity required. The situation \cite{Aihara:1995iq} is somewhat
improved at the LHC, as shown in Figure \ref{Fig7}, or at a 500 or 1500
GeV NLC, as shown in Figure \ref{Fig8}.

\begin{figure}[tbp]
\centering
\epsfxsize 10cm \centerline{\epsffile{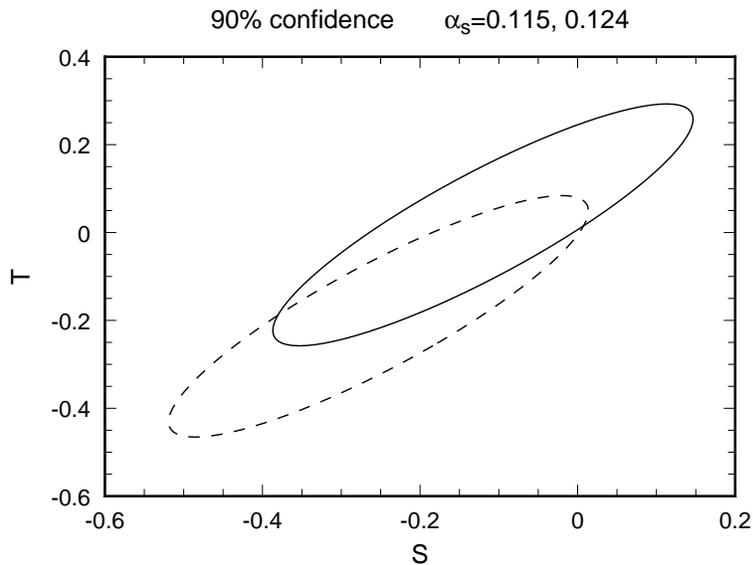}}
\caption{90\% confidence level bounds \protect\cite{Terning1996} on
the oblique parameters $S$ and $T$ for 
$\alpha_S = 0.115$ (solid) and 0.124 (dashed).}
\label{Fig9}
\end{figure}

The corrections
\cite{Peskin:1990zt,Peskin:1992sw,Golden:1991ig,Holdom:1990tc,Dobado:1991zh}
to the 2-point functions give rise to contributions to the ``oblique
parameters'' $S$
\beqa
&S&\equiv 16\pi \left[ \Pi'_{33}(0) - \Pi'_{3Q}(0)\right] \nonumber\\
&=& -\pi {\it l}_{10}\approx
4\pi\left({F^2_{\rho_{TC}}\over M^2_{\rho_{TC}}}
-{F^2_{A_{TC}}\over M^2_{A_{TC}}}\right) N_D~,
\eeqa
and $T$
\beq
\alpha T \equiv {g^2 \over {\cos^2\theta M_Z^2}} \left[\Pi_{11}(0) -
\Pi_{33}(0)\right] = \rho-1~.
\eeq
$S$ and $T$ measure the size of custodial-symmetry conserving and
violating radiative corrections to the gauge boson propagators {\it
  beyond} the corrections present in the standard model.  Current
experimental constraints \cite{Terning1996} imply the bounds shown in
Figure \ref{Fig9}, at 95\% confidence level for different values of
$\alpha_S$.  Scaling from QCD, we expect a contribution to S of order
\beq
S \approx 0.28 N_D (N_{TC}/3)~,
\eeq
for an $SU(N_{TC})$ technicolor theory with $N_D$ technidoublets.
From these we see that, with the possible exception of 
$N_D=1$ and $N_{TC}=2$ or 3, {\it QCD-like} technicolor is in
conflict with precision weak measurements.

In summary, dynamical electroweak symmetry breaking provides a natural
and attractive mechanism for producing the $W$ and $Z$ masses. {\it
  Generically} models of this type predict strong $WW$-Scattering,
signals of which may be observable at the LHC. While the simplest
QCD-like models serve as a useful starting point, they are excluded
(except, perhaps, for an $SU(2)_{TC}$ model with one doublet) since they
would give rise to unacceptably large contributions to the $S$
parameter. In the next section we will discuss the additional
interactions and features required in more realistic models to give
rise to the masses to the ordinary fermions.

\section{Flavor Symmetry Breaking and ETC}

\subsection{Fermion Masses \& ETC Interactions}

In order to give rise to masses for the ordinary quarks and leptons, we
must introduce interactions which connect the chiral-symmetries of
technifermions to those of the ordinary fermions. The most popular
choice \cite{Eichten:1979ah,Dimopoulos:1979es} is to introduce new
broken gauge interactions, called {\it extended technicolor
  interactions} (ETC), which couple technifermions to ordinary fermions.
At energies low compared to the ETC gauge-boson mass, $M_{ETC}$, these
effects can be treated as local four-fermion interactions
\beq
{\lower15pt\hbox{\epsfysize=0.5 truein \epsfbox{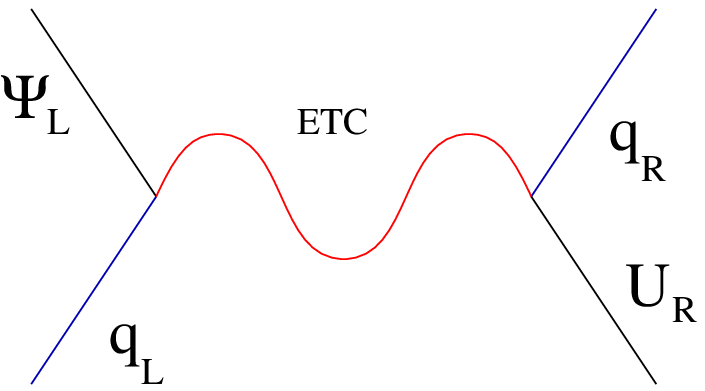}}}
\ \ \rightarrow\ \  {{g_{ETC}^2\over M^2_{ETC}}}(\overline{\Psi}_L U_R)
({\overline{q}_R q_L})~.
\label{etcint}
\eeq
After technicolor chiral-symmetry breaking and the formation of a
$\langle \bar{U} U \rangle$ condensate, such an interaction gives rise
to a mass for an ordinary fermion
\beq
m_q \approx {{g_{ETC}^2\over M^2_{ETC}}} \langle\overline{U} U\rangle_{ETC}~,
\label{fmass}
\eeq
where $\langle \overline{U} U\rangle_{ETC}$ is the value of the
technifermion condensate evaluated at the ETC scale (of order
$M_{ETC}$).  The condensate renormalized at the ETC scale in eq.
(\ref{fmass}) can be related to the condensate renormalized at the
technicolor scale as follows
\beq
\langle\overline{U} U\rangle_{ETC} = \langle\overline{U} U\rangle_{TC}
\exp\left(\int_{\Lambda_{TC}}^{M_{ETC}} {d\mu \over \mu}
\gamma_m(\mu)\right)~,
\eeq
where $\gamma_m(\mu)$ is the anomalous dimension of the
fermion mass operator and $\Lambda_{TC}$ is the analog of $\Lambda_{QCD}$
for the technicolor interactions.

For QCD-like technicolor (or any theory which is ``precociously''
asymptotically free), $\gamma_m$ is small over in the range between
$\Lambda_{TC}$ and $M_{ETC}$ and using dimensional analysis
\cite{Weinberg:1979kz,Georgi1984,Manohar:1984md,Gasser:1984yg,Gasser:1985gg}
we find
\beq
\langle\overline{U} U\rangle_{ETC} \approx \langle\overline{U} U\rangle_{TC}
\approx 4\pi F^3_{TC}~.
\eeq
In this case eq. (\ref{fmass}) implies that
\beq
{{M_{ETC}\over g_{ETC}}} \approx 40 \tev 
\left({F_{TC}\over 250\gev}\right)^{3\over 2}
\left({100 \mev \over m_q}\right)^{1\over 2}~.
\eeq

In order to orient our thinking, it is instructive to consider a simple
``toy'' extended technicolor model. The model is based on an
$SU(N_{ETC})$ gauge group, with technicolor as an extension of flavor.
In this case $N_{ETC} = N_{TC} + N_F$, and we add the (anomaly-free)
set of fermions
\medskip
\begin{center}
$
\begin{array}{l@{\extracolsep{15pt}}l}
Q_L=(N_{ETC},3,2)_{1/6} & L_L=(N_{ETC},1,2)_{-1/2} \\
U_R=(N_{ETC},3,1)_{2/3} & E_R=(N_{ETC},1,1)_{-1} \\
D_R=(N_{ETC},3,1)_{-1/3} & N_R=(N_{ETC},1,1)_{0}~,
\end{array}
$
\end{center}
\medskip\noindent
where we display their quantum numbers under $SU(N_{ETC})\times
SU(3)_C \times SU(2)_W \times U(1)_Y$. We break the
ETC group down to technicolor in three stages
\medskip
\begin{center}
{$SU(N_{TC}+3)$}
\end{center}
\begin{center}
$\Lambda_1\ \ \ \ \ \downarrow \ \ \ \ \  
m_1\approx{4\pi F^3\over \Lambda^2_1}$
\end{center}
\begin{center}
{$SU(N_{TC}+2)$}
\end{center}
\begin{center}
$\Lambda_2\ \ \ \ \ \downarrow \ \ \ \ \  
m_2\approx{4\pi F^3\over \Lambda^2_2}$
\end{center}
\begin{center}
{$SU(N_{TC}+1)$}
\end{center}
\begin{center}
$\Lambda_3\ \ \ \ \ \downarrow \ \ \ \ \  
m_3\approx{4\pi F^3\over \Lambda^2_3}$
\end{center}
\begin{center}
{$SU(N_{TC})$}
\end{center}
\medskip\noindent
resulting in three isospin-symmetric families of degenerate
quarks and leptons, with $m_1 < m_2 < m_3$. Note that the
{\it heaviest} family is related to the {\it lightest} ETC 
scale!

Before continuing our general discussion, it is worth noting
a couple of points. First,
in this example the ETC gauge bosons do not carry color
or weak charge
\beq
[G_{ETC},SU(3)_C]=[G_{ETC},SU(2)_W]=0~.
\label{commute}
\eeq
Furthermore, in this model there is one technifermion for each type of
ordinary fermion: that is, this is a ``one-family'' technicolor model
\cite{Farhi:1979zx}.  Since there are eight left- and right- handed
technifermions, the chiral symmetry of the technicolor theory is (in the
limit of zero QCD and weak couplings) $SU(8)_L \times SU(8)_R \to
SU(8)_V$. Such a theory would yield $8^2-1=63$ (pseudo-)Goldstone
bosons. Three of these Goldstone bosons are unphysical --- the
corresponding degrees of freedom become the longitudinal components of
the $W^\pm$ and $Z$ by the Higgs mechanism.  The remaining 60 must
somehow obtain a mass. This will lead to the condition in eq.
(\ref{commute}) being modified in a realistic model.  We will return to
the issue of pseudo-Goldstone bosons below.

The most important feature of this or any ETC-model is that a successful
extended technicolor model will provide a {\it dynamical theory of
  flavor}! As in the toy model described above and as explicitly
shown in eq. (\ref{etcint}) above, the masses of the ordinary fermions
are related to the masses and couplings of the ETC gauge-bosons. A successful
and complete ETC theory would predict these quantities and, hence, the
ordinary fermion masses. 

Needless to say, constructing such a theory is very difficult. No
complete and successful theory has been proposed.  Examining our toy
model, we immediately see a number of shortcomings of this model that
will have to be addressed in a more realistic theory:

\begin{narrower}
\begin{itemize}
\item What breaks ETC?
\item Do we require a {separate} scale for each family?
\item How do the $T_3 = \pm {1\over 2}$ fermions of a given generation
receive {\it different} masses?
\item How do we obtain quark mixing angles?
\item What about right-handed technineutrinos and $m_\nu$?
\end{itemize}
\end{narrower}

\subsection{Flavor-Changing Neutral-Currents}

Perhaps the single biggest obstacle to constructing a realistic ETC
model (or any dynamical theory of flavor) is the potential for
flavor-changing neutral currents \cite{Eichten:1979ah}.  Quark mixing implies
transitions between different generations: $q \to \Psi \to q^\prime$,
where $q$ and $q'$ are quarks of the same charge from different
generations and $\Psi$ is a technifermion. Consider the commutator of
two ETC gauge currents:
\beq
[\overline{q}\gamma \Psi, \overline{\Psi}\gamma q^\prime]\, \supset\, 
\overline{q}\gamma q^\prime\, .
\eeq
Hence we expect there are ETC gauge bosons which couple to
flavor-changing neutral currents. In fact, this argument is slightly too
slick: the same applies to the charged-current weak interactions!
However in that case the gauge interactions, $SU(2)_W$ respect a global
$(SU(5) \times U(1))^5$ chiral symmetry \cite{Chivukula:1987py} leading
to the usual {GIM} mechanism.

Unfortunately, the ETC interactions {cannot} respect GIM exactly;
they must distinguish between the various generations in order to give
rise to the masses of the different generations. Therefore, flavor-changing
neutral-current interactions are (at least at some level) unavoidable.

The most severe constraints come from possible $|\Delta S| = 2$
interactions which contribute to the $K_L$-$K_S$ mass
difference. In particular, we would expect that in order to produce
Cabibbo-mixing the same interactions which give rise to the $s$-quark
mass could cause the flavor-changing interaction
\beq
{\cal L}_{\vert \Delta S \vert = 2} = {g^2_{ETC} \, \theta^2_{sd} \over
{M^2_{ETC}}} \,\, \left(\overline{s} \Gamma^\mu d\right) \,\, 
\left(\overline{s} \Gamma'_\mu d\right) + {\rm
h.c.}~,
\eeq
where $\theta_{sd}$ is of order the Cabibbo angle. Such an interaction
contributes to the neutral kaon mass splitting
\beq 
(\Delta M^2_K)_{ETC} =  {g^2_{ETC} \, \theta^2_{sd} \over
{M^2_{ETC}}} \, \langle \overline{K^0} \vert \overline{s} \Gamma^\mu d \, \overline{s}
\Gamma'_\mu d \vert K^0 \rangle + {\rm c.c.}
\eeq
Using the vacuum insertion approximation we find
\beq
(\Delta M^2_K)_{ETC} \simeq {g^2_{ETC} \, 
{\rm Re}(\theta^2_{sd}) \over {2 M^2_{ETC}}} \,
f^2_K M^2_K ~.
\eeq
Experimentally we know that
$\Delta M_K <   3.5 \times 10^{-12}\,\mev $ and, hence, that
\beq
{M_{ETC} \over {g_{ETC} \, \sqrt{{\rm Re}(\theta^2_{sd})}}} >  600\,\tev
\eeq
Using eq. (\ref{fmass}) we find that
\beq
m_{q, \ell} \simeq {g_{ETC}^2 \over {M_{ETC}^2}}
\langle\overline{T}T\rangle_{ETC}  < {0.5\,\mev\over{N_D^{3/2} \, \theta_{sd}^2}} \,
\eeq
showing that it will be difficult to produce the {$s$}-quark
mass, let alone the {$c$}-quark!

\subsection{Pseudo-Goldstone Bosons}

As illustrated by our toy model above, a ``realistic'' ETC theory may
require a technicolor sector with a chiral symmetry structure bigger
than the $SU(2)_L \times SU(2)_R$ discussed in detail in the previous
lecture. The prototypical model of this sort is the one-family model
incorporated in our toy model. As discussed there, the theory has an
$SU(8)_L \times SU(8)_R \to SU(8)_V$ chiral symmetry
breaking structure resulting in 63 Goldstone bosons, 3 of which
are unphysical. The quantum numbers of the 60 remaining Goldstone
bosons are shown in table \ref{pgbtab}. Clearly, these objects
cannot be massless in a realistic theory!

\begin{table}[htbp]
\caption{Quantum numbers of the 60 physical Goldstone bosons (and the
corresponding vector mesons) in a one-family technicolor model. Note
that the mesons that transform as  3's of QCD are complex fields.}
\begin{tabular}{@{}lll@{}}   \hline
SU$(3)_C$   & SU$(2)_{V}$ &Particle         \\ \hline
$1$      &$1$  &$P^{0 \prime} \;,\; \omega_T$  \\
$1$      &$3$  &$P^{0,\pm} \;,\; \rho^{0,\pm}_T$  \\
$3$      &$1$  &$P^{0 \prime}_{3} \;,\; \rho^{0 \prime }_{T 3}$  \\
$3$      &$3$  &$P^{0,\pm}_{3} \;,\; \rho^{0,\pm}_{T 3}$  \\
$8$      &$1$  &$P^{0 \prime}_{8} (\eta_T) \;,\; \rho^{0 \prime }_{T 8}$  \\
$8$      &$3$    &$P^{0,\pm}_{8} \;,\; \rho^{0,\pm}_{T 8}$  \\ \hline
\end{tabular}
\label{pgbtab}
\end{table}

In fact, the ordinary gauge interactions break the full $SU(8)_L \times
SU(8)_R$ chiral symmetry explicitly. The largest effects are due to QCD
and the color octets and triplets mesons get masses of order 200 -- 300
GeV, in analogy to the electromagnetic mass splitting
$m_{\pi^+}-m_{\pi^0}$ in QCD. Unfortunately, the others
\cite{Eichten:1979ah} are {massless} to {\cal O}($\alpha$)!

Luckily, the ETC interactions (which we introduced in order to give
masses to the ordinary fermions) are capable of explicitly breaking the
unwanted chiral symmetries and producing masses for these mesons. This
is because in addition to coupling technifermions to ordinary fermions,
some ETC interactions also couple technifermions to themselves
\cite{Eichten:1979ah}. Using Dashen's formula \cite{Dashen:1969eg}, we
can estimate that such an interaction can give rise to an effect of
order
\beq
F^2_{TC} M^2_{\pi_T} \propto {g^2_{ETC} \over M^2_{ETC}}
\langle (\overline{T}T)^2\rangle_{ETC}~.
\label{dashen}
\eeq
In the vacuum insertion approximation for a theory with small
$\gamma_m$, we may rewrite the above formula using eq. (\ref{fmass}) and
find that
\beq
M_{\pi_T} \simeq 55\gev 
\sqrt{m_f \over 1\gev} \sqrt{250 \gev \over F_{TC}}~.
\eeq
It is unclear that this large enough.

In addition, there is a particularly troubling chiral symmetry
in the one-family model. The 
$SU(8)$-current $\overline{Q}\gamma_\mu \gamma_5 Q - 3 \overline{L} \gamma_\mu
\gamma_5 L$ is {spontaneously broken} and {has a color
  anomaly}. Therefore, we have a potentially
dangerous {weak scale axion \cite{Peccei:1977hh,Peccei:1977ur,Weinberg:1978ma,Wilczek:1978pj}}!
An ETC-interaction of the form
\beq
{g^2_{ETC} \over M^2_{ETC}} \left(\overline{Q}_L \gamma^\mu L_L \right)
\left(\overline{L}_R \gamma^\mu Q_R \right)~,
\eeq
is required to give to an axion mass, and
{we must \cite{Eichten:1979ah} embed $SU(3)_C$ in $ETC$.}

Finally, before moving on I would like to note that there is an implicit
assumption in the analysis of gauge-boson scattering presented in the
last section. We have assumed that {\it elastic} scattering dominates.
In the presence of many pseudo-Goldsone bosons, $WW$ scattering could
instead be dominated by {\it inelastic} scattering.  This effect has
been illustrated \cite{Chivukula:1991bx} in an $O(N)$-Higgs model with many
pseudo-Goldstone Bosons, solved in large-N limit.  Instead of the
expected resonance structure at high energies, the scattering can be
{small and structureless} at all energies.

\subsection{ETC etc.}

There are other model-building constraints \cite{Lane:1993wz} on a
realistic TC/ETC theory. A realistic ETC theory:

\begin{narrower}
\begin{itemize}

\item must be asymptotically free,

\item cannot have gauge anomalies,

\item must produce small (or zero) neutrino masses, 

\item cannot give rise to extra massless (or even light) gauge bosons,

\item should generate weak CP-violation without producing unacceptably
  large amounts of strong CP-violation,

\item must give rise to isospin-violation in fermion masses without
  large contributions to $\Delta\rho$ and,

\item must accomodate a large $m_t$ while giving rise to only small
  corrections to $Z\to \overline{b}b$ and $b\to s\gamma$.

\end{itemize}
\end{narrower}

Clearly, building a fully realistic ETC model will be quite difficult!
However, as I have emphasized before, this is because an ETC theory must
provide a complete dynamical explanation of flavor.  In the next
section, I will concentrate on possible solutions to the flavor-changing
neutral-current problem(s). As I will discuss in detail in sections 11
and 12, I believe the outstanding obstacle in ETC or any theory of
flavor is providing an explanation for the top-quark mass, {\it i.e.}
dealing with the last three issues listed above.

\subsection{Technicolor with a Scalar}

At this point, it would be easy to believe that it is impossible to
construct a model of dynamical electroweak symmetry breaking.
Fortunately, there is at least an existence
\cite{Simmons:1989fu,Samuel:1990dq,Kagan:1990az} proof of such a theory:
technicolor with a scalar.\footnote{Such a theory is also the effective
  low-energy model for a ``strong-ETC'' theory in which the ETC
  interactions {\it themselves} participate in electroweak symmetry
  breaking \cite{Appelquist:1989as,Miranskii:1988gk,Matumoto:1989hf}.
  There are also theories in which ordinary fermions get mass through
  their coupling to a {\it technicolored} scalar
  \cite{Kagan:1995qg,Dobrescu:1997kt}.} Admittedly, while electroweak
symmetry breaking has a dynamical origin in this theory, the
introduction of a scalar reintroduces the hierarchy and naturalness
problems we had originally set out to solve.

\begin{figure}[tbp]
\begin{center}
\epsfxsize=10cm\epsfbox{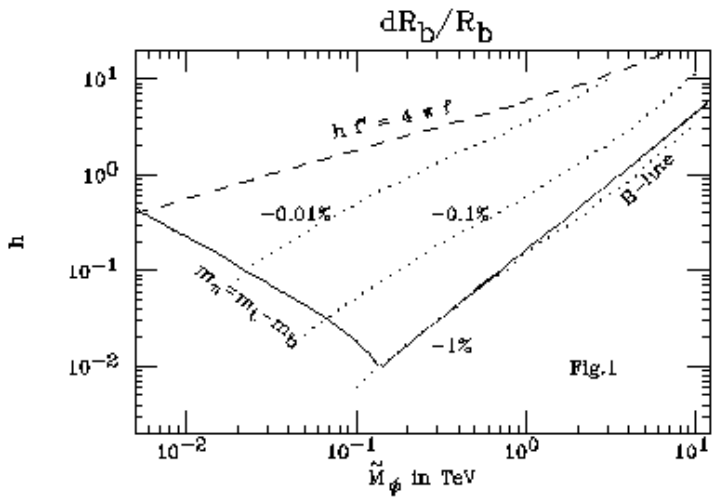}
\end{center}
\caption{Plot of allowed parameter space in model of technicolor
with a scalar \protect\cite{Carone:1995mx}. $h$ is the Yukawa coupling
of the scalar to technifermions and $\tilde{M}_\phi$ is the neutral scalar
mass. The ``triangular'' region formed by the solid and dashed
lines is allowed.}
\label{Fig10}
\end{figure}

In the simplest model one starts with a one doublet technicolor theory, 
and couples the chiral-symmetries of technifermions
to ordinary fermions through {\it scalar} exchange:
\beq
\hbox{\epsfxsize=2.5cm\epsfbox{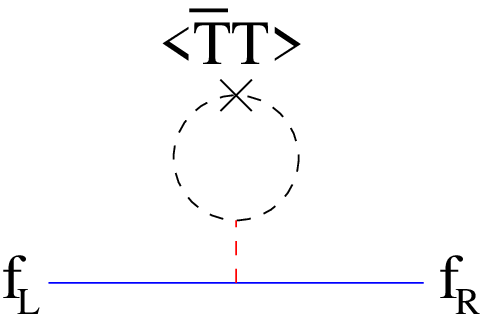}}
\eeq
The phenomenology of this model has been studied in detail \cite{Carone:1993rh},
and the allowed region is shown in Figure \ref{Fig10}.

\section{Walking Technicolor}

\subsection{The Gap Equation}

Up to now we have assumed that technicolor is, like QCD, precociously
asymptotically free and $\gamma_m(\mu)$ is small for $\Lambda_{TC} < \mu
< M_{ETC}$. However, as discussed above it is difficult to construct an
ETC theory of this sort without producing dangerously large
flavor-changing neutral currents. On the other hand, if $\beta_{TC}$ is
{\it small}, $\alpha_{TC}$ can remain large above the scale
$\Lambda_{TC}$ --- {\it i.e.} the technicolor coupling would ``walk''
instead of running. In this same range of momenta, $\gamma_m$ may
be large and, since
\beq
\langle\overline{T} T\rangle_{ETC} = \langle\overline{T} T\rangle_{TC}
\exp\left(\int_{\Lambda_{TC}}^{M_{ETC}} {d\mu \over \mu}
\gamma_m(\mu)\right) 
\eeq
this could enhance $\langle \overline{T} T\rangle_{ETC}$ and fermion
masses
\cite{Holdom:1981rm,Holdom:1985sk,Yamawaki:1986zg,Appelquist:1986an,Appelquist:1987tr,Appelquist:1987fc}.

\begin{figure}[tbp]
\hskip80pt\epsfxsize=6cm\epsfbox{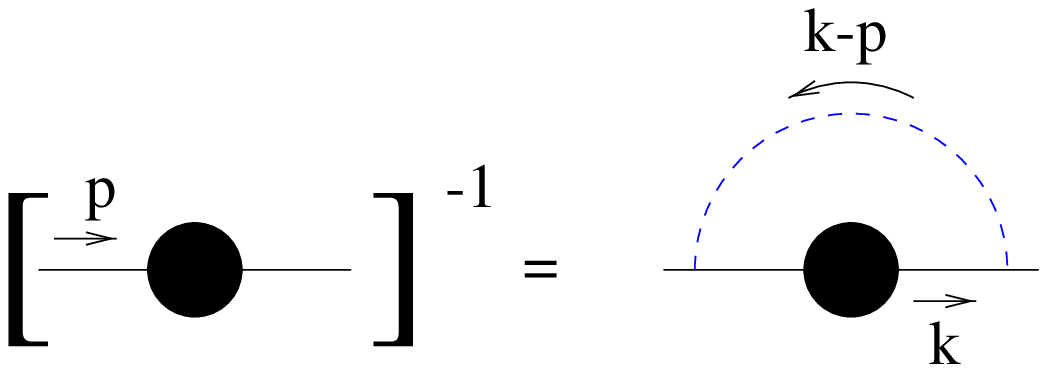}
\caption{Schwinger-Dyson equation for the
fermion self-energy function $\Sigma(p)$ in the rainbow
approximation.}
\label{Fig11}
\end{figure}

In order to proceed further, however, we need to understand how large
$\gamma_m$ can be and how walking affects the technicolor chiral symmetry
breaking dynamics.  These questions cannot be addressed in perturbation
theory.  Instead, what is conventionally done is to use a 
nonperturbative aproximation for $\gamma_m$ and chiral-symmetry breaking
dynamics based on the ``rainbow'' approximation \cite{Pagels:1975se,Peskin:1982mu}
to the Schwinger-Dyson equation shown in Figure \ref{Fig11}.
Here we write the full, nonperturbative, 
fermion propagator in momentum space as
\beq
iS^{-1}(p) = Z(p)(\slashchar{p} - \Sigma(p))~.
\eeq

The linearized form of the gap equation in
Landau gauge (in which $Z(p) \equiv 1$ in the rainbow
approximation) is
\beq
\Sigma(p) = 3 C_2(R)\, \int {d^4 k \over {(2 \pi)^4}}
\, {\alpha_{TC}((k-p)^2) \over {(k-p)^2}} \, {\Sigma(k) \over {k^2}}~.
\eeq
Being separable, this integral equation can be converted to a
differential equation which has the approximate (WKB)
solutions \cite{Fukuda:1976zb,Higashijima:1984gx}
\beq
\Sigma(p) \propto p^{-\gamma_m(\mu)}\, ,\, \, p^{\gamma_m(\mu)-2}\, .
\eeq
Here $\alpha(\mu)$ is assumed to run slowly, as
will be the case in walking technicolor, and
the anomalous dimension of the fermion mass operator is
\beq
\gamma_m(\mu)=1-\sqrt{1-{\alpha_{TC}(\mu)\over\alpha_C}}\, ; \ \ \alpha_C
\equiv {\pi \over 3 C_2(R)}\, .
\label{crit}
\eeq

One can give a physical interpretation of these two
solutions \cite{Lane:1974he,Politzer:1976tv}. Using the operator product expansion, we find
\beq
\lim_{p\to\infty} \Sigma(p) \ \ \propto \ \ 
\lower15pt\hbox{\epsfxsize=4cm\epsfbox{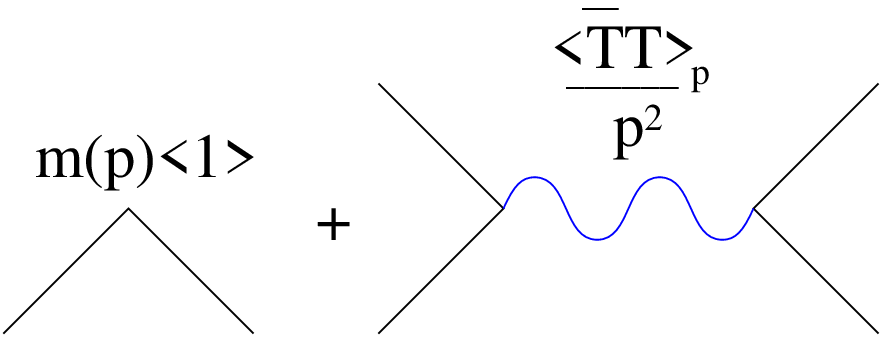}}~,
\eeq
and hence the first solution corresponds to a ``hard mass'' or explicit
chiral symmetry breaking, while the second solution corresponds to a
``soft mass'' or spontaneous chiral symmetry breaking.  If we let $m_0$
be the explicit mass of a fermion, dynamical symmetry breaking occurs
only if
\beq
\lim_{m_0 \to 0} \Sigma(p) \neq 0\, .
\eeq
A careful analysis of the gap equation, or equivalently the appropriate
effective potential \cite{Cornwall:1974vz}, implies that this happens only if
$\alpha_{TC}$ reaches a critical value of chiral
symmetry breaking, $\alpha_C$ defined in eq. (\ref{crit}).
Furthermore, the chiral symmetry breaking scale $\Lambda_{TC}$ is
defined by the scale at which
\beq
\alpha_{TC}(\Lambda_{TC})=\alpha_C 
\eeq
and hence, at least in the rainbow approximation, at which
\beq
\gamma_m(\Lambda_{TC})=1.
\eeq
In the rainbow approximation, then, chiral symmetry breaking occurs when
the {``hard''} and {``soft''} masses scale the same way. It is
believed that even beyond the rainbow approximation $\gamma_m =1$ at the
critical coupling \cite{Appelquist:1988yc,Cohen:1989sq,Mahanta:1989rb}.

\subsection{Implications of Walking: Fermion and PGB Masses, $S$}

If $\beta(\alpha_{TC}) \simeq 0$ all the way from $\Lambda_{TC}$
to $M_{ETC}$, then  $\gamma_m(\mu) \cong 1$ in this
range. In this case, eq. (\ref{fmass}) becomes
\beq
m_{q,l} = {g^2_{ETC} \over {M^2_{ETC}}} \times
\left(\langle\overline{T}T\rangle_{ETC} \cong 
\langle\overline{T}T\rangle_{TC} \, {M_{ETC} \over {\Lambda_{TC}}} \right)~.
\eeq
We have previously estimated that flavor-changing
neutral current requirements imply that the
ETC scale associated with the second generation must
be greater than of order 100 to 1000 TeV. In the case of walking
technicolor the enhancement of the technifermion condensate implies that
\beq
m_{q,l} \simeq {{50\, -\, 500\mev}\over N^{3/2}_D \theta^2_{sd}}~,
\eeq
arguably enough to accomodate the strange and charm quarks.

While this is very encouraging, two {caveats} should be kept in mind.
First, the estimates given are for limit of {``extreme walking''}, {\it
  i.e.} assuming that the technicolor coupling walks all the way from
the technicolor scale $\Lambda_{TC}$ to the relevant ETC scale
$M_{ETC}$. To produce a more complete analysis, ETC-exchange must be
incorporated into the gap-equation technology in order to estimate
ordinary fermion masses. Studies of this sort are encouraging, it
appears possible to accomodate the first and second generation masses
without necessarily having dangerously large flavor-changing neutral
currents
\cite{Holdom:1981rm,Holdom:1985sk,Yamawaki:1986zg,Appelquist:1986an,Appelquist:1987tr,Appelquist:1987fc}.
The second issue, however, is what about the third generation quarks,
the {top} and {bottom}?  As we will see in the next section, because
of the large top-quark mass, further refinements or modifications will
be necessary to produce a viable theory of dynamical electroweak
symmetry breaking.

In addition to modifying our estimate of the relationship
between the ETC scale and ordinary fermion masses, walking
also influences the size of pseudo-Goldstone boson masses. 
In the case of walking, Dashen's formula for the
size of pseudo-Goldstone boson masses in the presence
of chiral symmetry breaking from ETC interactions, eq. (\ref{dashen}),
reads:
\beqa
F^2_{TC} M^2_{\pi_T} & \propto & {g^2_{ETC} \over M^2_{ETC}}
\langle \left(\overline{T}T\right)^2\rangle)_{ETC} \nonumber \\
&\approx& {g^2_{ETC} \over M^2_{ETC}} 
\left(\langle\overline{T}T\rangle_{ETC}\right)^2 \nonumber \\
&\simeq&{g^2_{ETC} \over M^2_{ETC}}
{M^2_{ETC}\over \Lambda^2_{TC}}
\left(\langle\overline{T}T\rangle_{TC}\right)^2\, .
\eeqa
Consistent with the rainbow approximation, we have used the vacuum-insertion 
to estimate the strong matrix element.
Therefore we find
\beqa
M_{\pi_T} & \simeq &  g_{ETC} 
\left({4\pi F^2_{TC} \over \Lambda_{TC}}\right) \nonumber\\
&\simeq &g_{ETC} \left({750\gev \over N_D}\right)
\left({1\tev\over \Lambda_{TC}}\right)~,
\eeqa
{\it i.e.} walking also enhances the size of pseudo-Goldstone
boson mases!

Finally, {{what about S?}} As emphasized by Lane \cite{Lane:1993wz}, the
assumptions of previous estimate of $S$ included
\cite{Peskin:1990zt,Peskin:1992sw,Golden:1991ig,Holdom:1990tc,Dobado:1991zh}
that:

\begin{narrower}
\begin{itemize}

{
\item techni-isospin is a good symmetry, and

\item {technicolor is QCD-like, {\it i.e.}.}}

\begin{narrower}
\begin{enumerate}

\item Weinberg's sum rules are valid,

\item the spectral functions are saturated by the lowest resonances,

\item the masses and couplings of resonances can be scaled from QCD.

\end{enumerate}
\end{narrower}
\end{itemize}
\end{narrower}
\noindent A ``realistic'' walking technicolor theory would be very unlike QCD:
\begin{narrower}
\begin{itemize}

{\item Walking leads to a different behavior of the spectral functions.

\item Many flavors and PGBs, as well as possible non-fundamental
  representations makes scaling from QCD suspect.  }
\end{itemize}
\end{narrower}
For these reasons the analysis given previously does not apply, and a
walking theory could be phenomenologically acceptable.  Unfortunately,
technicolor being a strongly-coupled theory, it is not possible to give
a compelling argument that the value of $S$ in a walking technicolor
theory {\it is} definitely acceptable.

\section{Top in Models of Dynamical Symmetry Breaking}

\subsection{The ETC of $m_t$}

Because of its large mass, the top quark poses a particular
problem in models of dynamical electroweak symmetry breaking.
Consider an ETC interaction ({\it c.f.} eq. (\ref{fmass}))
giving rise to the top quark mass
\beq
{\lower15pt\hbox{\epsfysize=0.5 truein \epsfbox{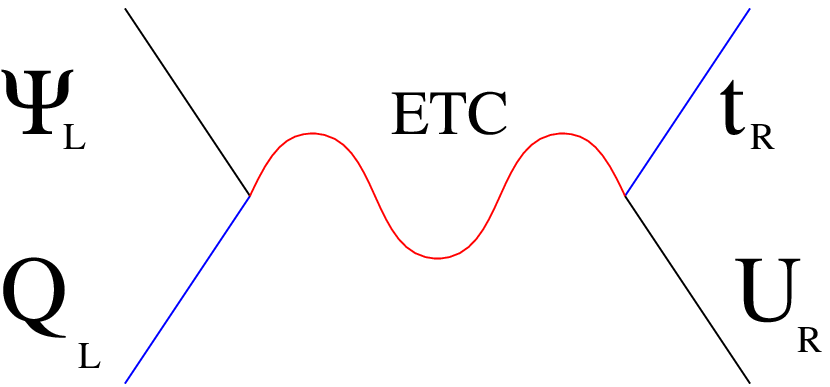}}}
\ \ \rightarrow\ \  {{g_{ETC}^2\over M^2_{ETC}}}(\overline{\Psi}_L U_R)
({\overline{t}_R Q_L})~,
\eeq
yielding
\beq
m_t \approx {{g_{ETC}^2\over M^2_{ETC}}} \langle\overline{U} U\rangle_{ETC}~.
\eeq
In conventional technicolor, using
\beq
\langle\overline{U} U\rangle_{ETC} \approx \langle\overline{U} U\rangle_{TC}
\approx 4\pi F^3_{TC}
\eeq
we find
\beq
{{M_{ETC}\over g_{ETC}}} \approx 1 \tev 
\left({F_{TC}\over 250\gev}\right)^{3\over 2}
\left({175 \gev \over m_t}\right)^{1\over 2}~.
\label{tmass}
\eeq
That is, {the scale of top-quark ETC-dynamics is {\it very} low.}
Since $M_{ETC} \simeq \Lambda_{TC}$ and
\beq
\langle\overline{U} U\rangle_{ETC} = \langle\overline{U} U\rangle_{TC}
\exp\left(\int_{\Lambda_{TC}}^{M_{ETC}} {d\mu \over \mu}
\gamma_m(\mu)\right)~,
\eeq
we see that {walking} cannot alter this conclusion \cite{Chivukula:1993tz}.
As we will see in the next few sections, a low ETC scale for the
top quark is problematic.

\subsection{ETC Effects on $Z \rightarrow b \overline{b}$}

For ETC models of the sort discussed in the last lecture, in which the
ETC gauge-bosons do not carry weak charge, the gauge-boson responsible
for the top-quark mass couples to the current 
\beq \xi
(\overline{\Psi}^{i\alpha}_L \gamma^\mu {Q^i_L}) + \xi^{-1}
(\overline{U}^\alpha_R \gamma^\mu {t_R})~,
\label{current}
\eeq                                                   
(or $h.c.$) where $\alpha$ is the technicolor index and the
contracted $i$ are weak indices.  The part of the exchange interaction
coupling left- and right-handed fermions leads to the top-quark mass.

Additional interactions arise from the same dynamics, including
\beq
-{{g^2_{ETC}\over M^2_{ETC}}}
({\overline{t}_R}\gamma^\mu U^\alpha_R)
(\overline{U}^\alpha_R\gamma_\mu {t_R})
\eeq
and
\beq
-{{g^2_{ETC}\over M^2_{ETC}}}
({\overline{Q}^i_L}\gamma^\mu \Psi^{i\alpha}_L)
(\overline{\Psi}^{j\alpha}_L\gamma_\mu {Q^j_L})~.
\eeq
The last interaction involves both {$b_L$} and the {technifermions}.
After a Fierz transformation, the left-handed operator becomes the
product of weak triplet currents
\beq
-{1\over 2}{{g^2_{ETC}\over M^2_{ETC}}}
({\overline{Q}^i_L \gamma^\mu \tau_a^{ij} Q^j_L})
(\overline{\Psi}^k_L \gamma_\mu \tau_a^{kl} \Psi^l_L)~,
\eeq
where the $\tau$ are the Pauli matrices, plus terms involving
weak singlet currents (which will not concern us here).

The exchange of this ETC gauge-bosons produces
a correction \cite{Chivukula:1992ap}
to the coupling of the $Z$ to $b\bar{b}$
\beq
{\lower25pt\hbox{
\epsfysize=0.75 truein \epsfbox{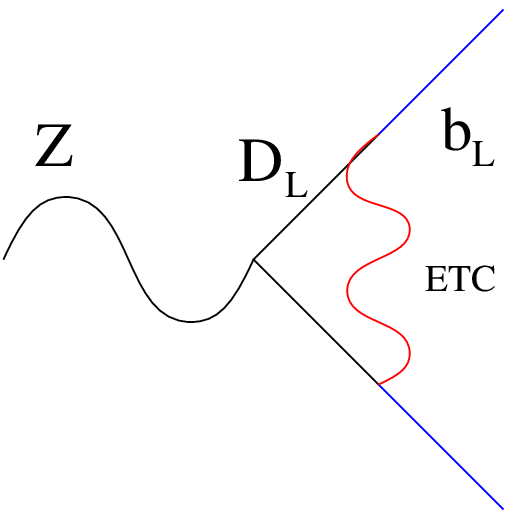}}
\hskip10pt
\ \ \rightarrow\ \   
\hskip10pt
{\lower25pt\hbox{
\epsfysize=0.75 truein \epsfbox{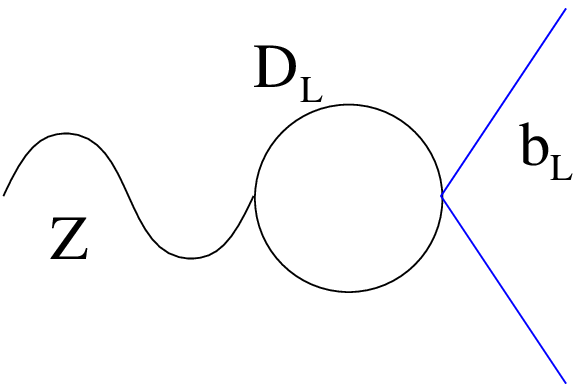}}}
}~.
\eeq
The size of this effect can be calculated by comparing it
to the technifermion weak vacuum-polarization diagrm
\beq
{\lower7pt\hbox{
\epsfxsize=1.0 truein \epsfbox{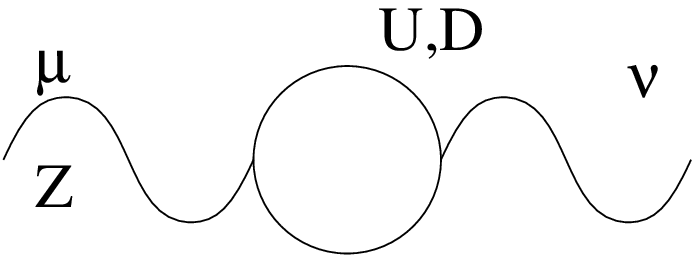}}
\ \ \rightarrow\ \   \pi^{\mu\nu}_{ij} = 
\left(q^2 g^{\mu\nu} - q^\mu q^\nu\right) \delta_{ij} \pi(q^2)\, ,}
\label{zbbetc}
\eeq
which, by the Higgs mechanism yields
\beq
\pi(q^2) = \frac{e^2 v^2}{4 \sin^2_\theta \cos^2\theta} {1\over q^2}\, .
\eeq

Therefore, exchange of the ETC gauge-boson responsible
for the top-quark mass leads to a low-energy effect which
can be summarized by the operator
\beq
-{e\over 2 \sin\theta \cos\theta} {g^2_{ETC} v^2\over M^2_{ETC}}
\xi^2 (\overline{Q}_L \slashchar{Z} \tau_3 Q_L)~.
\eeq
Hence this effect results
in a change in the $Zb\overline{b}$ coupling 
\beq
\delta g_L = +{1\over 4}
{e \over \sin\theta \cos\theta} 
\, \xi^2 {g^2_{ETC} v^2 \over M^2_{ETC}}\, ,
\eeq
which, using the relation in eq. (\ref{tmass}), 
results in
\beq
{\delta \Gamma_b \over \Gamma_b} \approx 
{2 g_L \delta g_L \over  g^2_L + g^2_R} \approx
-6.5\% \cdot \xi^2 \cdot \left({m_t \over 175 \gev}\right)\, .
\label{dgamma}
\eeq
It is convenient to form the ratio $R_b = \Gamma_b/\Gamma_h$, where
$\Gamma_b$ and $\Gamma_h$ are the width of the $Z$ boson to $b$-quarks
and to all hadrons, respectively, since this ratio is largely
independent of the ``oblique'' corrections $S$ and $T$. The
shift in eq. (\ref{dgamma}) results in a shift in $R_b$ of
approximately
\beq
{{\delta R_b \over R_b }} \approx
{\delta \Gamma_b \over \Gamma_b}(1-R_b) \approx 
{-5.1\%} \cdot \xi^2 \cdot \left({m_t \over 175 \gev}\right)\, .
\eeq

\begin{figure}[tbp]
\begin{center}
\epsfxsize=7cm
{\epsffile{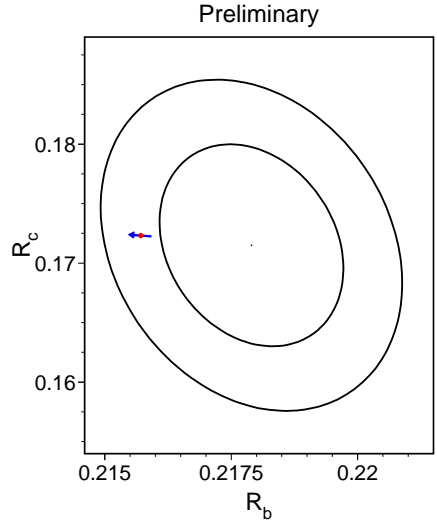}}
\end{center}
\caption{Contours in the $R_b$-$R_c$ plane from LEP data \protect\cite{LEPEWWG},
corresponding to 68\% and 95\% confidence levels assuming Gaussian systematic
errors. The Standard Model prediction for $m_t$=175$\pm$6 GeV is
also shown. The arrow points in the direction of increasing values of 
$m_t$.}
\label{Fig12}
\end{figure}

Recent LEP results \cite{LEPEWWG} on $R_b$ are shown in Figure
\ref{Fig12}.  As we see, the current experimental value of $R_b$ is
about 1.8$\sigma$ {\it above} the standard model prediction, while a
shift of -5.1\% would \footnote{Given the current experimental plus
  systematic experimental error \cite{LEPEWWG} one $\sigma$ corresponds
  to a shift of approximately 0.7\%.} {\it lower} $R_b$ by approximately
7$\sigma$! Clearly, conventional ETC generation of the top-quark mass is
ruled out.

It should be noted, however, that there are nonconventional ETC models
in which $R_b$ {\it may not} be a problem. The analysis leading to the
result given above assumes that (see eq. (\ref{current})) the ETC
gauge-boson responsible for the top-quark mass {\it does not} carry
weak-$SU(2)$ charge.  It is possible to construct models \cite{Chivukula:1994mn,Chivukula:1996gu}
where this is not the case.  Schematically, the group-theoretic
structure of such a model would be as follows
\medskip
\begin{center}

$ETC  \times SU(2)_{light}$

\vspace{5pt}

$\ \ \ \ \ \downarrow\ \ \ \ \ f $

\vspace{5pt}

$TC \times SU(2)_{heavy}  \times SU(2)_{light}$

\vspace{5pt}

$\ \ \ \ \ \downarrow\ \ \ \ \ u $

\vspace{5pt}

$TC  \times SU(2)_{weak} $

\end{center}
\medskip\noindent
where ETC is extended technicolor, $SU(2)_{light}$ is essentially
weak-$SU(2)$ on the light fermions nad $SU(2)_{heavy}$ (originally embedded
in the ETC group) is weak-$SU(2)$ for the heavy fermions, and where 
$SU(2)_{light} \times SU(2)_{heavy}$ break to their diagonal subgroup
(the conventional weak-interactions, $SU(2)_{weak}$) at scale $u$.

In this case a {\it weak-doublet, technicolored} ETC boson 
coupling to
\beq
\xi\overline{Q}_L \gamma^\mu U_L + {1\over \xi}\bar t_R \gamma^\mu \Psi_R~,
\eeq
is responsible for producing $m_t$.
A calculation analogous to the one above yields a correction
\beq
{\lower25pt\hbox{
\epsfysize=0.75 truein \epsfbox{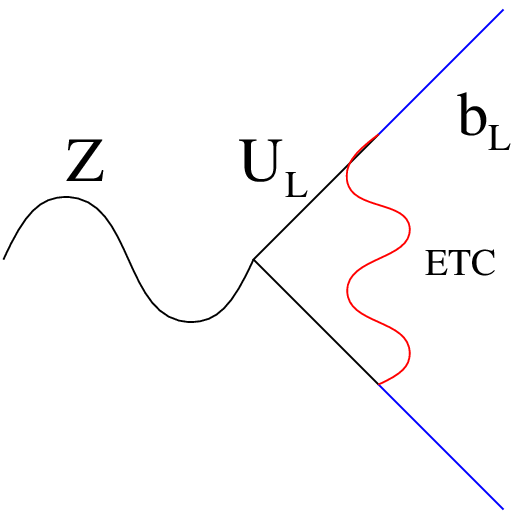}}
\ \ \rightarrow\ \   
\delta g_L = -{1\over 4}
{e \over \sin\theta \cos\theta} 
\, \xi^2 {g^2_{ETC} v^2 \over M^2_{ETC}}
}
\eeq
of the {\it opposite} sign. In fact, the situation is
slightly more complicated: there is an extra
{ $Z$-boson} which also contributes. The 
total contribution is found \cite{Chivukula:1994mn,Chivukula:1996gu} to be
\beq
{\delta R_b \over R_b} \approx +5.1\% \cdot \xi^2 \cdot 
\left({m_t\over 175 {\rm GeV}}\right)\left( 1 - 
{\sin^2 \alpha \over \xi^2} {f^2\over u^2}\right)
\eeq
where $\tan\alpha=g^\prime/g$ is the ratio of the $SU(2)_{light}$ and
$SU(2)_{heavy}$ coupling constants. Since $\sin\alpha$, $\xi$, and $f/u$
are all expected to be of order one, the overall contribution to $R_b$
is very model-dependent but can lie within the experimentally allowed
window.

\subsection{Isospin Violation: $\Delta\rho$}

\medskip

\noindent\underline{``Direct'' Contributions}

\medskip

ETC interactions {\it must} violate weak isospin in order to give rise
to the mass splitting between the top and bottom quarks. This could
induce dangerous $\Delta I=2$ {technifermion}
operators \cite{Appelquist:1984nc,Appelquist:1985rr}
\beq
{\lower10pt\hbox{\epsfxsize=1.5in \epsffile{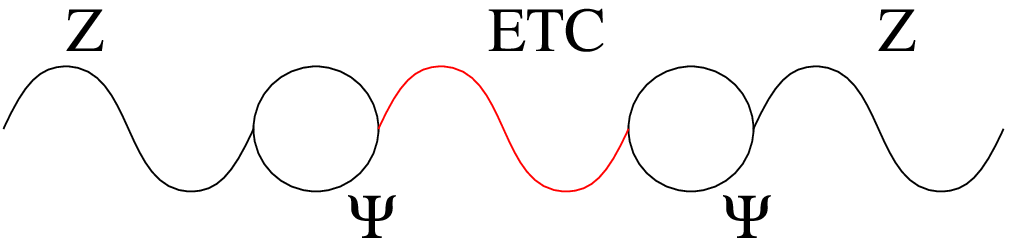}}}
\ \ \rightarrow \ \ {g^2_{ETC}\over M^2_{ETC}}
\left(\overline{\Psi}_R \gamma_\mu \tau_3 \Psi_R\right)^2\, .
\eeq
We can estimate the contribution of these operators to $\Delta\rho$
using the vacuum-insertion approximation
\beq
\Delta\rho   \simeq  {2 g^2_{ETC}\over M^2_{ETC}}
{N^2_D F^4_{TC} \over v^2} 
\eeq
which yields
\beq
\Delta\rho \approx  12\% \cdot 
\left({\sqrt{N_D} F_{TC} \over 250 \gev}\right)^2
\cdot \left({1 \tev \over M_{ETC}/g_{ETC}}\right)^2\, .
\eeq
If we require that $\Delta \rho \le 0.4\%$, we find
\beq
{M_{ETC} \over g_{ETC}} > 5.5\tev \cdot
\left({\sqrt{N_D} F_{TC} \over 250 \gev}\right)^2\, ,
\eeq
{\it i.e.} $M_{ETC}$ must be {greater} 
than required for $m_t \simeq 175 \gev$.

There is {another possibility}.  It is possible that $N_D F^2_{TC}
\ll (250 \gev)^2$, {if} the sector responsible for the top-quark
mass {does not} give rise to the bulk of electroweak symmetry
breaking.  In this scenario,
the constraint is
\beq
F_{TC}< {105 \gev \over N^{1/2}_D} \cdot
\left({M_{ETC}/g_{ETC} \over 1 \tev}\right)^{1/2}\, .
\eeq
However, this modification would {\it enhance} the effect of
ETC exchange in $Z \to b \overline{b}$.

\medskip

\underline{``Indirect'' Contributions} to $\Delta\rho$

\medskip

\begin{figure}[tbp]
\begin{center}
\epsfxsize=8cm
\epsffile{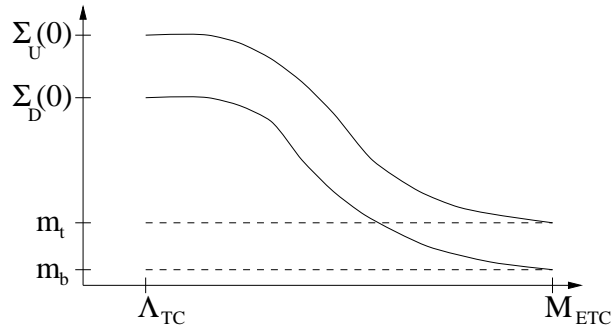}
\end{center}
\caption{Momentum dependent dynamical masses of the technifermions
responsible for the $t$- and $b$-quark masses, based on an
a gap-equation analysis.}
\label{Fig13}
\end{figure}

Isospin violation in the ordinary fermion masses suggests
the existence of isospin violation in the technifermion dynamical
masses. Indeed, an analysis of the gap equation shows that
if the top- and bottom-quarks get masses from technifermions in
the same technidoublet the dynamical masses of the corresponding
technifermions are as shown in Figure \ref{Fig13}. At a scale
of order $M_{ETC}$ the technifermions and ordinary fermions are
unified into a single gauge group, so it is not surprising
that their masses are approximately equal at that scale. Below
the ETC scale, the technifermion dynamical mass runs (because of the
technicolor interactions), while the ordinary fermion masses
do not. As shown in Figure \ref{Fig13}, therefore,
we expect that $\Sigma_U(0) - \Sigma_D(0)$
\gae$m_t - m_b$.

We can estimate the contribution of this effect to $\Delta\rho$
\beq
\lower10pt\hbox{\epsfxsize=1.0in\epsffile{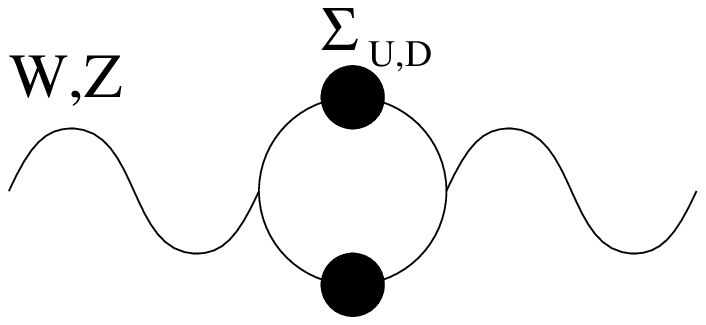}}
\ \ \propto\ \  {N_D d \over 16\pi^2}{(\Sigma_U(0) - \Sigma_D(0))^2\over v^2}\, ,
\eeq
where $N_D$ is the number of technidoublets 
and $d$ is the dimension of the TC representation. If we
require $\Delta\rho \le 0.4\%$, this yields
\beq
d\, N_D  \left({\Delta\Sigma(0)\over 175 \gev}\right)^2
\le 2.7~.
\eeq
This is perhaps possible if $N_D=1$ and $d=2$ ({\it i.e.}
$N_{TC}=2$), but is generally problematic.

\subsection{Evading the Unavoidable}

\begin{figure}[tbp]
\begin{center}
\epsfxsize=8cm
\epsffile{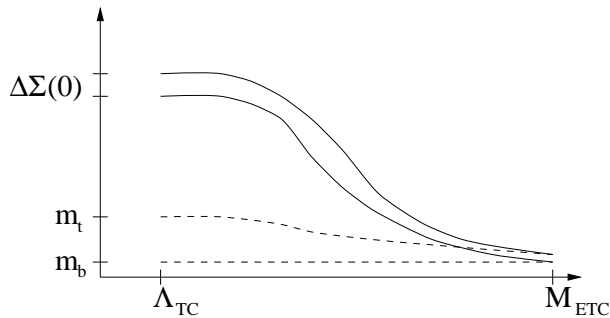}
\end{center}
\caption{Momentum dependent dynamical masses of the technifermions
  which couple to the $t$- and $b$-quarks in a theory where additional
  strong interactions (other than techicolor) are responsible for the
  bulk of the top and bottom quark masses. These additional strong
  interactions allow for the quark masses to run significantly below the
  ETC scale.}
\label{Fig14}
\end{figure}

The problems outlined in the last two sections, namely potentially
dangerous ETC corrections to the branching ratio of $Z\to b\overline{b}$
and to the $\rho$ parameter, rule out the possibility of generating the
top-quark mass using conventional extended technicolor interactions.  A
close analysis of these problems, however, suggests a framework for
constructing an acceptable model: arrange for the $t$- and $b$-quarks to
get the majority of their masses from interactions {other than
  technicolor}. If this is the case, the top- and bottom-quark
masses can run substantially below the ETC scale as shown in
Figure \ref{Fig14}, allowing for 
\beq
{\Delta \Sigma(0) \simeq m_t(M_{ETC})-m_b(M_{ETC}) \ll m_t}~.
\eeq
Since the technicolor and ETC interactions would only be responsible for a
{\it portion} of the top-quark mass in this type of model, the problems
outlined in the previous two sections are no longer relevant.  In order
to produce a substantial running of the third-generation quark masses,
the third-generation fermions must have an additional {
  strong-interaction} not shared by the first two generations of
fermions or (at least in an isospin-violating way) by the
technifermions.

\section{Top-Condensate Models and Topcolor}

\subsection{Top-Condensate Models}

Before constructing a model of the sort proposed in last section, we
should pause to consider another possibility.  Having entertained the
notion that the top-quark mass may come from a strong interaction felt
(at least primarily) by the third generation, one should ask if there is
any longer a need for technicolor! After all, any interaction that gives
rise to a quark mass {\it must} break the weak interactions.
Furthermore, recall that $m_t \simeq M_W,\, M_Z$; the top-quark is much
heavier than other fermions it must be more strongly coupled to
the symmetry-breaking sector.  Perhaps all
\cite{Miranskii:1989ds,Miranskii:1989xi,Nambu:1989jt,Marciano:1989xd,Bardeen:1990ds,Hill:1991at,Cvetic:1997eb}
of electroweak-symmetry breaking is due to a condensate of top-quarks,
$\langle \bar{t}t\rangle \neq 0$.

\begin{figure}[tbp]
\begin{center}
\epsfxsize=8cm
\epsffile{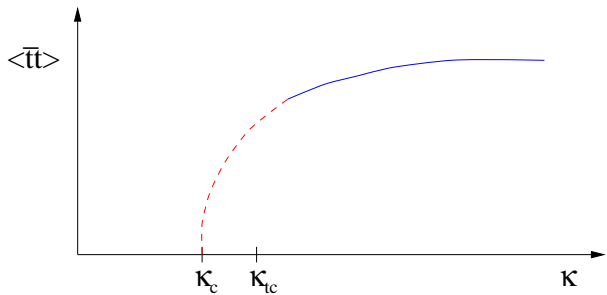}
\end{center}
\caption{Behavior of the condensate in a top-color
model as a function of the top-color coupling assuming
a {\it continuous} transition.}
\label{Fig15}
\end{figure}

Consider a {spontaneously broken strong gauge-interaction},
e.g. top-color:
\beq
{SU(3)_{tc}} \times {SU(3)}
\ \ \stackrel{M}{\to}\ \  SU(3)_{QCD}~,
\eeq
where $SU(3)_{tc}$ is a new, strong, topcolor interaction coupling
to the third-generation quarks and the other $SU(3)$ is a weak color
interaction coupling to the first two generations. At scales below
$M$, there remains ordinary QCD plus interactions which
couple primarily to the third generation quarks 
and can be summarized by an operator of the form
\beq
{\cal L} \ \ \supset\ \  - {4\pi\kappa \over M^2}
\left(\overline{Q}\gamma_\mu {\lambda^a\over 2} Q\right)^2~,
\eeq
where $\kappa \approx g^2_{tc}/4\pi$ is related to the top-color
coupling constant. Consider what happens as, for fixed $M$, we
vary $\kappa$. For small $\kappa$, the interactions are perturbative
and there is no chiral symmetry breaking. For large $\kappa$, since the
new interactions are attractive in the spin-zero, isospin-zero
channel, we expect chiral symmetry breaking with
$\langle \bar{t}t\rangle \propto M^3$. If the transition between these
two regimes is {\it continuous}, as it is in the bubble \cite{Nambu:1961er}
or mean-field approximation, we expect that the condensate
will behave as shown in Figure \ref{Fig15}.

In order to produce a realistic model of electroweak symmetry breaking
based on these considerations, one must introduce extra interactions to
split the top- and bottom-quark masses.  A careful analysis then shows
that it is not possible\footnote{An interesting alternative, in which
  the top mass arises from the seesaw mechanism, which may lead to a
  phenomenologically acceptible theory with $M\sim 40$ TeV has recently
  been proposed \cite{Dobrescu:1997nm}.} to achieve a phenomenologically
acceptable theory unless
\cite{Miranskii:1989ds,Miranskii:1989xi,Nambu:1989jt,Marciano:1989xd,Bardeen:1990ds,Hill:1991at,Cvetic:1997eb}
the scale $M \gg v$. Since the weak scale $v$ is fixed, this implies
that the condensate $\langle \bar{t}t\rangle \ll M^3$, and the top-color
coupling $\kappa$ must be finely tuned
\beq
{\Delta\kappa \over \kappa_c} \equiv
{{\kappa-\kappa_c}\over \kappa_c}
\propto {\langle\bar{t}t\rangle \over M^3}~.
\eeq
In this region, one has simply reproduced the
standard model \cite{Miranskii:1989ds,Miranskii:1989xi,Nambu:1989jt,Marciano:1989xd,Bardeen:1990ds,Hill:1991at,Cvetic:1997eb},
with the Higgs-boson $\phi$ produced dynamically as a
${\bar{t}_R}Q_L$ bound state!

\subsection{Topcolor-Assisted Technicolor (TC2)}

Recently, Chris Hill has proposed \cite{Hill:1995hp} a theory which
combines technicolor and top-condensation. Features of this type of
model include technicolor dynamics at 1 TeV, which dynamically generates
{most} of electroweak symmetry breaking, and extended technicolor
dynamics at scales much higher than 1 TeV, which generates the light
quark and lepton masses, as well as small contributions to the third
generation masses ($m_{t,b,\tau}^{ETC}$) of order 1 GeV. The top quark
mass arises predominantly from topcolor dynamics at a scale of order 1
TeV, which generates $\langle \bar{t} t \rangle \neq 0$ and $m_t \sim
175$ GeV. Topcolor {\bf cannot} be allowed to generate a large $b$-quark
mass, and therefore there must be isospin violation. This may be
acceptable because topcolor contributes a small amount to EWSB (with an
``F-constant'' $f_t \sim 60$ GeV). The extended symmetry-breaking sector
gives rise to extra pseudo-Goldstone bosons (``Top-pions'') which get
mass from ETC interactions which allow for mixing of third generation to
first two.

\medskip

{\underline{Hill's Simplest TC2 Scheme}}

\medskip

The simplest scheme \cite{Hill:1995hp} which realizes these features has the
following structure:
\medskip
\begin{center}
$G_{TC}  \times  SU(2)_{EW} \times 
SU(3)_{tc} \times SU(3) \times U(1)_H \times U(1)_L$
\end{center}
\begin{center}
$\downarrow \ \ M$ \gae 1 TeV
\end{center}
\begin{center}
$G_{TC} \times SU(3)_C  \times SU(2)_{EW} \times U(1)_Y $
\end{center}
\begin{center}
$\downarrow\ \ \ \Lambda_{TC}\sim 1{\ \rm TeV} $
\end{center}
\begin{center}
$SU(3)_C \times U(1)_{EM}$
\end{center}
\medskip
Here $U(1)_H$ and $U(1)_L$ are $U(1)$ gauge groups coupled to the
(standard model) hypercharges of the third-generation and first-two generation
fermions respectively.
Below $M$, this leads to the effective interactions:
\beq
-{{4\pi \kappa_{tc}}\over{M^2}}\left[\overline{\psi}\gamma_\mu 
{{\lambda^a}\over{2}} \psi \right]^2~,
\eeq
from top-color exchange and the isospin-violating interactions
\beq
-{{4\pi \kappa_1}\over{M^2}}\left[{1\over3}\overline{\psi_L}\gamma_\mu  \psi_L
+{4\over3}\overline{t_R}\gamma_\mu  t_R
-{2\over3}\overline{b_R}\gamma_\mu  b_R
\right]^2\, ,
\label{hyper}
\eeq
from exchange of the ``heavy-hypercharge'' ($Z^\prime$) 
gauge boson.

Since the interactions in eq. (\ref{hyper}) are attractive in the
$\bar{t}t$ channel, but repulsive in the $\bar{b}b$ channel, the
couplings $\kappa_{tc}$ and $\kappa_1$ can be chosen to produce $\langle
\bar{t} t \rangle \neq 0$ and a large $m_t$, but leave $\langle
\bar{b} b \rangle = 0$. In the Nambu--Jona--Lasinio approximation \cite{Nambu:1961er},
we require
\beq
 \kappa^t = \kappa_{tc} +{1\over3}\kappa_1 \ >\ 
\kappa_c \left( = {{3\pi}\over{8}} \right)_{NJL}\ >\ 
\kappa^b=\kappa_{tc} -{1\over 6}\kappa_1~ .
\label{fine}
\eeq

\vspace{5pt}

\subsection{$\Delta \rho$ in TC2}

\indent\underline{Direct Contributions}

\medskip

Couplings of the (potentially strong) $U(1)_H$ group are isospin
violating, at least in regard to the third generation.  Isospin
violating couplings to technifermions could be very dangerous
\cite{Chivukula:1995dc}, as shown above.  For example, in the one-family
technicolor model, if the $U(1)_H$ charges of the technifermions are
proportional to $Y$, the result is:
\beq
\Delta \rho^{\rm T} \approx 152\% \
\kappa_1 \left({{1\ {\rm TeV}}\over{M}}\right)^2~.
\eeq
If $M\simeq 1\tev$, we must have $\kappa_1 \ll 1$. From eq. (\ref{fine})
above, this implies a {fine tuning} of $\kappa_{tc}$.  In order to avoid
this problem, one must construct a ``Natural TC2'' model in which the
$U(1)_H$ couplings to technifermions are isospin symmetric
\cite{Lane:1995gw}.

\medskip

\underline{Indirect/Direct Contribution}

\medskip

Since there are additional (strong) interactions felt by the
third-generation of quarks, there are new 
``two-loop'' contributions \cite{Chivukula:1995dc} to $\Delta \rho$:
\beq
\lower7pt\hbox{\epsfxsize 3cm {\epsffile{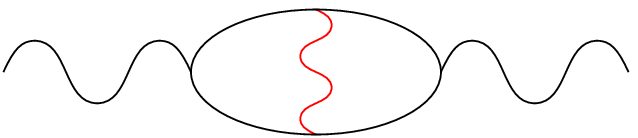}}}
\ \ \leftrightarrow\ \ 
\lower7pt\hbox{\epsfxsize 3cm {\epsffile{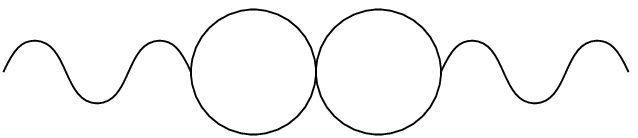}}}~.
\eeq
This contribution yields
\beq
{\Delta \rho^{\rm tc} \approx 0.53\%
\left({\kappa_{tc}\over \kappa_c}\right)
\left(1\ {\rm TeV}\over M\right)^2
\left(f_t \over 64\ {\rm GeV}\right)^4\, .}
\eeq
From this we find that $M$ \gae 1.4 TeV.

\subsection{Electroweak Constraint on Natural TC2}

\begin{figure}[tbp]
\begin{center}
\epsfxsize=8cm
\centerline{\epsfbox{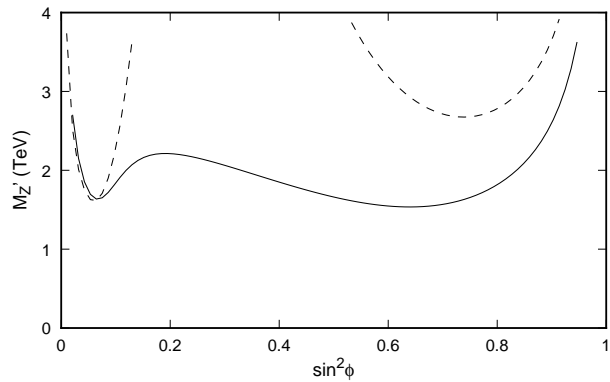}}
\end{center}
\caption{Bounds \protect\cite{Chivukula:1996cc}
on the mass of the $Z^\prime$ in natural
TC2 models as a function of the angle $\phi$ 
where $\tan\phi = g^L_1/g^H_1$.
Bounds are shown for $\alpha_s(M_Z)=0.115$ (solid), 0.124 (dashed).}
\label{Fig16}
\end{figure}

If the $U(1)_H$ couplings to technifermions are isospin-symmetric,
electroweak phenomenology is specified by $M^2_{Z^\prime}$, $\tan\phi =
g^L_1/g^H_1$, and the charges $Y_H$ of ordinary fermions. To get a
feeling for the size of constraints on these models from electroweak
phenomenology, consider a ``baseline'' model: $Y_H = Y$.  While this may
be unrealistic, it is flavor universal. In this case the third generation
is picked out by its couplings to $SU(3)_H$.

Constraints (arising from $Z$-$Z^\prime$ mixing as well
as $Z^\prime$ exchange) from all precision electroweak data
are shown in Figure \ref{Fig16}. We see that, even in light of
current LEP data, natural TC2 with a $Z^\prime$ mass of order
1-2 TeV is allowed.

\section{Where have we come from, where are we going?}

In these lectures I have tried to provide an introduction to the full range
of theories that have been proposed to explain electroweak symmetry
breaking. We have come a long way, and it is worth reviewing the logical
progression that has brought us here:

\medskip
\begin{narrower}
\begin{itemize}

\item In the absence of supersymmetry, models of electroweak
symmetry breaking with fundamental scalars suffer from the
naturalness/hierarchy and triviality problems.

\item Triviality implies that any theory with fundamental scalars
is at best an effective theory below some high-energy
scale $\Lambda$, and {\it lower bounds} on $\Lambda$ give rise to
{\it upper bounds} on scalar masses.

\item The search for a {\bf natural} explanation of electroweak
  symmetry breaking leads to models with weak-scale supersymmetry or models
  of dynamical electroweak symmetry breaking.

\item Weak-scale supersymmetry protects the Higgs mass, but does not
  {\it explain} the weak scale ({\it i.e.} there remains the
  $\mu$-problem) and requires extra dynamics to give rise to
  supersymmetry breaking.

\item Technicolor provides a dynamical explanation for electroweak
  symmetry breaking. Accommodating and explaining the $u,\, d,\, s$ and
  $c$ masses in such theories without large {\bf flavor-changing
    neutral-currents} leads us to consider ``walking'' technicolor.
  Accommodating the bottom- and, especially, the top-quark mass without
  large corrections to {$\Delta \Gamma_b$} and {$T$} leads us to
  consider top-color assisted technicolor.

\end{itemize}
\end{narrower}
\medskip

Despite the progress that has been made, no complete and consistent
model exists.  Ultimately, these problems are not likely to be solved
without {\bf experimental} direction. With continuing experiments at LEP
II and the Tevatron and the construction of the LHC, the next decade
promises to bring us some answers.


\section*{Acknowledgments}
I thank the organizers, especially Rajan Gupta, for arranging a
stimulating summer school and the students for their active
participation and interest. I also thank my collaborators in work
discussed in this review, especially Bogdan Dobrescu, Elizabeth Simmons,
and John Terning for comments on the manuscript. {\em This work was
  supported in part by the Department of Energy under grant
  DE-FG02-91ER40676.}

\bibliography{leshouches}

\begin{thebibliography}{100}

\bibitem{Chivukula:1996rz}
R.S. Chivukula, E.H. Simmons and B.A. Dobrescu,
\newblock (1996), hep-ph/9703206.

\bibitem{Chivukula:1996uy}
R.S. Chivukula,
\newblock (1996), hep-ph/9701322.

\bibitem{Weinberg:1967pk}
S. Weinberg,
\newblock Phys. Rev. Lett. 19 (1967) 1264.

\bibitem{Salam:1968rm}
A. Salam,
\newblock Weak and electromagnetic interactions,
\newblock originally printed in Svartholm: Elementary Particle Theory,
  Proceedings Of The Nobel Symposium Held 1968 At Lerum, Sweden, Stockholm
  1968, 367-377.

\bibitem{Weinstein:1973gj}
M. Weinstein,
\newblock Phys. Rev. D8 (1973) 2511.

\bibitem{Sikivie:1980hm}
P. Sikivie et~al.,
\newblock Nucl. Phys. B173 (1980) 189.

\bibitem{Buchmuller:1986jz}
W. Buchmuller and D. Wyler,
\newblock Nucl. Phys. B268 (1986) 621.

\bibitem{Grinstein:1991cd}
B. Grinstein and M.B. Wise,
\newblock Phys. Lett. B265 (1991) 326.

\bibitem{Hinchliffe:1996we}
I. Hinchliffe and J. Womersley,
\newblock (1996), hep-ex/9612006.

\bibitem{Bagley:1996ne}
P.P. Bagley et~al.,
\newblock Summary of the tev33 working group,
\newblock DPF / DPB Summer Study on New Directions for High-Energy Physics
  (Snowmass 96).

\bibitem{Haber:1996qb}
H.E. Haber et~al.,
\newblock (1996), hep-ph/9703391.

\bibitem{Honma:1997}
A. Honma,
\newblock Lepc presentation 11/11/97,
\newblock 1997.

\bibitem{LEPEWWG:1997}
LEP Electroweak Working Group,
\newblock www.cern.ch/LEPEWWG/plots/summer97/.

\bibitem{Bomestar:1995mu}
D. Bomestar et~al.,
\newblock Fizika B4 (1995) 273.

\bibitem{Veltman:1977rt}
M. Veltman,
\newblock Acta Phys. Polon. B8 (1977) 475.

\bibitem{Lee:1977yc}
B.W. Lee, C. Quigg and H.B. Thacker,
\newblock Phys. Rev. Lett. 38 (1977) 883.

\bibitem{Chanowitz:1987vj}
M. Chanowitz, M. Golden and H. Georgi,
\newblock Phys. Rev. D36 (1987) 1490.

\bibitem{Chanowitz:1986hu}
M. Chanowitz, M. Golden and H. Georgi,
\newblock Phys. Rev. Lett. 57 (1986) 2344.

\bibitem{Cornwall:1974km}
J.M. Cornwall, D.N. Levin and G. Tiktopoulos,
\newblock Phys. Rev. D10 (1974) 1145.

\bibitem{Vayonakis:1976vz}
C.E. Vayonakis,
\newblock Nuovo Cim. Lett. 17 (1976) 383.

\bibitem{Chanowitz:1985hj}
M.S. Chanowitz and M.K. Gaillard,
\newblock Nucl. Phys. B261 (1985) 379.

\bibitem{Weinberg:1976pe}
S. Weinberg,
\newblock Phys. Rev. Lett. 36 (1976) 294.

\bibitem{Linde:1977mm}
A.D. Linde,
\newblock Phys. Lett. 70B (1977) 306.

\bibitem{Witten:1981ez}
E. Witten,
\newblock Nucl. Phys. B177 (1981) 477.

\bibitem{Coleman:1973tz}
S. Coleman and E. Weinberg,
\newblock Phys. Rev. D7 (1973) 1888.

\bibitem{Yamagishi:1981qq}
H. Yamagishi,
\newblock Phys. Rev. D23 (1981) 1880.

\bibitem{Quiros:1997vk}
M. Quiros,
\newblock (1997), hep-ph/9703412.

\bibitem{'tHooft:1980xb}
G. 't~Hooft,
\newblock Recent Developments in Gauge Theories. Proceedings NATO Advanced
  Study Institute, Cargese, France, Aug. 26 - Sept. 8, 1979, edited by e..
  G.~'t Hooft et~al., New York, Usa: Plenum ( 1980) 438 P. ( Nato Advanced
  Study Institutes Series: Series B, Physics, 59), 1980.

\bibitem{Wilson:1971dh}
K.G. Wilson,
\newblock Phys. Rev. B4 (1971) 3184.

\bibitem{Wilson:1974dg}
K.G. Wilson and J. Kogut,
\newblock Phys. Rept. 12 (1974) 75.

\bibitem{Wilson:1971bg}
K.G. Wilson,
\newblock Phys. Rev. B4 (1971) 3174.

\bibitem{Miranskii:1989ds}
V.A. Miranskii, M. Tanabashi and K. Yamawaki,
\newblock Mod. Phys. Lett. A4 (1989) 1043.

\bibitem{Miranskii:1989xi}
V.A. Miranskii, M. Tanabashi and K. Yamawaki,
\newblock Phys. Lett. B221 (1989) 177.

\bibitem{Nambu:1989jt}
Y. Nambu,
\newblock Bootstrap symmetry breaking in electroweak unification,
\newblock EFI-89-08.

\bibitem{Marciano:1989xd}
W.J. Marciano,
\newblock Phys. Rev. Lett. 62 (1989) 2793.

\bibitem{Bardeen:1990ds}
W.A. Bardeen, C.T. Hill and M. Lindner,
\newblock Phys. Rev. D41 (1990) 1647.

\bibitem{Hill:1991at}
C.T. Hill,
\newblock Phys. Lett. B266 (1991) 419.

\bibitem{Cvetic:1997eb}
G. Cvetic,
\newblock (1997), hep-ph/9702381.

\bibitem{Kaplan:1984fs}
D.B. Kaplan and H. Georgi,
\newblock Phys. Lett. 136B (1984) 183.

\bibitem{Kaplan:1984sm}
D.B. Kaplan, H. Georgi and S. Dimopoulos,
\newblock Phys. Lett. 136B (1984) 187.

\bibitem{Dugan:1985hq}
M.J. Dugan, H. Georgi and D.B. Kaplan,
\newblock Nucl. Phys. B254 (1985) 299.

\bibitem{Cabibbo:1979ay}
N. Cabibbo et~al.,
\newblock Nucl. Phys. B158 (1979) 295.

\bibitem{Dashen:1983ts}
R. Dashen and H. Neuberger,
\newblock Phys. Rev. Lett. 50 (1983) 1897.

\bibitem{Kuti:1988nr}
J. Kuti, L. Lin and Y. Shen,
\newblock Phys. Rev. Lett. 61 (1988) 678.

\bibitem{Luscher:1989uq}
M. Luscher and P. Weisz,
\newblock Nucl. Phys. B318 (1989) 705.

\bibitem{Hasenfratz:1987eh}
A. Hasenfratz et~al.,
\newblock Phys. Lett. 199B (1987) 531.

\bibitem{Hasenfratz:1989kr}
A. Hasenfratz et~al.,
\newblock Nucl. Phys. B317 (1989) 81.

\bibitem{Bhanot:1990zd}
G. Bhanot et~al.,
\newblock Nucl. Phys. B343 (1990) 467.

\bibitem{Bhanot:1991ai}
G. Bhanot et~al.,
\newblock Nucl. Phys. B353 (1991) 551.

\bibitem{Weinberg:1979kz}
S. Weinberg,
\newblock Physica 96A (1979) 327.

\bibitem{Manohar:1984md}
A. Manohar and H. Georgi,
\newblock Nucl. Phys. B234 (1984) 189.

\bibitem{Georgi:1993dw}
H. Georgi,
\newblock Phys. Lett. B298 (1993) 187, hep-ph/9207278.

\bibitem{Chivukula:1992nw}
R.S. Chivukula, M.J. Dugan and M. Golden,
\newblock Phys. Lett. B292 (1992) 435, hep-ph/9207249.

\bibitem{Barnett:1996hr}
Particle Data Group, R.M. Barnett et~al.,
\newblock Phys. Rev. D54 (1996) 1.

\bibitem{Chivukula:1995dc}
R.S. Chivukula, B.A. Dobrescu and J. Terning,
\newblock Phys. Lett. B353 (1995) 289, hep-ph/9503203.

\bibitem{Eichten:1979ah}
E. Eichten and K. Lane,
\newblock Phys. Lett. 90B (1980) 125.

\bibitem{Dimopoulos:1979es}
S. Dimopoulos and L. Susskind,
\newblock Nucl. Phys. B155 (1979) 237.

\bibitem{Appelquist:1984nc}
T. Appelquist et~al.,
\newblock Phys. Rev. Lett. 53 (1984) 1523.

\bibitem{Appelquist:1985rr}
T. Appelquist et~al.,
\newblock Phys. Rev. D31 (1985) 1676.

\bibitem{Gunion:1989we}
J.F. Gunion et~al.,
\newblock The Higgs Hunter's Guide (Addison-Wesley, 1990).

\bibitem{Glashow:1977nt}
S.L. Glashow and S. Weinberg,
\newblock Phys. Rev. D15 (1977) 1958.

\bibitem{Coleman:1973sx}
S. Coleman and D.J. Gross,
\newblock Phys. Rev. Lett. 31 (1973) 851.

\bibitem{Kominis:1993zc}
D. Kominis and R.S. Chivukula,
\newblock Phys. Lett. B304 (1993) 152, hep-ph/9301222.

\bibitem{Georgi:1985nv}
H. Georgi and M. Machacek,
\newblock Nucl. Phys. B262 (1985) 463.

\bibitem{Weinberg:1976gm}
S. Weinberg,
\newblock Phys. Rev. D13 (1976) 974.

\bibitem{Georgi:1986df}
H. Georgi, D.B. Kaplan and L. Randall,
\newblock Phys. Lett. 169B (1986) 73.

\bibitem{Peccei:1977hh}
R.D. Peccei and H.R. Quinn,
\newblock Phys. Rev. Lett. 38 (1977) 1440.

\bibitem{Peccei:1977ur}
R.D. Peccei and H.R. Quinn,
\newblock Phys. Rev. D16 (1977) 1791.

\bibitem{Weinberg:1978ma}
S. Weinberg,
\newblock Phys. Rev. Lett. 40 (1978) 223.

\bibitem{Wilczek:1978pj}
F. Wilczek,
\newblock Phys. Rev. Lett. 40 (1978) 279.

\bibitem{Kim:1979if}
J.E. Kim,
\newblock Phys. Rev. Lett. 43 (1979) 103.

\bibitem{Shifman:1980if}
M.A. Shifman, A.I. Vainstein and V.I. Zakharov,
\newblock Nucl. Phys. B166 (1980) 493.

\bibitem{Adler:1969av}
S.L. Adler,
\newblock Phys. Rev. 177 (1969) 2426.

\bibitem{Bell:1969re}
J.S. Bell and R. Jackiw,
\newblock Nuovo Cim. 60A (1969) 47.

\bibitem{Bardeen:1969md}
W.A. Bardeen,
\newblock Phys. Rev. 184 (1969) 1848.

\bibitem{Bagger:1996ka}
J.A. Bagger,
\newblock (1996), hep-ph/9604232.

\bibitem{Witten:1982fp}
E. Witten,
\newblock Phys. Lett. B117 (1982) 324.

\bibitem{Cohen:1996vb}
A.G. Cohen, D.B. Kaplan and A.E. Nelson,
\newblock Phys. Lett. B388 (1996) 588, hep-ph/9607394.

\bibitem{Weinberg:1979bn}
S. Weinberg,
\newblock Phys. Rev. D19 (1979) 1277.

\bibitem{Susskind:1978ms}
L. Susskind,
\newblock Phys. Rev. D20 (1979) 2619.

\bibitem{Dawson:1996cq}
S. Dawson,
\newblock (1996), hep-ph/9612229.

\bibitem{Kane:1994td}
G.L. Kane et~al.,
\newblock Phys. Rev. D49 (1994) 6173, hep-ph/9312272.

\bibitem{Froidevaux}
D. Froideveaux et~al.,
\newblock  preprint ATLAS Note PHYS-No-74.

\bibitem{Chivukula:1995dt}
R.S. Chivukula et~al.,
\newblock (1995), hep-ph/9503202.

\bibitem{Donoghue:1988xa}
J.F. Donoghue, C. Ramirez and G. Valencia,
\newblock Phys. Rev. D38 (1988) 2195.

\bibitem{'tHooft:1974jz}
G. 't~Hooft,
\newblock Nucl. Phys. B72 (1974) 461.

\bibitem{Bagger:1994zf}
J. Bagger et~al.,
\newblock Phys. Rev. D49 (1994) 1246, hep-ph/9306256.

\bibitem{Bagger:1995mk}
J. Bagger et~al.,
\newblock Phys. Rev. D52 (1995) 3878, hep-ph/9504426.

\bibitem{Golden:1995xv}
M. Golden, T. Han and G. Valencia,
\newblock (1995), hep-ph/9511206.

\bibitem{Kawarabayashi:1966kd}
K. Kawarabayashi and M. Suzuki,
\newblock Phys. Rev. Lett. 16 (1966) 255.

\bibitem{Riazuddin1966}
. {Riazuddin and Fayyazuddin},
\newblock Phys. Rev. 147 (1966) 1071.

\bibitem{Kroll:1967it}
N.M. Kroll, T.D. Lee and B. Zumino,
\newblock Phys. Rev. 157 (1967) 1376.

\bibitem{Barklow1996}
T. Barklow,
\newblock International Symposium on Vector Boson Self-Interactions, , AIP
  Conference Proceedings No.  350, 1995.

\bibitem{Georgi1984}
H. Georgi,
\newblock Weak Interactions and Modern Particle Theory (Benjamin/Cummings,
  1984).

\bibitem{Gasser:1984yg}
J. Gasser and H. Leutwyler,
\newblock Ann. Phys. 158 (1984) 142.

\bibitem{Gasser:1985gg}
J. Gasser and H. Leutwyler,
\newblock Nucl. Phys. B250 (1985) 465.

\bibitem{Hagiwara:1987vm}
K. Hagiwara et~al.,
\newblock Nucl. Phys. B282 (1987) 253.

\bibitem{Aihara:1995iq}
H. Aihara et~al.,
\newblock (1995), hep-ph/9503425.

\bibitem{Peskin:1990zt}
M.E. Peskin and T. Takeuchi,
\newblock Phys. Rev. Lett. 65 (1990) 964.

\bibitem{Peskin:1992sw}
M.E. Peskin and T. Takeuchi,
\newblock Phys. Rev. D46 (1992) 381.

\bibitem{Golden:1991ig}
M. Golden and L. Randall,
\newblock Nucl. Phys. B361 (1991) 3.

\bibitem{Holdom:1990tc}
B. Holdom and J. Terning,
\newblock Phys. Lett. B247 (1990) 88.

\bibitem{Dobado:1991zh}
A. Dobado, D. Espriu and M.J. Herrero,
\newblock Phys. Lett. B255 (1991) 405.

\bibitem{Terning1996}
J. Terning,
\newblock Private Communication.

\bibitem{Farhi:1979zx}
E. Farhi and L. Susskind,
\newblock Phys. Rev. D20 (1979) 3404.

\bibitem{Chivukula:1987py}
R.S. Chivukula and H. Georgi,
\newblock Phys. Lett. 188B (1987) 99.

\bibitem{Dashen:1969eg}
R. Dashen,
\newblock Phys. Rev. 183 (1969) 1245.

\bibitem{Chivukula:1991bx}
R.S. Chivukula and M. Golden,
\newblock Phys. Lett. B267 (1991) 233.

\bibitem{Lane:1993wz}
K. Lane,
\newblock (1993), hep-ph/9401324.

\bibitem{Simmons:1989fu}
E.H. Simmons,
\newblock Nucl. Phys. B312 (1989) 253.

\bibitem{Samuel:1990dq}
S. Samuel,
\newblock Nucl. Phys. B347 (1990) 625.

\bibitem{Kagan:1990az}
A. Kagan and S. Samuel,
\newblock Phys. Lett. B252 (1990) 605.

\bibitem{Appelquist:1989as}
T. Appelquist et~al.,
\newblock Phys. Lett. B220 (1989) 223.

\bibitem{Miranskii:1988gk}
V.A. Miranskii and K. Yamawaki,
\newblock Mod. Phys. Lett. A4 (1989) 129.

\bibitem{Matumoto:1989hf}
K. Matumoto,
\newblock Prog. Theor. Phys. 81 (1989) 277.

\bibitem{Kagan:1995qg}
A.L. Kagan,
\newblock Phys. Rev. D51 (1995) 6196, hep-ph/9409215.

\bibitem{Dobrescu:1997kt}
B.A. Dobrescu and J. Terning,
\newblock Phys. Lett. B416 (1998) 129, hep-ph/9709297.

\bibitem{Carone:1993rh}
C.D. Carone and E.H. Simmons,
\newblock Nucl. Phys. B397 (1993) 591, hep-ph/9207273.

\bibitem{Carone:1995mx}
C.D. Carone, E.H. Simmons and Y. Su,
\newblock Phys. Lett. B344 (1995) 287, hep-ph/9410242.

\bibitem{Holdom:1981rm}
B. Holdom,
\newblock Phys. Rev. D24 (1981) 1441.

\bibitem{Holdom:1985sk}
B. Holdom,
\newblock Phys. Lett. 150B (1985) 301.

\bibitem{Yamawaki:1986zg}
K. Yamawaki, M. Bando and K. iti Matumoto,
\newblock Phys. Rev. Lett. 56 (1986) 1335.

\bibitem{Appelquist:1986an}
T.W. Appelquist, D. Karabali and L.C.R. Wijewardhana,
\newblock Phys. Rev. Lett. 57 (1986) 957.

\bibitem{Appelquist:1987tr}
T. Appelquist and L.C.R. Wijewardhana,
\newblock Phys. Rev. D35 (1987) 774.

\bibitem{Appelquist:1987fc}
T. Appelquist and L.C.R. Wijewardhana,
\newblock Phys. Rev. D36 (1987) 568.

\bibitem{Pagels:1975se}
H. Pagels,
\newblock Phys. Rept. 16 (1975) 219.

\bibitem{Peskin:1982mu}
M.E. Peskin.

\bibitem{Fukuda:1976zb}
R. Fukuda and T. Kugo,
\newblock Nucl. Phys. B117 (1976) 250.

\bibitem{Higashijima:1984gx}
K. Higashijima,
\newblock Phys. Rev. D29 (1984) 1228.

\bibitem{Lane:1974he}
K. Lane,
\newblock Phys. Rev. D10 (1974) 2605.

\bibitem{Politzer:1976tv}
H.D. Politzer,
\newblock Nucl. Phys. B117 (1976) 397.

\bibitem{Cornwall:1974vz}
J.M. Cornwall, R. Jackiw and E. Tomboulis,
\newblock Phys. Rev. D10 (1974) 2428.

\bibitem{Appelquist:1988yc}
T. Appelquist, K. Lane and U. Mahanta,
\newblock Phys. Rev. Lett. 61 (1988) 1553.

\bibitem{Cohen:1989sq}
A. Cohen and H. Georgi,
\newblock Nucl. Phys. B314 (1989) 7.

\bibitem{Mahanta:1989rb}
U. Mahanta,
\newblock Phys. Rev. Lett. 62 (1989) 2349.

\bibitem{Chivukula:1993tz}
R.S. Chivukula et~al.,
\newblock Phys. Lett. B311 (1993) 157, hep-ph/9305232.

\bibitem{Chivukula:1992ap}
R.S. Chivukula, S.B. Selipsky and E.H. Simmons,
\newblock Phys. Rev. Lett. 69 (1992) 575, hep-ph/9204214.

\bibitem{LEPEWWG}
{LEP Electroweak Working Group},
\newblock {\tt http://www.cern.ch/LEPEWWG/stanmod/ppe96.ps.gz}.

\bibitem{Chivukula:1994mn}
R.S. Chivukula, E.H. Simmons and J. Terning,
\newblock Phys. Lett. B331 (1994) 383, hep-ph/9404209.

\bibitem{Chivukula:1996gu}
R.S. Chivukula, E.H. Simmons and J. Terning,
\newblock Phys. Rev. D53 (1996) 5258, hep-ph/9506427.

\bibitem{Nambu:1961er}
Y. Nambu and G. Jona-Lasinio,
\newblock Phys. Rev. 122 (1961) 345.

\bibitem{Dobrescu:1997nm}
B.A. Dobrescu and C.T. Hill,
\newblock (1997), hep-ph/9712319.

\bibitem{Hill:1995hp}
C.T. Hill,
\newblock Phys. Lett. B345 (1995) 483, hep-ph/9411426.

\bibitem{Lane:1995gw}
K. Lane and E. Eichten,
\newblock Phys. Lett. B352 (1995) 382, hep-ph/9503433.

\bibitem{Chivukula:1996cc}
R.S. Chivukula and J. Terning,
\newblock Phys. Lett. B385 (1996) 209, hep-ph/9606233.

\end{thebibliography}
\bibliographystyle{h-elsevier}

\end{document}